\documentclass[pre,aps,reprint,nofootinbib,eqsecnum, showpacs]{revtex4-1}

\usepackage{amsmath,amssymb,bm}
\usepackage[colorlinks]{hyperref}
\usepackage[all]{hypcap}
\usepackage{graphicx,psfrag,float,epsfig}
\usepackage{mathrsfs,wasysym}
\usepackage{epstopdf}
\usepackage{comment}
\usepackage{pbox}
\usepackage{array}
\usepackage{xspace}
\usepackage[usenames,dvipsnames]{color}

\DeclareMathAlphabet\mathbfcal{OMS}{cmsy}{b}{n}

\DeclareSymbolFontAlphabet{\mathrsfs}{rsfs}
\DeclareMathAlphabet{\mathcal}{OMS}{cmsy}{m}{n}

\DeclareSymbolFont{bbold}{U}{bbold}{m}{n}
\DeclareSymbolFontAlphabet{\mathbbold}{bbold}

\newcommand{\be}{\begin{equation}}
\newcommand{\ee}{\end{equation}}
\newcommand{\nn}{\nonumber}

\newcommand{\pd}{\partial}
\newcommand{\cd}{\nabla}

\def\ba{ {\bm a} }

\def\bd{ {\bm d} }
\def\bF{ {\bm F} }

\def\bx{ {\bm x} }

\def\bq{ {\bm q} }
\def\bQ{ {\bm Q} }

\def\bs{ {\bm s} }

\def\bv{ {\bm v} }
\def\bw{ {\bm w} }

\def\calE{ {\cal E} }
\def\calJ{ {\cal J} }
\def\calL{ {\cal L} }
\def\calK{ {\cal K} }

\def\calO{ {\cal O} }
\def\calQ{ {\cal Q} }
\def\calS{ {\cal S} }
\def\calT{ {\cal T} } 
\def\calV{ {\cal V} }

\begin{document}

\title{The principle of stationary nonconservative action \\for classical mechanics and field theories}

\author{Chad R.\!~Galley}
\email{crgalley@tapir.caltech.edu}
\affiliation{Theoretical Astrophysics, California Institute of
  Technology, Pasadena, California 91125, USA}

\author{David Tsang}\email{dtsang@physics.mcgill.ca}
\affiliation{Department of Physics, McGill University,
  Montreal, QC, H3A 2T8, Canada}

\author{Leo C. Stein}
\email{leostein@astro.cornell.edu}
\thanks{Einstein fellow}
\affiliation{Center for Radiophysics and Space Research, Cornell
  University, Ithaca, NY 14853, USA}


\begin{abstract}
  We further develop a recently introduced variational principle of stationary action 
  for problems in nonconservative classical mechanics and extend it to classical field theories. The variational calculus
  used is consistent with an initial value formulation of physical problems
  and allows for time-irreversible processes, such as dissipation, to be
  included at the level of the action.
In this formalism, the equations of motion are generated by
  extremizing a nonconservative action $\calS$, which is a
  functional of a doubled set of degrees of freedom.
  The corresponding nonconservative Lagrangian contains a ``potential'' $K$ which
  generates nonconservative forces and interactions.
  Such a nonconservative potential
  can arise in several ways, including from an open system
  interacting with inaccessible degrees of freedom or from
  integrating out or coarse-graining a subset of variables in
  closed systems. 
  We generalize Noether's theorem to show how
   Noether currents
  are modified and no longer conserved when $K$ is non-vanishing.
  Consequently, the nonconservative aspects of a physical system are
  derived solely from $K$.
  We show how to use the formalism with several examples of
  nonconservative actions for discrete systems including forced damped
  harmonic oscillators, radiation reaction on an accelerated charge,
  and RLC circuits. 
  We also present several examples for nonconservative classical field
  theories. We demonstrate how our
  approach
  naturally allows for irreversible thermodynamic processes to be
  included in an unconstrained variational principle for problems in fluid dynamics. We present the
  nonconservative action for a Navier-Stokes fluid including the
  effects of viscous dissipation and heat diffusion, as well as an
  action that generates the Maxwell model for viscoelastic materials,
  which can be easily generalized to more realistic rheological
  models.
  We also show that the nonconservative action has a fundamental 
  origin and can be derived as
  the classical limit of a more complete quantum theory.
\end{abstract}

\maketitle

\section{Introduction}\label{sec:Intro}

Hamilton's variational principle of stationary
action~\cite{hamilton1834general, Goldstein} is a cornerstone of
mathematical physics that allows one to find equations of motion for
many problems of varying degrees of complexity, from the simple
harmonic oscillator to complicated gauge quantum field theories.  An
action principle is a particularly useful way to formulate physical
theories for several reasons.

First, it is usually straightforward to accommodate extra degrees of 
freedom and interactions simply by adding the desired energy terms
 into the action. The interactions and forces that govern the evolution 
 of the degrees of freedom are simply determined through Hamilton's
  principle in a formulaic and practical manner. 

Second, if the system admits some set of symmetries, this must also be
reflected in the action being invariant under those transformations.
In fact, there is a very important connection between the symmetries
of an action and quantities that are conserved through the system's
evolution, which is expressed in Noether's
theorem~\cite{Noether1918}.

Third, approximations made at the level of the action\footnote{
By ``level of the action'' we mean the action, Lagrangian, and Hamiltonian 
and manipulations performed on them directly as opposed to the 
``level of the equations of motion.''} 
are often easier to implement than in the equations of motion themselves. 
For example, the effective field theory paradigm takes great advantage 
of this observation (together with the second advantage above) to help 
efficiently organize perturbative calculations that are often tedious 
when performed within the equations of motion themselves. 

Fourth, problems involving constraints are more naturally handled at
the level of the action. From the perspective of Hamiltonian
mechanics, Dirac developed an elegant theory that modifies
the Poisson brackets to preserve the symplectic structure
when constraints are applied to a system~\cite{dirac1950generalized}.

Fifth, there is  a close relationship between actions in classical and 
quantum mechanics. For example, canonical quantization entails 
promoting generalized coordinates, momenta, and the Hamiltonian to operators and
transforming Poisson brackets to the Dirac commutator while path 
integral quantization considers the action divided by Planck's constant 
$\hbar$ as the phase of unit-amplitude wavefunctions that are summed over.

Finally, all of the information about the system is contained at the level of the action, 
which is a single invariant quantity, and gives rise to a certain elegance 
to the framework of classical mechanics.

Unfortunately Hamilton's principle, actions, Lagrangians, and
Hamiltonians are generally unable to account for generic interactions
and often irreversible processes that arise from dissipation, damping,
causal history-dependence, coarse-graining, etc. These effects, and others,
are {\it nonconservative} because they cannot be derived generically
from a potential function $V$; those that can are {\it
  conservative}~\cite{Goldstein}. Nonconservative effects can include, 
  but are not limited to,
irreversible processes\footnote{See, e.g.,~\cite{CalzettaHu} for an
  excellent discussion of irreversible processes and the subtleties
  distinguishing them.} in mechanics and non-equilibrium
thermodynamics~\cite{degroot1984non}.
For example, a conservative action with unconstrained variations cannot be 
found to describe the flow of a viscous fluid nor can one be found to describe 
the radiation reaction on the accelerated motion of a charge. 

This shortcoming has led to the development of arguably {\it ad hoc} 
methods to augment Hamilton's principle for the purpose of accommodating 
some nonconservative effects. One of the notable first attempts was made by Lord Rayleigh 
in~\cite{rayleigh1896theory} who introduced a dissipation ``potential'' that is 
quadratic in the system's velocity $\vec{v}$, the derivative of which gives a damping 
force on the system that is linear in $\vec{v}$. Rayleigh's dissipation potential is not part of the 
Lagrangian or action formulation but its velocity gradient is inserted at the 
equation of motion level. However, it is well known that Rayleigh's approach 
is not sufficiently general to be useful for generic problems in nonconservative 
mechanics.\footnote{Recent work has extended Rayleigh's approach to certain nonlinear damping forces \cite{2014arXiv1406.6906V}.
}

Bauer~\cite{bauer1931dissipative} showed that a linear dissipative system 
with constant coefficients has equations of motion that cannot be derived from Hamilton's principle. 
Whereas Bauer assumed that only a single equation of motion could
arise from Hamilton's principle, Bateman~\cite{bateman1931dissipative}
allowed for a second equation of motion to appear for linear
dissipative systems.\footnote{An alternative is to adopt fractional
  derivatives (see e.g., \cite{DreisigmeyerYoung:2003}),
although we find this approach less physically intuitive and it may not be easily generalizable to generic problems.
} This second equation resulted from the
introduction of a second degree of freedom that evolved backward in
time and effectively absorbed the energy lost by the first degree of
freedom so that the two variables, considered as a whole, were energy
conserving and thus amenable to Hamilton's principle. Unfortunately,
Bateman's work seems to have been largely unexplored in part because
of the appearance of an unphysical degree of freedom that evolves
acausally as well as its specificity to linear dissipative systems.
Staruszkiewicz~\cite{1970AnP...480..362S} as well as Jaranowski and
Sch\"afer~\cite{1997PhRvD..55.4712J} have also incorporated a second
degree of freedom to be able to describe the dissipative effects of
radiation reaction on moving charges and masses,
respectively. However, they provide little justification for doing
so other than the resulting manipulations yielding the
correct damping forces.

The reason that Hamilton's action principle is not suitable for non-conservative systems 
was recently identified in~\cite{Galley:2012hx}. The underlying issue can be seen most 
transparently in conservative systems with multiple degrees of freedom (see Sec.~\ref{sec:Illustrative}
for an illustrative example). In such problems, eliminating a subset of the variables (by 
solving their respective equations of motion with initial data and substituting the corresponding 
solutions back into the action\footnote{One could simply do these manipulations at the level 
of the equations of motion but this would provide no insight into the underlying issues because 
Hamilton's principle will have already been exhausted.}) reveals the problem.
It is a generic feature of actions that the interactions they describe are time-symmetric
and necessarily energy-conserving (if the
Lagrangian does not depend explicitly on time). 
Consequently, Hamilton's principle in its current form is not compatible with systems 
displaying time-irreversible processes and, more generally, non-conservative interactions. 
Therefore, a new variational principle is required allowing one to ``break'' the time-symmetry 
manifest in the action while also being applicable to generic systems. 
Such a principle was presented in~\cite{Galley:2012hx}. 
The formalism in~\cite{Galley:2012hx} corresponds to a variational principle  specified
by initial data, which is to be compared with Hamilton's that fixes the configuration
of the system at the initial {\it and} final times.
What is remarkable is that the variational principle of~\cite{Galley:2012hx} naturally  provides a framework
 to write down actions, Lagrangians, and Hamiltonians for generic nonconservative systems 
 thus filling a long-standing gap in classical Lagrangian and Hamiltonian mechanics.

This principle of stationary nonconservative action is designed to accommodate the fact that
in many problems of physical interest there is either a natural hierarchy or a choice of observable degrees
of freedom that are accessible either to observation and measurement or to calculation.
In particular, our variational principle is able to describe the dynamics of an ``accessible,'' effectively open
subset of the degrees of freedom in a system, while still including the influence of the ``non-accessible''
variables that are not explicitly modeled by the action. The accessible degrees of
freedom may consist of the macroscopic or collective variables describing emergent structure (e.g.,~thermodynamic
variables), or degrees of freedom that are explicitly tracked or observed (e.g.,~a particle's position or
oscillator's amplitude). The non-accessible degrees of freedom may
consist of microscopic variables that have been coarse-grained away (e.g.,~the individual
positions and velocities of molecules in a fluid), untracked or unknown degrees of freedom
that couple to the accessible variables (e.g.,~the thermal degrees of freedom in a mechanical
damper), or degrees of freedom that have been integrated out due to the requirements of the
problem (e.g.,~the electromagnetic field when considering radiation reaction on a charge), or
the imprecision of measurements (e.g.,~high frequency modes in low frequency observations).
See~\cite{CalzettaHu} for more details about such separations, and
recent work in~\cite{machta2013parameter}
who refer to these as ``stiff'' and ``sloppy'' degrees of freedom.

Having an action, Lagrangian, and Hamiltonian at one's disposal for 
nonconservative problems can be extremely useful and important. 
Effective field theories~\cite{Burgess:EFT, Manohar:1996cq, Rothstein:EFT1, Goldberger:LesHouches}
are often constructed at the level of the action because
one uses the action's invariance under a set of appropriate symmetries to parametrize 
unknown interactions. 
Hence, studying the real-time dissipative processes of, for example, radiative 
systems is not possible without a framework able to incorporate generic nonconservative
interactions at the level of the action. As another example, one of our main results
in this paper is writing down actions for viscous fluids and viscoelastic flows in thermodynamic 
nonequilibrium, which includes effects from heat diffusion. From our actions, one may study important aspects 
of viscous fluid flows from a more unifying starting point using familiar methods from mechanics.

More generally, the nonconservative mechanics formalism
should be useful for any method or technique that normally uses or could
benefit from using actions, Lagrangians, and Hamiltonians. This includes studying 
the phase space of nonlinear dissipative dynamical systems, developing new
variational numerical integrators for systems with physical dissipation \cite{GalleyFuture1}, generating
partition functions for nonconservative statistical systems, as well as for 
variational problems in optimal control theory, engineering, and other non-idealized applications. In addition, the nonconservative
mechanics formalism provides what we believe is an elegant and natural
generalization of classical Lagrangian
and Hamiltonian mechanics to nonconservative systems such that
many of the tools and techniques learned for conservative problems can be
carried over to this new framework. In fact, we show in an appendix that nonconservative 
mechanics can be derived properly as the classical limit from a more fundamental quantum 
theory, which firmly roots our approach within a ``first principles'' context.

In this paper we further develop the formalism introduced in~\cite{Galley:2012hx} 
for discrete systems and extend it to classical field theories. 

We begin by reviewing the nonconservative action formalism in
Sec.~\ref{sec:Review}, using an illustrative example of coupled
oscillators to help motivate our approach (Sec.~\ref{sec:Illustrative}
\& Sec.~\ref{sec:IllustrativeRevisited}). In
Sec.~\ref{sec:NonconsDiscLag} we discuss the nonconservative potential
$K$ which, in addition to the conservative Lagrangian $L$, can be used
to formulate the nonconservative Lagrangian
$\Lambda$. The modified Euler-Lagrange equations are then derived, which
accommodate nonconservative forces from the action, $\calS = \int \Lambda dt$.

In Sec.~\ref{sec:discreteNoether} we generalize Noether's theorem and
develop a new result that shows how Noether currents evolve in the
presence of nonconservative processes. We find that the conservative
Noether currents are shifted by the effect of nonconservative
interactions and evolve in time due
to contributions from the nonconservative potential $K$. In
Sec.~\ref{sec:discreteClosure} we discuss closure conditions dictated
by the physics of each particular problem that may be needed to close
the system of equations. We consider, in particular, ``open'' system
closure conditions, for systems where energy removed from the
accessible degrees of freedom does not change the system parameters,
and ``closed'' system closure conditions for isolated systems,
which enforces energy conservation when the energy of the inaccessible
degrees of freedom can be included in the action and may
modify the system parameters (e.g., via the entropy). We then discuss systematic ways to choose
appropriate nonconservative potentials $K$ in Sec.~\ref{sec:choosingK}. 

In Sec.~\ref{sec:discreteExamples}, in order to demonstrate its broad
applicability, we apply the nonconservative formalism to several
examples including the forced damped harmonic oscillator
(Sec.~\ref{sec:ForcedDampedOscillator}), the Maxwell Element
(Sec.~\ref{sec:maxw-elem-spring}) with a closure condition (Sec.~\ref{sec:discreteMaxwellClosure}), radiation reaction of an
accelerated charge (Sec.~\ref{sec:emrr}), and RLC circuits
(Sec.~\ref{sec:Circuits}).

We generalize the nonconservative variational principle to classical
field theories in Sec.~\ref{sec:ContinuumMechanics} beginning with
the Lagrangian mechanics described by a Lagrangian density $\calL$ 
and nonconservative potential density $\calK$
(Sec.~\ref{sec:continuumLagrange}) and then generalizing Noether's
theorem to nonconservative field theories
(Sec.~\ref{sec:ContinuumNoether}). We find new expressions for the
Noether currents and their divergences that depend on contributions
from $\calK$. In Sec.~\ref{sec:continuumClosure} we discuss
closure conditions for continuum systems.

In Sec.~\ref{sec:ContinuumExamples} we provide examples of
nonconservative classical field theories, focusing in particular on
examples in continuum mechanics. We first consider two coupled
relativistic scalar fields in Sec.~\ref{sec:phichi} as a basic example
of the theory. 
We then develop
several examples in non-equilibrium hydrodynamics
(Sec.~\ref{sec:Hydro}), culminating in an action for a
Navier-Stokes fluid (Sec.~\ref{sec:Navier-Stokes}) including the
effects of both viscous dissipation and heat diffusion. We also examine 
the Stokes regime of viscous hydrodynamics (Sec.~\ref{sec:microhydro}) 
and accommodate dynamical boundaries from the surfaces of
particles composing the suspension microstructure of the fluid. Lastly, we
present an action principle for a Maxwell model of a viscoelastic fluid in
Sec.~\ref{sec:Viscoelastic}, which may be easily generalized to represent more
realistic rheological fluids. 
Mathematical notations used in these examples are
developed in App.~\ref{sec:ContinuumMechPrelim}.

In Sec.~\ref{sec:Discussion} we review our main results
and discuss the natural connection of nonconservative mechanics to the classical limit of
non-equilibrium quantum theories. We consider future avenues of
research including further development of non-conservative Hamiltonian
mechanics, numerical computing applications of this approach, and
further examples and applications in non-conservative continuum
mechanics and classical field theories. Two appendices are included
that present the incorporation of dynamical boundaries in classical field theories (App.~\ref{sec:boundaries})
as well as the fundamental observation that the action of our nonconservative mechanics
is the classical limit of a more complete quantum framework (App.~\ref{sec:quantum}).

We have intentionally prepared a rather lengthy paper to show, with some pedagogy, how
to derive the basic equations in this framework and how to apply them
to some example problems of varying difficulty. We relegate technical details
and further discussion to appendices when appropriate. The paper is written
to allow one to read the parts of interest to the reader without necessarily
having to read all of the preceding material.

\section{Review of nonconservative discrete mechanics} \label{sec:Review}

The variational principle introduced in~\cite{Galley:2012hx} is
consistent with specifying only initial data. The principle involves
formally doubling the degrees of freedom in the problem
and thus involves varying two sets of paths, one for each 
of the doubled sets variables. 
A new action can then be defined as the time integral of the Lagrangian along each path such that 
the coordinates and velocities of the two paths are
equal to each other at the final time but, importantly, not fixed to
any particular values. In the usual formulation of Hamilton's principle one is free
to add an arbitrary potential function $V$ to the Lagrangian.
Doubling the degrees of freedom has the important consequence that one
is free to include a second arbitrary function, $K$, 
that {\it couples} the two paths together. The $K$ function was shown in~\cite{Galley:2012hx}
to be responsible for generating arbitrary generalized forces in the
Euler-Lagrange equations and for determining the energy lost or gained
by the system.
Here, we expand on~\cite{Galley:2012hx} with more details and new results.

\subsection{An illustrative example} \label{sec:Illustrative}

The motivation for doubling the degrees of freedom for nonconservative systems
can be seen most clearly in the simple, solvable example of two coupled harmonic oscillators~\cite{Galley:2012hx}.
Let $q(t)$, $m$, and $\omega$ be the amplitude, mass, and natural frequency of the first
oscillator, respectively, and likewise $Q(t)$, $M$, and $\Omega$ for the second. The usual conservative action $S$
will be given by
\begin{equation}
\label{eq:twooscaction1}
\begin{aligned}
  S[ q, Q] = \int_{t_i}^{t_f} \!\!\! dt \bigg\{ &\frac{m}{2} \left( \dot{q}^2 - \omega^2 q^2 \right) + \lambda q Q \\
		& {}+ \frac{M}{2} \left( \dot{Q}^2 - \Omega^2 Q^2 \right)  \bigg\}  .
\end{aligned}
\end{equation}

The total system clearly conserves energy since
the corresponding Lagrangian is explicitly independent of time.
However, if we are only interested in the dynamics of $q(t)$,
because one lacks access to $Q$, either through ignorance or choice,
then $q$ itself is an open system that may gain or lose energy in a
nonconservative manner.

Accounting for the physical effects of $Q$ on the evolution of $q(t)$ amounts to finding solutions
to the equations of motion for $Q$ and substituting them into the action in \eqref{eq:twooscaction1}.
This process is called {\it integrating out} and results in an {\it effective action}\footnote{The effective action
is sometimes called a Fokker-type action~\cite{WheelerFeynman}.} that only depends
on $q(t)$,
\begin{equation}
\label{eq:effactiontwoosc1}
\begin{aligned}
	S_{\rm eff} [q] = {} \int_{t_i}^{t_f } & \!\!\! dt \, \bigg\{ \frac{m}{2} \left( \dot{q}^2 - \omega^2 q^2 \right) + \lambda q Q^{(h)}(t) \\
		& {} + \frac{ \lambda^2 }{ 2 M } \int_{t_i}^{t_f} \!\!\! dt' \, q(t) G_{\rm ret} (t-t') q(t') \bigg\}  .
\end{aligned}
\end{equation}

Upon integrating out $Q$ from the action, we have used initial data in the form of the retarded Green's function $G_{\rm ret}(t-t')$ and a homogeneous solution $Q^{(h)}(t)$, the precise form of which is irrelevant for our purposes here. 

Importantly, we see the factor of $q(t) q(t')$ is symmetric under $t \leftrightarrow t'$ so that only the \emph{time-symmetric} piece of the retarded Green's function contributes to the last term in the effective action of \eqref{eq:effactiontwoosc1},
\begin{align}
	 \frac{\lambda^2 }{ 2M } \int_{t_i}^{t_f} \!\!\! dt dt' \, q(t) \left[ \frac{ G_{\rm ret} (t-t') + G_{\rm adv}(t-t') }{ 2 } \right] q(t')  .
\label{eq:badterm1}
\end{align}
Here, we have used the identity $G_{\rm ret}(t'-t) = G_{\rm adv}(t-t')$ where $G_{\rm adv}(t-t')$ is the advanced Green's function. The resulting equation of motion for $q$ results from an application of Hamilton's principle of stationary action giving
\begin{align}
	m \ddot{q} + m \omega^2 q = {} & \lambda Q^{(h)} (t) + \frac{ \lambda^2}{ 2M } \int_{t_i }^{t_f} \!\!\! dt' \Big[ G_{\rm ret} (t-t') \nonumber \\
		& {\hskip0.8in} + G_{\rm adv}(t-t') \Big] q(t')  .
\end{align}
Thus we see that knowledge of the state of $q$ in the future, as
indicated by the presence of the advanced Green's function, spoils any
causal description of the oscillator. In particular, solving the
equation of motion cannot be accomplished with initial data
alone. Furthermore, the sum of the retarded and advanced Green's
functions is symmetric in time and implies that the integral accounts
for energy-conserving (i.e., conservative) interactions between $q$ and $Q$. Indeed, any
action that can be expanded in the degree(s) of freedom, such as this
example, shows that the integration kernels (which may or may not be
local in time) manifest only in a time-symmetric way:
\begin{align}
	S [ q] = {} & \int_{t_i}^{t_f} \!\!\! dt \, q(t) A(t)
			+ \int_{t_i}^{t_f} \!\!\! dt \, dt' \, \frac{1}{2!} q(t) q(t') B(t,t') \nonumber \\
			& + \int_{t_i}^{t_f} \!\!\! dt \, dt' \, dt'' \, \frac{1}{3!} q(t) q(t') q(t'') C(t,t',t'')  \nonumber \\
			& + \cdots \, .
\end{align}
That is, the quantity $B(t,t')$ is automatically symmetrized by the prefactor of $q(t)q(t')$,
which is symmetric under the exchange $q(t)\leftrightarrow
q(t')$. The same is true order by order, under all interchanges of the
integration variables.
In other words, {\it the action is too restrictive to allow anything
but time-symmetric potentials and interactions for nonconservative systems}.

A more extreme example occurs with $N \gg 1$
oscillators. For many choices of parameters, the $q$ variable would exhibit truly dissipative
dynamics for $N \gtrsim 20$, where the Poincare recurrence time is
longer than the age of the universe~\cite{Weiss}. However, integrating
out any $N$ degrees of freedom at the level of the action still
results in time symmetric and energy-conserving motion.

It should be noted that we may of course choose to work at the level of the 
equations of motion and instead integrate out $Q$ by substituting its solution 
(from initial data) into the equation of motion for $q$. Doing so gives the 
expected, causal, nonconservative dynamics for $q$. The point with this 
example is to instead understand why Hamilton's principle cannot accommodate 
the causal interactions that should dictate the evolution of an initial value problem. 

The last term in~\eqref{eq:effactiontwoosc1} simultaneously
demonstrates the problem with the usual Hamilton's principle and
suggests a solution. The reason why the advanced Green's function
makes an appearance in the effective action and equation of motion for
$q$ is because the retarded Green's function couples to the oscillator
in a time-symmetric way via $q(t) q(t')$. The solution, as given
in~\cite{Galley:2012hx}, is to ``break'' this symmetry by formally
introducing two sets of variables, $q \to (q_1, q_2)$ so that $q_1(t)
q_2(t')$ couples to the full retarded Green's function. Varying with
respect to only $q_1$, say, then gives the correct force if we set
$q_2 = q_1$ after the variation is performed. The formalism introduced
in~\cite{Galley:2012hx} develops this procedure in a formal way that
applies to a general nonconservative system.

In the next section, we review the nonconservative classical 
mechanics framework of~\cite{Galley:2012hx} and include new results 
and discussion not included in that paper. We will return to the two-oscillator problem later.

\subsection{Lagrangian mechanics}
\label{sec:NonconsDiscLag}

\begin{figure}
	\includegraphics[width=0.9\columnwidth]{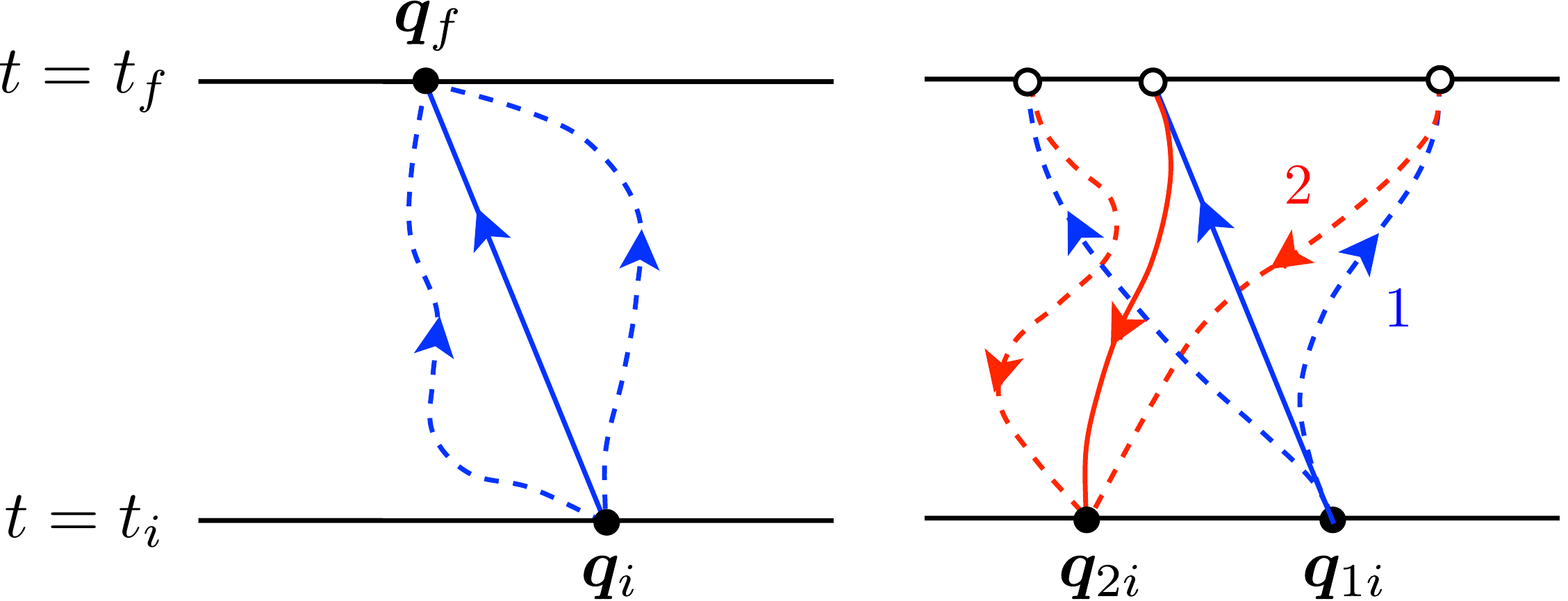}
	\caption{ 
		{\bf Left}: A schematic of a trajectory $\bq(t)$ for conservative mechanics. Dashed lines represent varied paths and the solid line indicates the path for which the action is stationary. Hamilton's principle requires variations with fixed endpoints.
		{\bf Right}: Same as the left schematic but for the doubled degrees of freedom from~\cite{Galley:2012hx}.  
		Using this variational principle allows for variations where only the initial data are specified, which is consistent with initial value problems 
		and ``breaks'' the time symmetry inherent in conservative actions.  
		In both figures, arrows indicate the direction for the time integration of the Lagrangian (i.e., the action).
\vspace{-1em}
	}
	\label{fig:histories}
\end{figure}

Let $\bq(t) = \{q^I(t)\}_{I=1}^N$ and $\dot{\bq}(t) = \{ \dot{q}^I(t)\}_{I=1}^N$ be a set of of $N$ generalized coordinates and velocities of a general dynamical system.
In the framework of~\cite{Galley:2012hx}, the degrees of freedom are formally doubled so that 
\begin{align}
	q^I(t) \to \big( q^I_1(t), q^I_2(t) \big)
\end{align}
and likewise for the velocities. In conservative mechanics, one considers 
the evolution of the system from some initial time $t=t_i$ to a final time $t=t_f$ 
so that the degrees of freedom trace out a trajectory in coordinate space, which 
is shown on the left side of~Fig.~\ref{fig:histories}. It is well-known (e.g., 
see~\cite{Goldstein}) that in some coordinate systems one can write the 
Lagrangian for the system as the difference between the kinetic energy 
and the potential energy $V$. Generally, the potential function is an arbitrary 
function of $\bq$, the gradient of which gives the conservative forces on the 
system and $L$ is an arbitrary function of $\bq$ and $\dot{\bq}$. 
The action is the time integral of the Lagrangian along a trajectory $\bq(t)$ 
that passes through $\bq(t_i) = \bq_i$ and $\bq(t_f) = \bq_f$ at the initial and final times, respectively,
\begin{align}
	S  [ \bq ] = \int_{t_i}^{t_f} \!\!\! dt \, L (\bq, \dot{\bq}, t)  .
\end{align}
The arrow on the trajectory in Fig.~\ref{fig:histories} indicates that 
the Lagrangian is integrated from the initial to the final time.

Doubling the degrees of freedom corresponds to the schematic on the 
right side of Fig.~\ref{fig:histories}. The interpretation is that the $\bq_1$ 
variable evolves from some initial value $\bq_{1i}$ at $t=t_i$ to the final 
time where upon $\bq_{1f}$ is a value determined by the  evolution, rather 
than specified with the problem. Likewise, for $\bq_2$ but taking on the value 
$\bq_{2i}$ at the initial time. The arrows on the paths, or {\it histories}, 
corresponds to the integration direction for the integral of the Lagrangian. 
In particular, the arrows should not be confused with the direction that the 
doubled variables evolve through time, which is not yet determined.

The time integral of the Lagrangian is, of course, the action. The new
action for the doubled variables, $\calS$, can be written as
\begin{align}
	\calS [ \bq_1, \bq_2 ] = {} & \int_{t_i}^{t_f} \!\!\! dt \, L(\bq_1, \dot{\bq}_1, t) + \int_{t_f} ^{t_i} \!\!\! dt \, L (\bq_2, \dot{\bq}_2, t )  \nonumber \\
		= {} & \int_{t_i}^{t_f} \!\!\! dt \, \Big[ L(\bq_1, \dot{\bq}_1, t) - L (\bq_2, \dot{\bq}_2, t ) \Big]  .
\end{align}
Notice that $\bq_1$ and $\bq_2$ are decoupled from each
other. However, as with $V$ we may add an arbitrary function, $K
(\bq_1, \bq_2, \dot{\bq}_1, \dot{\bq}_2, t)$, that {\it does} couple
the doubled variables. More generally, the action above is given by
\begin{equation}
\begin{aligned}
	\calS [ \bq_1, \bq_2 ] = {} \int_{t_i}^{t_f} dt \, \Big[ & L(\bq_1, \dot{\bq}_1, t) - L (\bq_2, \dot{\bq}_2, t ) \\
		& {} + K (\bq_1, \bq_2, \dot{\bq}_1, \dot{\bq}_2, t) \Big]\,.
\end{aligned}
\label{eq:newaction1}
\end{equation}
This form defines a Lagrangian $\Lambda (\bq_1, \bq_2, \dot{\bq}_1, \dot{\bq}_2, t )$ for the doubled degrees of freedom,
\begin{align}
	\Lambda  = L(\bq_1, \dot{\bq}_1, t) - L (\bq_2, \dot{\bq}_2, t ) + K (\bq_1, \bq_2, \dot{\bq}_1, \dot{\bq}_2, t)
\label{eq:Lambda141abc}
\end{align}
so that the action is
\begin{align}
	\calS [ \bq_1, \bq_2] = \int_{t_i}^{t_f} \!\!\! dt \, \Lambda (\bq_1, \bq_2, \dot{\bq}_1, \dot{\bq}_2, t)
\,.
\label{eq:newaction2}
\end{align}
As with conservative mechanics, the Lagrangian $\Lambda$ is an
arbitrary function of the doubled coordinates, the doubled velocities,
and possibly time. Just as one can write (in certain
coordinates~\cite{Goldstein}) the usual Lagrangian $L$ as the
difference of the kinetic and potential energies, $T - V$ where $V$ is
an arbitrary function, so one can write $\Lambda$ as the difference of
the conservative Lagrangians on the histories with an arbitrary
``potential,'' $K$, on both histories.

We can determine some basic properties that $K$ should satisfy. A
first property is that if $K$ were written as the difference of two
functions, $U(\bq_1) - U(\bq_2)$, then $U$ could be absorbed into the
potential $V$ for each doubled variable leaving $K$ zero. If $K$
vanishes for a system then there may be no need to double the
variables because the system is thus conservative.\footnote{That is,
  conservative up to any explicit time dependence in $L$.} However, one
can derive $K$ by starting from a closed system and integrating out 
(or coarse graining) some inaccessible or irrelevant
degrees of freedom to yield an open system for the accessible or
relevant variables. Therefore, $K$ describes generalized forces that
are {\it not} derivable from a potential energy (i.e., nonconservative
forces) and necessarily couples the two histories with each other. One
may thus regard $K$ as a {\it nonconservative potential}.

A second property is that $K$ must be anti-symmetric under interchanges of the labels $1 \leftrightarrow 2$. To see this, we note that the  labels $1$ and $2$ are arbitrarily assigned to the histories in the right picture of~Fig.~\ref{fig:histories}. Since the resulting physics cannot change then the action should remain invariant under this exchange up to an overall irrelevant minus sign. This implies that
\begin{align}
	\calS[ \bq_2, \bq_1] = \int_{t_i}^{t_f} dt \Big[ &- L(\bq_1, \dot{\bq}_1, t) + L (\bq_2, \dot{\bq}_2, t) \nonumber \\
		& {} + K (\bq_2, \bq_1, \dot{\bq}_2, \dot{\bq}_1, t) \Big]
\,,
\end{align}
which equals $-\calS[\bq_1, \bq_2]$ if
\begin{align}
	K (\bq_2, \bq_1, \dot{\bq}_2, \dot{\bq}_1, t) = - K (\bq_1, \bq_2, \dot{\bq}_1, \dot{\bq}_2, t)
\,.
\end{align}
Therefore, $K$ is an antisymmetric function of $\bq_1$ and $\bq_2$ and vanishes when $\bq_2 = \bq_1$.

We next find the conditions that ensure a well-defined variational principle under which the action $\calS$ is stationary. 
Both coordinate paths are parameterized as~\cite{Galley:2012hx}
\begin{align}
	\bq_{1,2} (t, \epsilon) = \bq_{1,2} (t, 0) + \epsilon \, \bm{\eta}_{1,2}(t)
\,,
\label{eq:virtual1}
\end{align}
where $\bq_{1,2}(t,0)$ are the coordinates of the two histories that makes the 
action stationary, $\bm{\eta}_{1,2}(t)$ are arbitrary functions of time denoting 
virtual displacements of the paths, and $\epsilon \ll 1$. In the usual formulation 
of Hamilton's principle, there is one path and two conditions on its displacements, 
namely, that the latter vanish at the initial and final times. Here, we have two 
paths and so we need a total of four conditions to ensure that the variational 
principle is uniquely specified. As only initial data can be given for nonconservative 
systems we require that the variation of each path vanishes at the initial time 
so that $\bm{\eta}_{1,2}(t_i) = 0$. The remaining two conditions will follow from 
the variation of the action itself. The action in~\eqref{eq:newaction2} is stationary 
under the variations in (\ref{eq:virtual1}) if 
\begin{align}
	0 = \left[ \frac{ d \calS }{ d \epsilon } \right]_{\epsilon = 0}
\,,
\label{eq:variation1}
\end{align}
which leads to 
\begin{align}
	0 = {} & \int_{t_i}^{t_f} \!\!\! dt \, \bigg\{ \eta^I_1  \bigg[ \frac{ \partial \Lambda }{ \partial q^I_1 } - \frac{ d \pi_{1I} }{ dt}  \bigg]_0 - \eta^I_2  \bigg[ \frac{ \partial \Lambda }{ \partial q^I_2 } - \frac{ d \pi_{2I} }{ dt}  \bigg] _0 \bigg\} \nonumber \\
		& +  \left[ \eta^I_1 \pi_{1I} - \eta^I_2 \pi_{2I} \right]_{t=t_f} 
\,.
\label{eq:variation2}
\end{align}
A subscript $0$ indicates that the enclosed quantity is evaluated at $\epsilon=0$.
The quantities $\bm{\pi}_{1,2}$ are the canonical momenta conjugate to the doubled coordinates $\bq_{1,2}$ and defined through the nonconservative Lagrangian $\Lambda$ to be
\begin{align}
	\pi_{1I} (\bq_{1,2}, \dot{\bq}_{1,2} ) \equiv {} & \frac{ \partial \Lambda }{ \partial \dot{q}_1^I (t) } = \frac{ \partial L(\bq_1, \dot{\bq}_1) }{ \partial \dot{q}_1^I (t) }  + \frac{ \partial K }{ \partial \dot{q}_1^I(t) }
\,,
\end{align}
where the first term on the far right side is the (conservative) conjugate momentum from $L$ (usually called $p$) and the second term is the part of the total momentum that comes from nonconservative interactions via $K$.
Similarly, the momentum for the second history is
\begin{align}
	\pi_{2I} (\bq_{1,2}, \dot{\bq}_{1,2} ) \equiv {} & - \frac{ \partial \Lambda }{ \partial \dot{q}_2^I (t) } = \frac{ \partial L(\bq_2, \dot{\bq}_2) }{ \partial \dot{q}_2^I (t) }  - \frac{ \partial K }{ \partial \dot{q}_2^I(t) } .
\end{align}

The last line in (\ref{eq:variation2}) comes from integration by parts and will vanish if 
\begin{align}
	\eta^I_1(t_f) \pi_{1I} (t_f)  = \eta^I_2(t_f) \pi_{2I} (t_f) \,.
\end{align}
From Fig.~\ref{fig:histories} we see that the variations at the final time are equal to each other so that
\begin{align}
	\bm{\eta}_1(t_f) = \bm{\eta}_2 (t_f)\,,
\label{eq:eqcond1a}
\end{align}
and thus the final conjugate momenta are also equal,
\begin{align}
	\bm{\pi}_1(t_f) = \bm{\pi}_2(t_f) \,.
\label{eq:eqcond1b}
\end{align}
These two conditions together constitute the {\it equality
  condition} \cite{Galley:2012hx}. The equality condition ensures that the boundary term
from integration by parts in (\ref{eq:variation2}) will vanish for
arbitrary variations provided only that the two histories agree with
each other at the final time. The actual values of the variations are
free, but whatever they are, the final states match at
$t_{f}$. Likewise, the conjugate momenta at the final time must agree
with each other but are otherwise unspecified. 
This leads to the key point that the equality condition ensures that our variational principle
is consistent with our ignorance of the final state of the accessible degrees of freedom.
Indeed, it is unsatisfactory to fix the final configuration of the system in order to determine the
equations of motion that are to be solved from initial data alone.
The equality condition is thus a crucial ingredient in extending Hamilton's principle to
nonconservative systems.

The equations for both histories then follow by setting the integrand in (\ref{eq:variation2}) to zero for arbitrary variations, $\eta_{1,2}^I$, which gives the two equations
\begin{align}
	\frac{d  }{ dt } \frac{ \partial \Lambda }{ \partial \dot{q}^I_a} = \frac{ \partial \Lambda }{ \partial q^I_a }
\label{eq:nonconsEL1}
\end{align}
where $a=1,2$. However, the resulting equations are not necessarily physical until we take the {\it physical limit} (PL) wherein the histories are identified
\begin{align}
	q^I_1 = q^I_2 = q^I ~, ~~ \dot{q}^I_1 = \dot{q}^I_2 = \dot{q}^I
\label{eq:physlim1}
\end{align}
after all variations and derivatives of the Lagrangian are taken. 
The PL of both equations in (\ref{eq:nonconsEL1}) reduces to 
\begin{align}
\begin{split}
	\frac{ d}{ dt} \frac{ \partial L }{ \partial \dot{q}^I } - \frac{ \partial L }{ \partial q^I } = {} &  \left[ \frac{ \partial K }{ \partial q^I_1} - \frac{ d }{ dt } \frac{ \partial K }{ \partial \dot{q}^I_1} \right]_{PL} {\hskip-0.2in}  \\
		= {} & -\left[ \frac{ \partial K }{ \partial q^I_2} - \frac{ d }{ dt } \frac{ \partial K }{ \partial \dot{q}^I_2} \right]_{\rm PL}
\label{eq:nonconsEL2}
\end{split}
\end{align}
where we have written the nonconservative Lagrangian $\Lambda$ in terms of its conservative piece $L$ and the $K$ function, and ``PL'' denotes taking the physical limit in (\ref{eq:physlim1}). 
The right side on the first line comes from the physical limit of the $a=1$ equations while the second line from $a=2$. Notice that these differ by an overall minus sign but are equal because $K$ is antisymmetric under interchanges of the labels $1 \leftrightarrow 2$.

A more convenient parametrization of the coordinates that yields some important physical insight is given by the average and relative difference of the two histories,
\begin{align}
	q_+^I & \equiv \frac{ q_1^I + q_2^I }{ 2 } , \\
	q_-^I & \equiv q_1^I - q_2^I .
\end{align}
The physical limit is then simply given by
\begin{align}
	q_+^I \to q^I ~, ~~ q_-^I \to 0
\label{eq:physlim2}
\end{align}
Therefore, the average history is the physically relevant one that survives the physical limit 
while the difference coordinate simply vanishes. In these coordinates, the nonconservative Lagrangian is
\begin{align}
	\Lambda = \Lambda ( \bq_+, \bq_-, \dot{\bq}_+, \dot{\bq}_-, t)  .
\end{align}
It should be noted that $\Lambda$ cannot be written in the $\pm$ parametrization 
 as in (\ref{eq:Lambda141abc}) but can be derived simply from (\ref{eq:Lambda141abc}).
The equality condition in (\ref{eq:eqcond1a}) and (\ref{eq:eqcond1b}) is simply
\begin{align}
	\bm{\eta}_-(t_f) = 0 ~, ~~ \bm{\pi}_{-} (t_f) = 0\,
\label{eq:eqcond2}
\end{align}
implying that the physically relevant average ($+$) quantities are {\it not specified} at the final time in order to have a well-defined variational principle. Here, 
\begin{align}
	\pi_{+I} & = \frac{ \pi_{1I} + \pi_{2I} }{ 2 } = \frac{ \pd \Lambda }{ \pd \dot{q}^I_- } ~~,~~
	\pi_{-I}  = \pi_{1I} - \pi_{2I} = \frac{ \pd \Lambda }{ \pd \dot{q}^I_+}  . \nn
\end{align}
The resulting equations of motion are easily found to be~\cite{Galley:2012hx}
\begin{align}
	\frac{ d }{ dt} \frac{ \partial \Lambda }{ \partial \dot{q}_a^I } = \frac{ \partial \Lambda }{ \partial q_a^I }\,,
\label{eq:nonconsEL3}
\end{align}
where now $a = +,-$. Notice that this expression has the same form as in the 
$1,2$ parametrization in (\ref{eq:nonconsEL1}). This reflects a more general 
result that the nonconservative Euler-Lagrange equations are covariant 
(which is also true in conservative Lagrangian mechanics~\cite{Goldstein, Scheck}) 
with respect to the history indices. Therefore, the form of the nonconservative 
Euler-Lagrange equations does not depend on the specific choice of history labels.

Taking the physical limit of (\ref{eq:nonconsEL3}) is trivial. For $a=+$ we have that (\ref{eq:nonconsEL3}) is identically zero in the physical limit while the $a=-$ equations survive
\begin{align}
	\bigg[ \frac{ d }{ dt} \frac{ \partial \Lambda }{ \partial \dot{q}_-^I } - \frac{ \partial \Lambda }{ \partial q_-^I }\,, \bigg]_{\rm PL} = 0  .
\end{align}
Expressed in terms of $L$ and $K$ this yields
\begin{align}
	\frac{ d }{ d t} \frac{ \partial L }{ \partial \dot{q}^I } - \frac{ \partial L }{ \partial q^I } = \left[ \frac{ \partial K }{ \partial q_-^I } - \frac{d }{dt} \frac{ \partial K }{ \partial \dot{q}_-^I }  \right]_{\rm PL}  {\hskip-0.15in} \equiv Q_I (q^J, \dot{q}^J, t)\,.
\label{eq:nonconsEL4}
\end{align}
Here, $Q^I$ is the generalized nonconservative force derived from $K$. The structure of $Q^I$ suggests that $K$ is a nonconservative potential function. Equation~(\ref{eq:nonconsEL4}) is the Euler-Lagrange equations of motion for the nonconservative dynamics of $\bq(t)$, as derived with $\calS$
from the variational principle introduced in~\cite{Galley:2012hx} consistent with giving {\it initial} data.
We remark that these equations of motion are the same whether we choose to parametrize the histories
by $\{1,2\}$, $\{+,-\}$, or another pair of labels because the action $\calS$ is invariant.
We end by pointing out that (\ref{eq:nonconsEL4}) can also be derived by computing
\begin{align}
	0 = \bigg[ \frac{ \delta \calS }{ \delta q_-^I (t) } \bigg]_{\rm PL}
\label{eq:nonconsEL151}
\end{align}
where $\delta / \delta q_-^I(t)$ is a functional derivative with respect to $q_-^I(t)$.
This expression is the more general form when higher than first derivatives of $\bq_-$ appear in a problem. 
We discuss in Appendix \ref{sec:higherders} how the variational principle of stationary nonconservative action changes when 
higher time derivatives are present.

\subsection{Illustrative example revisited} \label{sec:IllustrativeRevisited}

With the framework in place to properly incorporate initial data and
causal dynamics into a variational principle for the nonconservative action, let us
now revisit the example from Sec.~\ref{sec:Illustrative} and check
that this formalism gives the correct equations of motion for $q(t)$
after eliminating $Q(t)$ from the original action. We assume initial
conditions are given for both oscillators, namely,
\begin{align}
	q(t_i) = q_i ~~&{\rm and} ~~ \dot{q}(t_i) = v_i  , \\
	Q(t_i) = Q_i ~~&{\rm and} ~~ \dot{Q}(t_i) = V_i   .
\label{eq:initialQ1}
\end{align}
The two oscillators taken together form a closed system, which conserves
the total energy, with an action given in the usual mechanics
formalism by \eqref{eq:twooscaction1}. We next integrate out $Q$ from
the action. However, we first double the degrees of freedom in the
problem in order to ensure that the proper causal conditions on the
dynamics of $Q$ are respected and maintained. The resulting
nonconservative action in the $\pm$ basis is
\begin{align}
	{\hskip-0.08in} \calS [ q_\pm, Q_\pm ] = {} & \!\! \int_{t_i}^{t_f} \!\!\! dt \, \bigg\{ m \dot{q}_- \dot{q}_+ - m \omega^2 q_- q_+ + \lambda q_- Q_+ \nonumber \\
		& \!\! \!\!+ \lambda q_+ Q_- \! + \! M \dot{Q}_- \dot{Q}_+ \!\! - \! M \Omega^2 Q_- Q_+ \bigg\}\,.
\label{eq:Stwoosc1111}
\end{align}
The fact that the total system is closed means that $K$ vanishes. Integrating out $Q$ will turn out to generate a non-zero effective $K$ for the open system dynamics of $q$.

The effective action for the open dynamics of $q$ is found by eliminating the $Q_\pm$ variables from (\ref{eq:Stwoosc1111}). The $Q_\pm$ satisfy
\begin{align}
	M \ddot{Q}_\pm  + M \Omega^2 Q_\pm = \lambda q_\pm  .
\label{eq:Qpmeom1}
\end{align} 
We associate the initial conditions in (\ref{eq:initialQ1}) with the initial conditions for $Q_+$ because the physical limit of $Q_+(t_i)$ at the initial time is just given by (\ref{eq:initialQ1}). The equality condition at the final time gives ``final'' conditions for $Q_-$ given by
\begin{align}
	Q_- (t_f) = \dot{Q}_- (t_f) = 0
\end{align}
The resulting solutions to (\ref{eq:Qpmeom1}) are thus 
\begin{align}
	Q_+ (t) &= Q^{(h)} (t) + \frac{\lambda }{ M } \int_{t_i}^{t_f} \!\!\! dt' \, G_{\rm ret} (t-t') q_+(t') , 
\label{eq:Qplussolution1} \\
	Q_- (t) & = \frac{ \lambda }{ M } \int_{t_i}^{t_f} \!\!\! dt \, G_{\rm adv} (t-t') q_- (t')
\label{eq:Qminussolution1} 
\end{align}
where $Q^{(h)}(t) = Q_i \cos \Omega (t-t_i) + V_i / \Omega \sin \Omega (t-t_i)$ is the homogeneous solution to the $Q_+$ equation. The retarded Green's function is given by
\begin{align}
	G_{\rm ret} (t-t') = \theta(t-t') \frac{ \sin \Omega (t-t') }{ \Omega } 
\end{align}
where $\theta(t)$ is the Heaviside step function, and the advanced Green's function is related through $G_{\rm adv}(t-t') = G_{\rm ret}(t'-t)$.
 Notice that because the solution to the (physical) $Q_+$ equation satisfies initial data while the solution to the (unphysical) $Q_-$ equation satisfies final data then the former evolves {\it forward} in time while the latter evolves ${\it backward}$, hence the appearance of the advanced Green's function. This is a general feature of the $\pm$ parametrization. We comment that 
 $Q$ in the $1,2$ parametrization evolve \emph{both} forward and backward in time, as can be easily shown from (\ref{eq:Qplussolution1}) and (\ref{eq:Qminussolution1}).

Substituting (\ref{eq:Qplussolution1}) and (\ref{eq:Qminussolution1}) back into the action in (\ref{eq:Stwoosc1111}) yields the effective action for $q_\pm (t)$,
\begin{align}
	\calS_{\rm eff} [ q_\pm ] = {} & \int_{t_i}^{t_f} \!\!\! dt \, \bigg\{ m \dot{q}_- \dot{q}_+ - m \omega^2 q_- q_+ + \lambda q_- Q^{(h)} \nonumber \\
		& {\hskip0.15in} + \frac{\lambda^2 }{  M } \int_{t_i}^{t_f} \!\!\! dt' \, q_- (t) G_{\rm ret} (t-t') q_+(t') \bigg\}\,,  \nn
\end{align}
from which we  read off that
\begin{align}
	L  & = \frac{1}{2} m \dot{q}^2 - \frac{ 1}{2} m \omega^2 q^2 \\
	K & = \lambda q_- Q^{(h)}(t) + \frac{\lambda^2 }{  M } \int_{t_i}^{t_f} \!\!\! dt' \, q_- (t) G_{\rm ret} (t-t') q_+(t') . \nn
\end{align}
Comparing with the effective action constructed using the usual Hamilton's principle in (\ref{eq:effactiontwoosc1}) reveals that the last term above contains a factor $q_- (t) q_+(t')$ that is not symmetric in $t \leftrightarrow t'$ and thus couples to the full retarded Green's function, not just the time-symmetric piece. The resulting equations of motion for $q(t)$ follows 
from (\ref{eq:nonconsEL4}) or (\ref{eq:nonconsEL151}) and gives
\begin{align}
	m \ddot{q} + m \omega^2 q = \lambda Q^{(h)}(t) + \frac{ \lambda^2 }{ M} \int _{t_i}^{t_f} \!\!\! dt' \, G_{\rm ret} (t-t') q(t') .
\label{eq:qeomcorrect1}
\end{align}
This is the correct equation of motion for $q$, which can
be easily verified by eliminating $Q$ at the level of the equations of
motion instead of at the level of the action. 
Importantly, solutions to (\ref{eq:qeomcorrect1}) evolve
causally from initial data and only the retarded Green's function
appears in the equation. Therefore, we have demonstrated that the
nonconservative generalization of Hamilton's principle presented in
Sec.~\ref{sec:NonconsDiscLag} gives the proper causal evolution for
the open system given only initial data and does not require fixing
the configuration of the degrees of freedom at the final time in order
to define the variational principle~\cite{Galley:2012hx}.

\subsection{Noether's theorem generalized}
\label{sec:discreteNoether}

In conservative Lagrangian mechanics, 
Noether's theorem~\cite{Noether1918} states that there exists a quantity conserved 
in time for every continuous transformation that keeps the action invariant when the Euler-Lagrange equations are satisfied. 
For example, time translation invariance and rotational invariance  give rise to 
energy and angular momentum conservation, respectively. 
Quantities that are conserved in conservative mechanics may no longer be  
when considering open systems subject to nonconservative interactions.  
Therefore, Noether's theorem must be modified.
Nevertheless, the corresponding conservative action $S = \int dt \, L$ is still
invariant under the original continuous transformations, and thus generates
the same Noether currents. However, because the Euler-Lagrange equations
are generally sourced by nonconservative forces in \eqref{eq:nonconsEL4}, then
one will expect the Noether currents to change in a manner depending on $K$. 
In this section, we show that this is indeed generally the case. More importantly, 
because we know how the nonconservative forces are derived from $K$, we will 
find very useful expressions for the Noether currents and their changes in time that follow directly from $K$. 

Consider the conservative action given by
\begin{align}
	S = \int_{t_i}^{t_f} dt \, L \big( \bq(t), \dot{\bq}(t), t \big)  .
\label{eq:noetherAction1}
\end{align}
Let us assume that $S$ is invariant under
the following infinitesimal transformations
\begin{align}
	t & \to t' = t + \delta t 
\label{eq:dt1} \\
	q^I(t) & \to q'{}^I(t') = q^I (t) + \dot{q}^I(t) \delta t + \delta q^I(t) 
\label{eq:dq1}
\end{align}
with
\begin{align}
	\delta q^I(t) =  \epsilon^a \left[ \frac{ \partial q^I (t) }{ \partial \epsilon^a } \right]_{\epsilon^a = 0} \equiv \epsilon^a \, \omega^I_a (t) \,,
\label{eq:deltaq137}
\end{align}
where $\epsilon^a$ is a small parameter with index $a$ associated with (the Lie algebra of) the symmetry group, 
which should not be confused with the history labels that will not appear in this section.
Under these transformations the conservative action takes the form
\begin{align}
	S = \int_{t_i - \delta t} ^{t_f - \delta t } \!\!\! dt' \, L \big( \bq'(t'), \dot{\bq}' (t'), t' \big)
\,,
\label{eq:noetherAction2}
\end{align}
which, by assumption, equals to the right side of (\ref{eq:noetherAction1}) so that
through first order in $\delta t$ and $\epsilon^a$ the change in the conservative action is
\begin{align}
	\delta S = {} & 0 = \int dt \, \bigg\{ \delta t \left[ \frac{ \partial L }{ \partial t } + \dot{q}^I \frac{ \partial L }{ \partial q^I } + \ddot{q}^I \frac{ \partial L }{ \partial \dot{q}^I } - \frac{ d L}{ dt } \right]  \nonumber \\
		& {\hskip0.4in} + \epsilon^a \left[ \omega^I_a \frac{ \partial L }{ \partial q^I} + \dot{\omega}^I_a \frac{ \partial L}{ \partial \dot{q}^I }   \right] \bigg\}  .
\label{eq:deltaSnoether1}
\end{align}
Applying the product rule and rearranging gives
\begin{align}
	 & {\hskip-0.1in} 0 = \!\!  \int \! \! dt \bigg\{ \delta t \bigg[ \frac{ d }{ dt } \bigg( \! \dot{q}^I \frac{ \partial L }{ \partial \dot{q}^I }  - L \bigg) \! + \! \frac{ \partial L}{ \partial t} - \dot{q}^I \! \bigg( \frac{ d }{ dt} \frac{ \partial L}{ \partial \dot{q}^I } - \frac{ \partial L}{ \partial q^I } \bigg) \! \bigg]  \nonumber \\
		&~~~  + \epsilon^a \bigg[ \frac{ d}{dt} \bigg( \omega^I_a \frac{ \partial L}{ \partial \dot{q}^I } \bigg) - \omega^I_a \bigg( \frac{d }{ dt} \frac{ \partial L }{ \partial \dot{q}^I} - \frac{ \partial L }{ \partial q^I} \bigg) \bigg] \bigg\}  .
\label{eq:deltaSnoether3}
\end{align}
The first term is the total time derivative of the quantity
\begin{align}
	E (\bq, \dot{\bq}, t) \equiv \dot{q}^I \frac{ \partial L}{ \partial \dot{q}^I } - L \,,
\label{eq:Echarge1}
\end{align}
which is the value of the Hamiltonian and is called the {\it energy
  function}~\cite{Goldstein}. The first term on the second line of
(\ref{eq:deltaSnoether3}) is the current associated with the
transformation in (\ref{eq:dq1}),
\begin{align}
	J_a (\bq, \dot{\bq}, t) \equiv \omega^I_a \frac{ \partial L}{ \partial \dot{q}^I } = \omega^I_a p_I (\bq, \dot{\bq}, t) .
\label{eq:Jcurrent1}
\end{align}
We may now use the Euler-Lagrange equations in (\ref{eq:nonconsEL4})
to write (\ref{eq:deltaSnoether3}) in terms of $E$, $J_a$, and the
non-conservative forces $Q_{I}$ as
\begin{align}
	0 = {} & \int dt \bigg\{ \delta t \bigg[ \frac{ d E}{ dt} + \frac{ \partial L}{ \partial t} - \dot{q}^I Q_I \bigg] - \epsilon^a \bigg[ \frac{ d J_a }{ dt}  - \omega^I_a Q_I \bigg] \bigg\}  .
\end{align}
Finally, since $\delta t$ and $\epsilon^a$ are independent then each factor in square brackets must vanish for the whole integral to vanish. The result is 
\begin{align}
	\frac{ d E}{ dt} & = - \frac{ \partial L}{ \partial t} + \dot{q}^I Q_I 
\label{eq:dEdt1} \\
	\frac{ dJ_a }{ dt} & = \omega^I_a  Q_I   ,
\label{eq:dJdt1}
\end{align}
which follows from the invariance of the {\it conservative} action under the 
transformations in (\ref{eq:dt1}) and (\ref{eq:dq1}). In this sense, (\ref{eq:dEdt1}) 
and (\ref{eq:dJdt1}) can be viewed as a generalization of Noether's theorem 
where instead of a conserved set of currents we have a set of equations that 
determines how these currents change with time in the presence of nonconservative 
forces and interactions. If $K$ is nonzero and the Lagrangian has no explicit time 
dependence then the energy $E$ and current $J_a$ will necessarily change in time. 
If $K$ vanishes then the energy and current are conserved in time and we recover 
Noether's theorem for discrete mechanical systems. The derivation of (\ref{eq:dEdt1}) 
and (\ref{eq:dJdt1}) does not rely on the new framework discussed in 
Sec.~\ref{sec:Review} and is not new. What is new is that we
know how $Q_I$ depends on the nonconservative interactions in the
action via $K$ from (\ref{eq:nonconsEL4}). This relation allows us to
provide powerful alternative but equivalent expressions for
(\ref{eq:dEdt1}) and (\ref{eq:dJdt1}).

We start with the energy equation in (\ref{eq:dEdt1}) and write out the 
nonconservative force $Q_I$ from (\ref{eq:nonconsEL4}) explicitly in terms of $K$ as
\begin{align}
	\frac{ d E }{ dt } = - \frac{ \partial L}{ \partial t} + \dot{q}^I \bigg[ \frac{ \partial K }{ \partial q_-^I } - \frac{ d }{ dt } \frac{ \partial K }{ \partial \dot{q}^I_- }  \bigg] _{\rm PL}.
\label{eq:Edot113}
\end{align}
Next, we note that the total canonical momentum $\pi_I$ is given by
\begin{align}
	\pi_I (\bq, \dot{\bq}, t) \equiv [ \pi_{+I} ] _{\rm PL} = \bigg[ \frac{ \partial \Lambda }{ \partial \dot{q}^I_- } \bigg]_{\rm PL} = \frac{ \partial L}{ \partial \dot{q}^I } + \bigg[ \frac{ \partial K }{ \partial \dot{q}^I_- } \bigg]_{\rm PL}.
\end{align}
The first term on the far right side is familiar as the part of the total canonical 
momentum that is associated with conservative actions
while the second term is the part of $\pi_I$ that comes from the nonconservative interactions of the system,
\begin{align}
	\kappa_I (\bq, \dot{\bq}, t) \equiv [\kappa_{+I} ] _{\rm PL} = \bigg[ \frac{ \partial K }{ \partial \dot{q}^I_- } \bigg]_{\rm PL}.
\label{eq:kappa1}
\end{align}
Noting that $\dot{q}^I = [ \dot{q}_+^I ] _{\rm PL}$ we can bring the velocity in (\ref{eq:Edot113}) into the square brackets. Doing so and using the product rule for the time derivative in the last term gives 
\begin{align}
	\frac{ d}{dt} \left( E + \dot{q}^I \kappa_I  \right) = - \frac{ \partial L }{ \partial t} + \dot{q}^I \bigg[ \frac{ \partial K }{ \partial q_-^I } \bigg]_{\rm PL} {\hskip-0.15in} + \ddot{q}^I \kappa_I,
	\label{eq:dEdt2}
\end{align}
after some rearranging. 
Similar manipulations turn the current equation in (\ref{eq:dJdt1}) into
\begin{align}
	\frac{ d }{ dt } \left( J_a +  \omega^I_a \kappa_I \right) = \omega^I_a \bigg[ \frac{ \partial K }{ \partial q_-^I } \bigg]_{\rm PL} +  \dot{\omega}^I_a \kappa_I .
\label{eq:dJdt2}
\end{align}
The left sides of both (\ref{eq:dEdt2}) and (\ref{eq:dJdt2}) are total time derivatives of a shifted energy and current,
\begin{align}
	{\cal E} & \equiv E +  \dot{q}^I \kappa_I = \pi_I \dot{q}^I - L,
\label{eq:Echarge2} \\
	{\cal J}_a & \equiv J_a +  \omega^I_a \kappa_I = \pi_I \omega^I_a.
\label{eq:Jcharge2}
\end{align}
The contributions that come from $\kappa_I$ are corrections to the
energy and current that result from the open system's interaction with
the inaccessible or eliminated degrees of freedom. We can regard
${\cal E}$ and ${\cal J}_a$ as the total energy and current of the
accessible degrees of freedom including contributions from the
nonconservative interactions. We will see a familiar example from
radiation reaction in electrodynamics that confirms this
interpretation in Sec.~\ref{sec:emrr}.
Our alternative expressions of Noether's theorem generalized to nonconservative systems are thus given by
\begin{align}
	\frac{d {\cal E} }{ dt } & = - \frac{ \partial L}{ \partial t } + \dot{q}^I \bigg[ \frac{ \partial K }{ \partial q_-^I } \bigg]_{\rm PL} {\hskip-0.15in} + \ddot{q}^I \kappa_I,
\label{eq:Echarge3} \\
	\frac{ d {\cal J}_a }{ dt } & = \omega^I_a \bigg[ \frac{ \partial K }{ \partial q_-^I } \bigg]_{\rm PL} +  \dot{\omega}^I _a \kappa_I,
\label{eq:Jcharge3}
\end{align}
and results directly from the generalized nonconservative forces $Q_I$ being expressed in terms of a known and/or derived $K$. 

Equations (\ref{eq:Echarge2})--(\ref{eq:Jcharge3}) constitute some of the main results of this paper.
These expressions indicate several interesting consequences. The first indicates that the {\it total} energy and Noether current of the accessible degrees of freedom include contributions from the nonconservative momentum $\kappa_I$.  The second is that when $\kappa_I$ is non-zero, the change in energy necessarily depends on the acceleration of the accessible variables, as seen in the last term of  \eqref{eq:dEdt2}.
Another key point is that  (\ref{eq:Echarge2})--(\ref{eq:Jcharge3}) are computed {\it directly} from the nonconservative potential $K$. Therefore, once $K$ is known then one can directly calculate how the energy and, for example, angular momentum of the system changes in time without having to perform separate calculations to explicitly compute these quantities.

As a final comment, if $K$ depends explicitly on $\bq_a$
then there can be an ambiguity concerning the interpretation of (\ref{eq:Echarge2}) as the total 
energy of the accessible degrees of freedom. This is best seen with a simple example. 
Choose $K = \alpha q^I_- \dot{q}_{+I}$, for constant $\alpha$, and $K' = - \alpha \dot{q}^I_- q_{+I}$, 
which is related to $K$ through integration by parts in the nonconservative action $\calS$.
One can show that the equations of motion are the same and the content of Noether's theorem in (\ref{eq:Echarge3}) is the same using either $K$ or $K'$.
However, the nonconservative conjugate momentum from $K$ vanishes ($\kappa_I = 0$) while $K'$ has $\kappa'_I = - \alpha q_I$. 
Therefore, the corresponding energies $\calE = E$ and $\calE' = E + \dot{q}^I \kappa'_I$ are different.
The former is the correct expression for interpreting (\ref{eq:Echarge2}) as the total energy of the accessible degrees of freedom.
However, one should recall that $\bq_+(t)$ is often only just a coordinate, as opposed to a geometric quantity like a vector or a tensor. Hence, $K'$ may not have the correct transformation properties whereas $K$ will. 
Therefore, to alleviate any ambiguity we believe it is useful to use the form of $K$, not $K'$, for problems where the nonconservative potential depends explicitly on $\bq_-$ and/or $\bq_+$. Such systems are studied in Secs.~\ref{sec:ForcedDampedOscillator}--\ref{sec:discreteMaxwellClosure}. In all other cases when $K$ depends on time derivatives of $\bq_a$ then there is no such ambiguity (see Appendix \ref{sec:higherders} for including higher time derivatives in the formalism) as demonstrated in Sec.~\ref{sec:emrr}.
When integrating out a subset of variables from the full conservative problem, the resulting nonconservative potential $K$ will typically be of the form such that the shifted Noether currents are consistent with the appropriate physical interpretation.

\subsection{Internal energy and closure conditions}
\label{sec:discreteClosure}
Until now we have been considering discrete mechanical ``open'' systems 
where the energy dissipated into the inaccessible degrees of freedom does 
not feed back on the dynamics of the accessible variables. 
We may instead consider systems where such feedback could occur and 
affect the parameters of the accessible subsystem. A damped harmonic
oscillator, for example, may heat up through friction and its natural frequency 
of oscillation or damping rate may change with the oscillator's temperature. 
In this scenario, the energy lost from the accessible degrees of freedom by 
damping the oscillator goes into exciting the inaccessible microscopic degrees of freedom
composing the oscillator.

If the ``internal'' energy of the inaccessible degrees of freedom plays a role in 
the dynamics of the accessible variables, then the equations of motion alone 
do not provide a closed set of equations because the internal energy evolution 
remains undetermined. We require additional information specified by the problem 
at hand to close the system of equations. 

Systems where the energy of all inaccessible degrees of freedom can be accounted 
for by an internal energy carried within the Lagrangian can be considered ``closed'' 
such that the \emph{total energy} $\calE$ is conserved.
This {\it closure condition}, $d{\cal E}/dt = 0$, also closes the system of equations,  telling us how the internal 
energy must change with time so that all the energy transferred to and from the 
accessible degrees of freedom are accounted for via energy conservation. 

In examples in thermal systems, it is convenient to parametrize the
internal energy of the inaccessible degrees of freedom by a
time-dependent \emph{entropy}. We will discuss such closure relations
below for examples in both discrete
(Sec.~\ref{sec:discreteMaxwellClosure}) and continuum mechanics
(Sec.~\ref{sec:InsulatingFluid}). Such an approach is particularly
useful in closed fluid systems, as we will see in
Sec.~\ref{sec:Hydro}.

Other conditions to close the system of equations are possible, though 
these in practice will depend on the specifics of particular systems 
(see Sec. \ref{sec:discreteMaxwellClosure} for an example including
an external force). Since these closure conditions describe the energy evolution of 
the inaccessible degrees of freedom, they must be in general specified in 
\emph{addition} to the variational principle that describes the accessible dynamics. 
Closure conditions are best illustrated through examples, as they depend on the
the systems being modelled, as we will see in 
Sec.~\ref{sec:discreteExamples} and \ref{sec:ContinuumExamples}.

\subsection{On choosing $K$}
\label{sec:choosingK}

How does one choose or find the nonconservative potential $K$ for a problem 
of interest? There are several ways to answer this question. The particular answer 
one might choose will depend on the problem and its setup.

First, this question is similar to ``How does one choose the conservative potential 
$V$?'' in conservative mechanics. In some problems, one is either given a $V$ or 
one chooses the potential such that its gradient gives the desired force. Indeed, 
the situation is similar for $K$. One may blindly prescribe $K$, 
motivate the form of $K$ through some physical reasoning,
 or choose $K$ such that its derivatives in (\ref{eq:nonconsEL4}) 
give the desired nonconservative force.

Second, {\it if} one knows the nonconservative forces that appear in the equations of 
motion then one can reconstruct the corresponding $K$, at least partially. This 
approach is useful for ``non-Lagrangian'' (or ``non-Hamiltonian'') forces where a 
conservative Lagrangian (or Hamiltonian) cannot be found to generate some of the 
forces on the system. For example, note from (\ref{eq:nonconsEL4}) that $Q_I$ is given 
by the pieces of $K$ linear in $q_-^I$ and $\dot{q}_-^I$. 
In particular, if $Q(\bq, \dot{\bq}, t)$ is the nonconservative force on the system than simply writing $K$ as
\begin{align}
	K ( \bq_a, \dot{\bq}_a) = q_-^I Q_I (\bq_+, \dot{\bq}_+, t) + \calO(-^3)
\end{align}
guarantees that the correct force enters the Euler-Lagrange equations of motion in (\ref{eq:nonconsEL4}).
Notice that we can only gain partial information about $K$ because we do not 
determine the higher order terms in the ``$-$'' variables in this way. 

Third, one may interpret the $q_-^I$ variables (when they are small) as being 
like virtual displacements. One can then regard $K$ as the virtual work done 
on the system by the inaccessible/irrelevant variables (whatever they may be) 
when displacing the system (evaluated in $q_+$ variables) through $q_-$ 
and/or $\dot{q}_-$. We use this approach in several examples below. 
We recall here the ambiguity associated with interpreting $\calE$
as the total energy of the accessible degrees of freedom for systems where $K$
depends explicitly on the generalized coordinates, $\bq_a(t)$, discussed in Sec.~\ref{sec:discreteNoether}.

Fourth, $K$ can be {\it derived} by integrating out a subset of degrees of freedom 
from a larger closed system. We showed an example of this already with the two
coupled oscillators in Sec.~\ref{sec:IllustrativeRevisited}. In this case, one either 
knows what the full system is or has a sufficient model for it. However, it may be 
difficult to integrate out the irrelevant degrees of freedom exactly, in which case 
perturbative calculations in a suitable small parameter (e.g., coupling constant, 
ratios of length, time, speed, energy, or mass scales) tend to be useful. 

Lastly, one can try to parametrize $K$ in a systematic fashion by
imposing that the nonconservative action $\calS$ be invariant under
the symmetries appropriate to the problem. Therefore, one can restrict
terms into $K$ to those compatible with the symmetries. This is
well-motivated from the the procedure of integrating out inaccessible
degrees of freedom, where the resulting $K$ would have to be
consistent with the underlying symmetries of the full system. This
approach is inspired by the effective field theory framework (see
e.g.,~\cite{Goldberger:2005cd}). However, in order for the resulting
action to be predictive it is useful for there to be a naturally small
expansion parameter that ensures only a finite number of terms will be
included in $K$ for a given accuracy.

\section{Examples in discrete mechanics} \label{sec:discreteExamples}

While this new framework for nonconservative mechanics may seem unfamiliar it can 
be used in a similar way as the familiar action and Lagrangian for conservative systems. 
Perhaps the best way to see how to use the nonconservative mechanics formalism is 
through examples. In this section, we will apply the formalism to a range 
of discrete nonconservative systems including the familiar forced damped harmonic 
oscillator, RLC circuits, and radiation reaction on an accelerating charge.

\subsection{Forced damped oscillator}\label{sec:ForcedDampedOscillator}

\begin{figure}[tb]
	\includegraphics[width=\columnwidth]{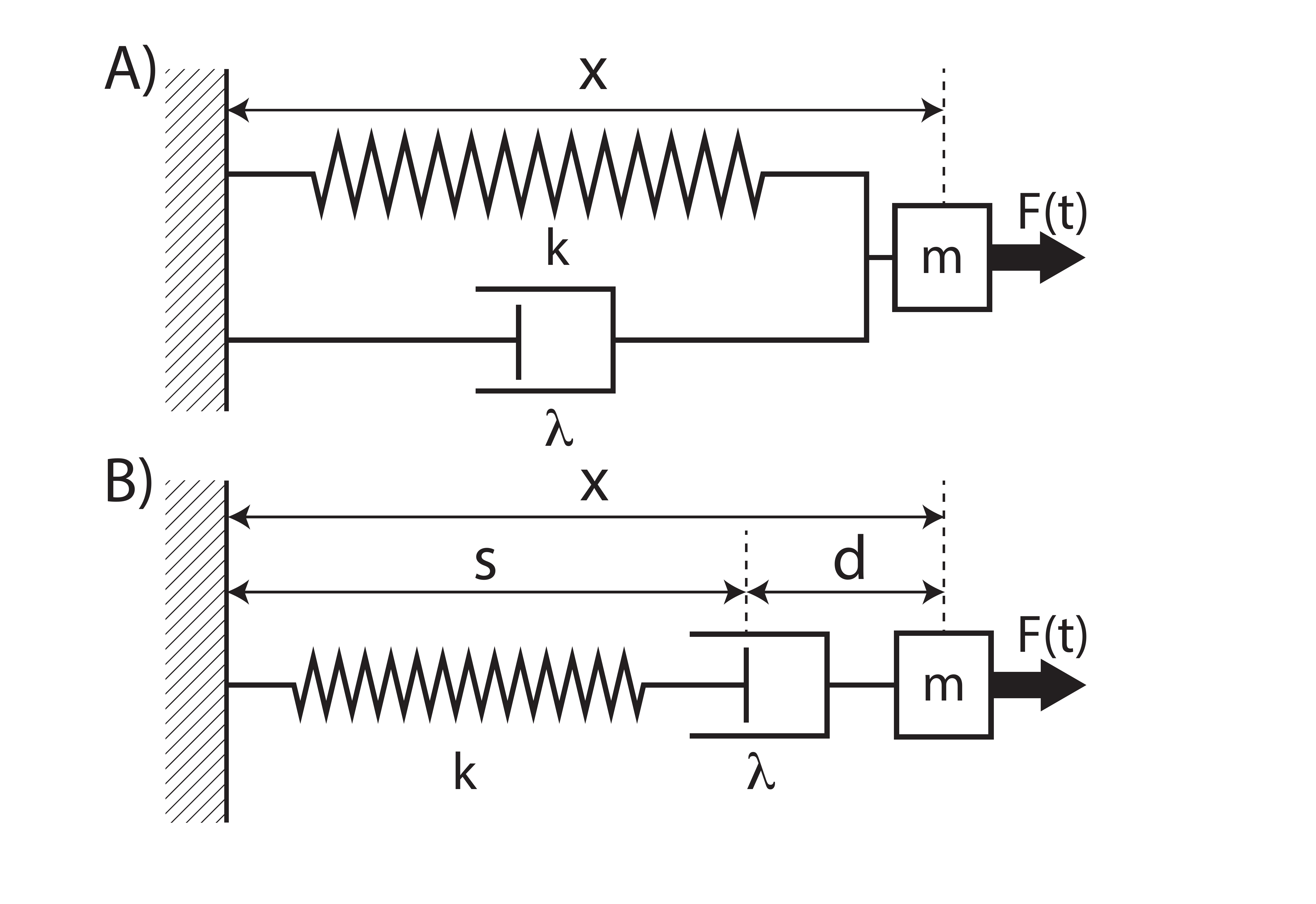}
	\caption{
		A) Schematic of a forced, damped harmonic oscillator. The oscillator mass is connected to a massless spring with spring constant $k$ and a massless dashpot with damping factor $\lambda$ in parallel.
		B) Schematic of the Maxwell element. The mass is connected to a spring and dashpot in series. Unlike the forced, damped harmonic oscillator above, the displacement of the center of mass is determined by the spring's displacement, which is elastic, but also by the ``plastic'' deformation of the dashpot.
\vspace{-1em}
		}
	\label{fig:maxwell}
\end{figure}

We first consider the familiar context of a
forced, damped, harmonic oscillator. One physical realization of this
system is shown in the top schematic in Fig.~\ref{fig:maxwell}.
Consider the following nonconservative Lagrangian $\Lambda$ where the conservative 
Lagrangian $L$ and nonconservative potential $K$ are given by
\begin{align}
	L = \frac{1}{2} m \dot{\bx}^2 - \frac{1}{2} k \bx^2 ~~ , ~~~ K = - \lambda \bx_- \cdot \dot{\bx}_+ + \bx_- \cdot {\bm F}(t)
\end{align}
for some external forcing function ${\bm F}(t)$. The conservative Lagrangian is simply that 
of a harmonic oscillator with mass $m$ and spring constant $k$. If we substitute $L$ 
and $K$ from above into the nonconservative Euler-Lagrange equations in 
\eqref{eq:nonconsEL4} then we find 
\begin{align}
	m \ddot{\bx} + \lambda \dot{\bx} + k \bx = {\bm F}(t)
\,,
\end{align}
which is the equation of motion for a forced, damped harmonic oscillator. 
The energy for the oscillator from \eqref{eq:Echarge2} is
\begin{align}
	\calE = \frac{1}{2} m \dot{\bx}^2 + \frac{1}{2} k \bx^2 
\end{align}
since, from (\ref{eq:kappa1}), the nonconservative part of the total conjugate momentum vanishes, $\kappa = 0$.
The change in the oscillator's energy is given by \eqref{eq:Echarge3} here as
\begin{align}
	\frac{ d \calE }{ dt} = \dot{\bx}\cdot {\bm F}(t) - \lambda \dot{\bx}^2  .
\end{align}
The first term corresponds to the power gained due to the external
force, while the last term is the power lost due to damping.

\subsection{The Maxwell element}
\label{sec:maxw-elem-spring}

In this example, we show how to use the nonconservative mechanics
formalism for a problem with multiple but coupled degrees of
freedom. This will also provide some background for the following
example in Sec.~\ref{sec:discreteMaxwellClosure}. This example is also
interesting for generating an equation of motion that is a first order
differential equation without implementing Lagrange multipliers.

Connecting a spring and a mechanical damper (e.g., a {\it dashpot}) in 
series yields a simple but ubiquitous description for modeling certain 
aspects in the rheology of visco-elastic materials under strain. This 
mechanical model, called a {\it Maxwell element}, is depicted by the 
lower schematic in Fig.~\ref{fig:maxwell}. A Maxwell element can be 
seen in an everyday example of a door closer, which prevents it from slamming shut.

When an external force $\bF(t)$ is applied the center of mass position, 
$\bx(t) = \bs(t) + \bd(t)$ changes. Since $\bF(t)$ is applied to the spring 
and the dashpot equally then the force that stretches the spring from 
its equilibrium position, $k \bs(t)$, also goes into forcing movement through 
the viscous fluid in the dashpot by an amount $\lambda \dot{\bd}(t)$. 
Therefore, we expect that 
\begin{align}
	k (\bx(t) - \bd(t) ) = \lambda \dot{\bd}(t)  .
\end{align}
We take $\bx(t)$ and $\bd(t)$ to be the accessible degrees of freedom in the problem.

We take the conservative Lagrangian to be
\begin{align}
	L = \frac{ 1 }{ 2} m \dot{\bx}^2 - \frac{1}{2} k (\bx - \bd)^2 .
\label{eq:MaxElL}
\end{align}
The nonconservative potential $K$ will contain the work done by the external force $\bF(t)$ in displacing the center of mass through $\bx_-(t)$ plus the amount of energy lost by the system due to heating the viscous fluid in the damper, which we will model as being linear in the velocity of $\bd$. Therefore, we propose
\begin{align}
	K = \bx_- \cdot \bF(t) - \lambda \bd_{-} \cdot \dot{\bd}_{+} .  \label{eq:MaxElK}
\end{align}
The equations of motion for $\bx(t)$ are found from \eqref{eq:nonconsEL4},
\begin{align}
	m \ddot{\bx} + k (\bx - \bd) = \bF(t) 
\end{align}
while that for $\bd(t)$ are
\begin{align}
	\lambda \dot{\bd} = k (\bx - \bd)
\,,
\end{align}
and together are the equations of motion that we expect. If we write 
$\bs = \bx - \bd$, as implied in Fig.~\ref{fig:maxwell}, then 
we get the equivalent equations of motion,
\begin{align}
	m \ddot{\bs} + \frac{ m k }{ \lambda} \dot{\bs} + k \bs & = \bF(t) , \\
		\lambda \dot{\bd} = k \bs  .
\end{align}
Notice that we did not have to introduce a Lagrange multiplier to get the 
second equation of motion, which is a first order differential equation. 
We can solve for $\bs(t)$ given the external force and some initial data, from 
which we can then construct the solution for $\bd(t)$ and thus $\bx(t)$,
\begin{align}
	\bd(t)-\bd(t_{i}) = \frac{ k }{ \lambda } \int_{t_i}^{t} dt' \bs(t')  .
\end{align}
Note that $\bd(t)$, and thus $\bx(t)$, depends on the integrated history of 
the elastic displacement of the spring alone. 

The energy of the Maxwell element from (\ref{eq:Echarge2}) is
\begin{align}
	\calE =  \frac{1}{2} m \dot{\bx}^2 + \frac{1}{2} k (\bx - \bd)^2
\label{eq:maxwellE1}
\end{align}
since ${\bm \kappa} = {\bm 0}$.
The time rate of change at which energy is changing is given by \eqref{eq:Echarge3},
\begin{align}
	\frac{ d \calE }{ dt}  = \dot{\bx} \cdot \bF(t) - \lambda \dot{\bd}^2  .
\end{align}
The energy changes due to the external force applied to the system as well 
as the energy dissipated by the viscous fluid in the dashpot.

\subsection{Closure condition for the Maxwell element}
\label{sec:discreteMaxwellClosure}

In this example, we show how to derive a closure condition (discussed
in Sec.~\ref{sec:discreteClosure}) for a Maxwell element to close the
system of equations when the evolution of the accessible variables is
affected by the internal energy of inaccessible degrees of
freedom. Recall that a Maxwell element is damped by a
viscous fluid in the dashpot. In the process, the damper may heat up 
and the damping coefficient $\lambda$ may change value. Therefore, 
the mechanical energy of the Maxwell element is transferred to the internal 
energy $U$ of the viscous fluid as heat. The internal energy of a viscous 
fluid is naturally parametrized by the thermodynamic entropy $S(t)$ assuming 
also an equation of state, $U(S)$, that does not depend on the generalized 
coordinates or velocities. With $\lambda(U(S)) = \tilde{\lambda}(S)$ a given 
function of entropy we can
write the nonconservative Lagrangian $\Lambda$ in (\ref{eq:MaxElL}) 
and (\ref{eq:MaxElK}) as
\begin{align}
	L & = \frac{1}{2} m \dot{\bx}^2 - \frac{ 1}{2} k (\bx - \bd) ^2 - U(S(t)) , \\
	K & = \bx_- \cdot \bF(t) - \tilde{\lambda} (S) \, \bd_- \cdot \dot{\bd}_+  .
\end{align}
The equations of motion are similar to those given in the previous example 
except that $\lambda$ depends on time through the entropy $S(t)$ of the 
viscous fluid in the dashpot,
\begin{align}
	m \ddot{\bs} + \frac{ m k }{ \tilde{\lambda}(S) } \, \dot{\bs} + k \bs & = \bF(t) 
\label{eq:maxwell13} \\
	\tilde{\lambda}(S) \dot{d}  = k \bs
\label{eq:maxwell14}
\end{align}
with $\bs = \bx - \bd$.
The energy of the Maxwell element is given by 
\begin{align}
	{\cal E}  = \frac{ 1 }{2} m \dot{\bx}^2 + \frac{ 1}{2} k (\bx - \bd)^2 + U( S( t))  .
\end{align} 
The change in the open subsystem's energy is given by
\eqref{eq:Echarge3} as
\begin{align}
	\frac{ d {\cal E}}{ dt } = \frac{ \partial U }{ \partial S } \dot{S} + \dot{\bx} \cdot \bF(t) - \tilde{\lambda}(S) \dot{\bd}^{\,2}  .
\label{eq:dEdtMaxwell10}
\end{align}
With the inclusion of the damper's internal energy, the total energy of the whole Maxwell element is given by $\calE$.
Therefore,
if all of the energy supplied by the external force $\bF(t)$ goes into changing ${\cal E}$ then 
\begin{align}
	\frac{ d {\cal E} }{ dt } = \dot{\bx} \cdot \bF(t) .
\label{eq:dEdtMaxwell11}
\end{align}
Thus the energy dissipated by the dashpot goes into changing the internal 
energy $U(S)$, which is natural since the Maxwell element is a completely 
closed system aside from the external force acting on it. Therefore, the entropy must 
change with time according to the closure condition implied from (\ref{eq:dEdtMaxwell10}) 
and (\ref{eq:dEdtMaxwell11}), namely,
\begin{align}
	\frac{dU}{dt} = T(S) \dot{S} = \tilde{\lambda}(S) \dot{\bd}^{\,2} 
\label{eq:entropyeqn1}
\end{align}
where we have defined temperature $T(S) \equiv \partial U / \partial S$ in the 
usual way. The closure relation thus gives the rate at which the damper is heated 
through viscous dissipation in the dashpot. Such a relation is needed if the internal 
energy of the dashpot changes with time in order to close the system of equations, 
which are given by (\ref{eq:maxwell13}), (\ref{eq:maxwell14}), and (\ref{eq:entropyeqn1}). 
Notice that if $\tilde{\lambda}(S) > 0$ that the entropy changes in a way that 
satisfies the second law of thermodynamics. Furthermore, the right side of 
(\ref{eq:entropyeqn1}) is the amount of energy from heating that we would expect.
That these results fall out naturally from our nonconservative
framework is a powerful feature when discussing dissipative fluids,
thermal diffusivity, and heat flow below in
Sec.~\ref{sec:ContinuumExamples}.

\subsection{Radiation reaction on an accelerating charge}
\label{sec:emrr}

Here we consider a system where the canonical momentum, $\kappa_I$ in 
(\ref{eq:kappa1}), associated with $K$ is non-zero and recover some well-known 
results in electrodynamical radiation reaction. As we shall see, being able to 
calculate $\kappa_I$ from $K$ directly as well as its effects on observable 
quantities like energy and angular momentum is a very powerful feature of 
nonconservative mechanics.

The nonconservative Lagrangian $\Lambda$ describing the relativistic motion 
of an extended charge experiencing radiation reaction was {\it derived} in 
Ref.~\cite{Galley:2010es} by integrating out the influence of the electromagnetic 
field on the motion of the charged body using effective field theory techniques. 

For simplicity and pedagogical purposes, but without loss of generality, we work here in the nonrelativistic limit where the nonconservative Lagrangian from Ref.~\cite{Galley:2010es} is expressed by 
\begin{align}
	\Lambda  & = m \bv_- \cdot \bv_+ - \frac{ e^2}{ 6 \pi } \, \bv_- \cdot \ba_+ + \bx_- \cdot \bF(t) + \calO(-^3)
\label{eq:EMRRlambda1}
\end{align}
where ${\bm F}(t)$ is an external force that sets the charge in motion, $\bv = \dot{\bx}$, and $\ba = \ddot{\bx}$.
Reading off from (\ref{eq:EMRRlambda1}), the conservative Lagrangian $L$ and nonconservative potential $K$ are
\begin{align}
	L &= \frac{ 1}{2} m \bv^{2} ~, \\
	K  &= - \frac{ e^2}{ 6 \pi } \, \bv_- \cdot \ba_+ + \bx_- \cdot \bF(t) + {\cal O} (-^3)
\end{align}
 The relativistic nonconservative Lagrangian with the leading order contributions 
 from the charge's finite size can be found in~\cite{Galley:2010es}.
The equations of motion in the physical limit are found from \eqref{eq:nonconsEL4} to be
\begin{align}
	m \ba = \bF(t) + \frac{ e^2 }{ 6 \pi } \dot{\ba}  ~,
\label{eq:ald1}
\end{align}
which is the well-known Abraham-Lorentz-Dirac equation of motion for the point-like charge \cite{jackson_classical_1999}. We will not discuss issues of runaway solutions associated with using a point-like charge to model the motion as this and related issues are outside the scope of this work. 
However, standard techniques for obtaining physically well-behaved solutions can be performed at the level of the equations of motion \cite{landau1986fieldtheory} and the action, which are commonly performed in effective field theories, through an order of reduction procedure that yields physically accurate solutions until quantum effects enter (see e.g., \cite{Koga:PRE70}). 

The momentum associated with $K$ is non-zero,
\begin{align}
	\kappa_i = \bigg[ \frac{ \partial K }{ \partial v^i_- } \bigg]_{\rm PL} = - \frac{e^2}{6\pi} a_i  .
\label{eq:kappaemrr1}
\end{align}
Therefore, the total energy for a radiating, accelerated charge is given by \eqref{eq:Echarge2} by
\begin{align}
	{\cal E} & = \frac{ 1}{2} m \bv^{2} - \frac{ e^2 }{ 6 \pi } \bv \cdot \ba \,.
\label{eq:totalEemrr1}
\end{align}
The second term in $\calE$ is the well-known Schott term and accounts for the 
energy of the near-zone part of the electromagnetic field that is not radiated to infinity. 
The Schott term arises precisely because of the charge's nonconservative 
self-interaction that results from eliminating the electromagnetic degrees of 
freedom from the action. The corresponding canonical momentum associated 
with the Schott term is given simply by ${\bm \kappa}$ in (\ref{eq:kappaemrr1}). 
Note that choosing instead $K' = +e^2/(6\pi) \ba_- \cdot \bv_+$ gives the
same expression for the nonconservative conjugate momentum in (\ref{eq:kappaemrr1}) upon
using the more general expression in (\ref{eq:higherderkappa1}). Therefore, either
expression for the nonconservative potential gives the same expressions for physical quantities.
The change in $\calE$ with time is given by \eqref{eq:Echarge3}, 
\begin{align}
	\frac{ d {\cal E}}{ dt }  & = - \frac{ e^2 }{ 6 \pi} \ba^{2}  + \bv \cdot \bF(t).
\label{eq:rrdhdt1}
\end{align}
The first term on the right side of \eqref{eq:rrdhdt1} is the power radiated 
by the accelerated charge (derived by Larmor~\cite{jackson_classical_1999}) 
and the second term is the power supplied by the external force to 
accelerate the charge in the first place. 

From the paragraph following (\ref{eq:Echarge3}) and (\ref{eq:Jcharge3}), 
the nonconservative part of the momentum ${\bm \kappa}$ will couple to the 
acceleration, which yields the Larmor contribution. However, this necessarily 
implies that the energy is shifted by $\kappa_i$ contracted with the charge's 
velocity. Therefore, the fact that the radiated power depends 
on the acceleration implies the existence of the Schott energy term as a 
contribution to the total energy associated with the dynamics of the charge.
A key point is that the change in the
energy of the system is derived directly from the new Lagrangian
formulation without needing to apply any additional arguments about
energy or balancing fluxes or performing a separate calculation for
the radiated field (see e.g.,~\cite{jackson_classical_1999}). 

The transformation of the conservative action under a rotation through
a small angle $| \bm{\theta} |$ about the direction $\bm{\theta} /
|\bm{\theta}|$,
\begin{align}
	\delta q^i = - \epsilon^i{}_{jk} q^j \theta^k ~~\Longrightarrow~~ \omega^i{}_k = - \epsilon^i{}_{jk} q^j
\end{align}
implies that the total angular momentum components $\calJ_k$ in (\ref{eq:Jcharge2}) are
\begin{align}
	{\cal J}_k & = \left[ \bx \times  \left( m \bv -\frac{e^2 }{ 6\pi} \ba  \right) \right]_k  . 
\label{eq:Jemrr1}
\end{align}
Like the total energy, the total angular momentum receives a correction 
from the interaction between the charge and the electromagnetic field, 
which can be interpreted as the angular momentum from the field in the 
near zone that is not radiated to infinity but carried along with the charge as a whole. 
The change in time of ${\cal J}_k$ is the torque on the charge and is given in (\ref{eq:Jcharge3}) by
\begin{align}
	\frac{ d {\cal J}_k }{ dt } = - \frac{e^2}{6\pi} \left( \bv \times \ba \right)_k + \left( \bx \times \bF(t) \right)_k
\,.  
\label{eq:dJdtemrr1}
\end{align}
The torque on the charge is thus driven by the acceleration and the external force.

\subsection{Linear RLC circuits}\label{sec:Circuits}

In this final example, we show that the nonconservative formalism
discussed in Sec.~\ref{sec:Review} can be applied to non-mechanical
systems such as circuits.

Circuits consisting of only energy-conserving elements (e.g.,
inductors and capacitors) may be described by a standard variational
formulation~\cite{OberBlobaum2013498}.
However, with the variational principle discussed in Sec.~\ref{sec:Review} we
may also describe dissipative elements (e.g., resistors). Though we focus
here on linear circuit elements there is no obstacle to including
nonlinear ones (e.g., transistors).

\begin{figure}[tb]
  \centering
\includegraphics[width=0.66\columnwidth]{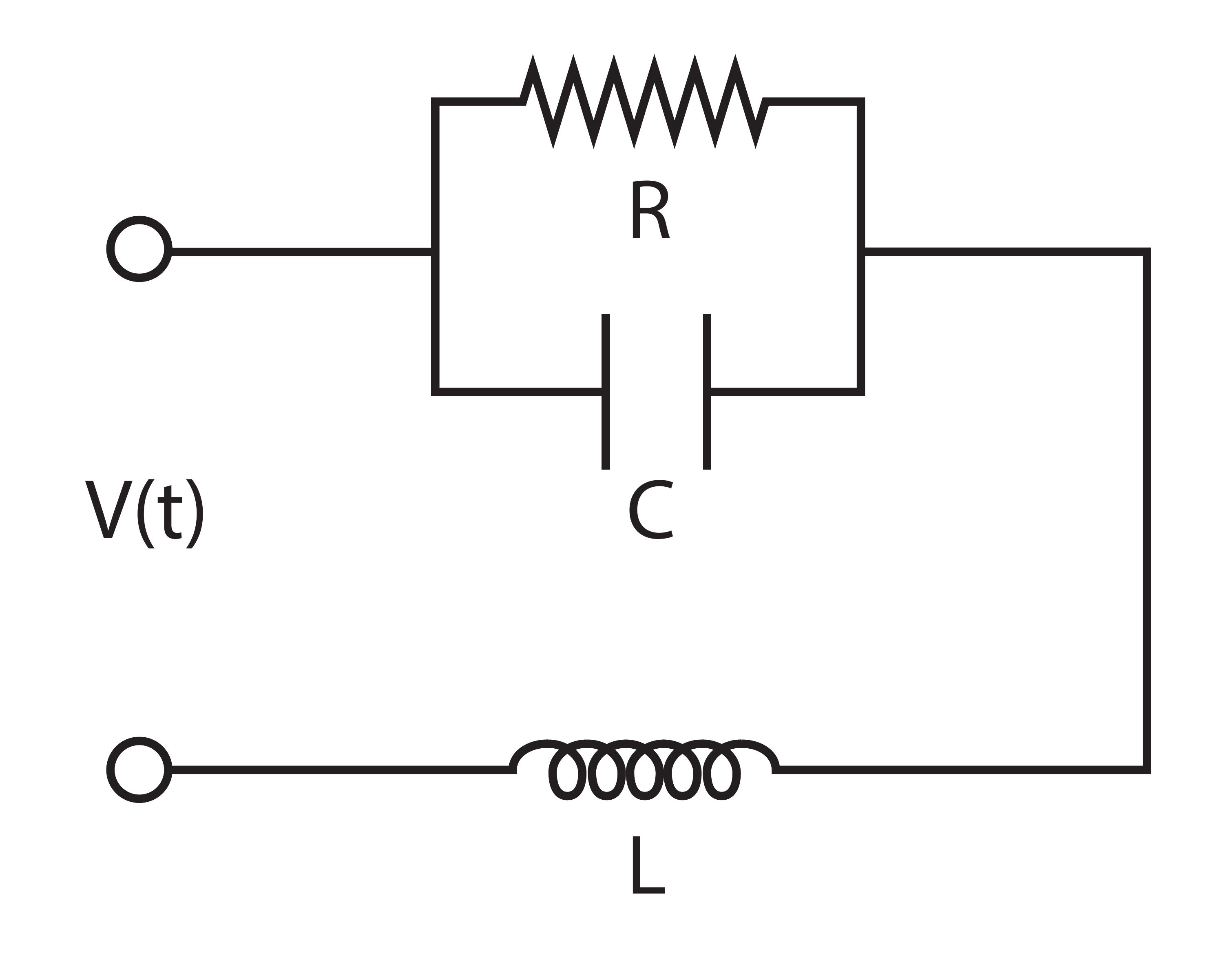}
  \caption{A simple circuit composed of a resistor (R), inductor (L), and capacitor (C). 
   An external voltage $V(t)$ is applied to complete the circuit.
    \label{fig:RLC-circuit}
\vspace{-1em}
  }
\end{figure}

Figure \ref{fig:RLC-circuit} shows an example circuit composed 
of a resistor, inductor, and capacitor. The degrees of freedom in this
problem are the net charges $Q_A(t)$ that have flowed through elements
$A=\{R, L, C\}$. The corresponding generalized velocities are the
currents, $\dot{Q}_A$. However, the variables $Q_A$ are not all 
independent because there are Kirchoff constraints to be applied.
The number of {\it independent} net-charges is actually the number $n$
of loop currents in the whole circuit. These independent degrees of freedom are
denoted by $q_a$ for $a=1,\ldots,n$. Then each $Q_A$ is 
a directed sum of the of the independent charges flowing through
the loop currents,
\begin{equation}
\label{eq:Qofq}
  Q_{A}(q_{a}) =  Q_{A0} + \sum_{a \cap A} (-)_{A,a} q_{a} \,,
\end{equation}
where $Q_{A0}$ is a constant of integration, $a\cap A$ denotes those
$q_{a}$ which go through $Q_{A}$, and
$(-)_{A,a}$ is negative if the currents $\dot{Q}_{A}$ and $\dot{q}_{a}$ are oppositely
oriented. 

There are then two approaches to constructing the action for a
circuit. One may impose the constraints of \eqref{eq:Qofq} by
inserting Lagrange multipliers. Alternatively, since we are here only
dealing with linear elements, we may directly write the $Q$'s as
functions of $q$'s and be assured of the same dynamics.

The circuit in Fig.~\ref{fig:RLC-circuit} requires two current loops
and thus two independent net-charges, $q_1(t)$ and $q_2(t)$.
Take current $\dot{q}_{1}$ to flow through
the $R$--$C$ loop in the clockwise direction and current
$\dot{q}_{2}$ to flow through the $C$--$L$--$V(t)$ loop in the
clockwise direction. This gives three relations,
\begin{align}
Q_{R} &= + q_{1} \label{eq:QR1a} \\
Q_{C} &= - q_{1} + q_{2} + Q_{C0} \label{eq:QC1a} \\
Q_{L} &= + q_{2} \label{eq:QL1a}
\end{align}
where $Q_{C0}$ is a constant of integration denoting the charge on the
capacitor at time $t=0$ (alternatively, this constant may be absorbed
into $q_{1}(0)$ and/or $q_{2}(0)$). 

The conservative Lagrangian $L'$ and nonconservative potential $K$ are taken to be
\begin{align}
	L' & = \frac{L}{2} \dot{Q}_L^2 - \frac{1}{2C} Q_C^2 \,,\\
		& = \frac{L}{2} \dot{q}_2^2 - \frac{1}{2C} ( -q_1 + q_2 + Q_{C0} )^2 \,, \\
	K & = q_{2-} V(t) - R q_{1-} \dot{q}_{1+} \,.
\end{align}
The ``kinetic energy'' in an inductor is $L \dot{Q}_{L}^{2}/2$, while
the potential energy stored in a capacitor is $Q_{c}^{2}/2C$. These
two terms constitute the conservative Lagrangian.
Meanwhile, $K$ is composed of two terms. The first comes from the energy
associated with the external voltage acting on the amount of charge
$q_{2-}$ flowing through the $C$--$L$--$V(t)$ loop. The second
is the amount of energy lost by heating the resistor.

A straightforward calculation shows that this clearly reproduces
Kirchhoff's voltage law about each loop. In particular, the variations with respect to $q_1$ and $q_2$ yield
\begin{align}
	R I_R = \frac{ Q_C }{ C }  \qquad {\rm and} \qquad L \dot{I}_L - \frac{ Q_C }{ C} = V(t)  
\end{align}
upon using (\ref{eq:QR1a})-(\ref{eq:QL1a}) where $I_R = \dot{Q}_R$ and $I_L = \dot{Q}_L$.
Note that this nonconservative 
action formalism can also be easily applied to circuits with
nonlinear elements and with more elements than we considered in this example.

\section{Nonconservative classical field theories} \label{sec:ContinuumMechanics}

We have seen that discrete nonconservative systems can be modeled
through nonconservative actions and the variational principle
described in Sec.~\ref{sec:Review}, and explored several examples of
such systems in Sec.~\ref{sec:discreteExamples}. In the remainder of
this paper we will extend this formalism to include nonconservative
continuum mechanics and classical field theories
(Sec.~\ref{sec:ContinuumMechanics}), followed by several example
applications (Sec.~\ref{sec:ContinuumExamples}), with particular focus
on continuum mechanics. In what follows, we will often use ``continuum
mechanics'' and ``field theories'' interchangeably.

Consider $N$ fields $\phi^I (x^\mu)$ where $I=1,\ldots,N$ labels
the components whose evolution is to be studied for times in $T=[t_i,
t_f]$ and within a spatial volume $V$, such that $x^\mu \in \{T \times
V\}$, $x^{\mu}=(x^{0},x^{1},x^{2},x^{3})$ where $x^0 = t$. For example, $\phi^I(x^\mu)$
may represent the electromagnetic vector potential $A_\mu(t, \bx)$, which is a relativistic field (where $I = \mu$ is the space-time indices)
that transforms as a vector under Poincare transformations.
For a field undergoing nonconservative interactions, we must double
the degrees of freedom as we did for discrete systems, in which case $\phi^I
\rightarrow (\phi_1^I, \phi_2^I)$, in order to capture the appropriate
nonconservative (e.g., dissipative) effects and account for the correct causal evolution of the open system dynamics.

\subsection{Lagrangian mechanics}\label{sec:continuumLagrange}

The action for doubled variables is given by
\begin{align}
  \calS [ \phi^I_a ] = &
  \int_{t_i}^{t_f} \!\!\! dt \, \int_V d^3x \, \Omega [ \phi^I_a ] =
  \int_\calV d^4x \, \Omega [ \phi^I_a ]
\label{eq:action0}
\end{align}
where $\calV = T \times V$ is the space-time volume of interest and
$\Omega$ is the nonconservative Lagrangian density and, as with the discrete conservative Lagrangian $L$,
is an arbitrary function of the doubled fields, their derivatives, and possibly the space-time coordinates $x^\mu$,
\begin{align}
	\Omega [ \phi^I_a] = \Omega(\phi^I_a, \partial_\mu \phi^I_a, x^\mu)
\end{align}
Recall that for certain coordinates the usual Lagrangian can be
written as the difference of the kinetic and potential energies.
Likewise, for certain history labels the new Lagrangian density can be
separated into its conservative and nonconservative pieces as
\begin{align}
    \Omega [ \phi^I_1, \phi^I_2 ] = {} &
    \calL ( \phi^I_1, \pd_\mu \phi^I_1, x^\mu ) - \calL ( \phi^I_2, \pd_\mu \phi^I_2, x^\mu ) \nn \\
    	& + \calK (\phi^I_1, \phi^I_2, \pd_\mu \phi^I_1, \pd_\mu \phi^I_2, x^\mu)  .
\label{eq:action1}
\end{align}
Note that, in general, $\Omega$ is an arbitrary function of the
doubled variables and does not necessarily have the form on the right
side of \eqref{eq:action1} in a general set of canonical coordinates.

One such set of coordinates is provided by the $\pm$ basis, defined by
\begin{align}
  \phi_+^I  = \frac{1}{2} ( \phi_1^I + \phi_2^I ) ~, ~~ \phi_-^I = \phi_1^I - \phi_2^I  .
\end{align}
For generality, we label the doubled fields by a lower case Roman
letter from the beginning of the alphabet, $\phi_a^I(x^\mu)$. To
vary the action we let
\begin{align}
  \phi_{a} ^I(x^\mu)  \to \phi_{a}^I (x^\mu, \epsilon) =
  \phi^I_{a} (x^\mu, 0) + \epsilon \eta_{a}^I (x^\mu)
\label{eq:fieldvary1}
\end{align}
where $\epsilon \ll 1$ and $\eta_{a}^I$ are arbitrary functions of
$x^\mu$. The fields evaluated at $\epsilon = 0$ are
taken to be the ones for which the action is stationary with respect
to changes in $\epsilon$. Substituting \eqref{eq:fieldvary1} into
\eqref{eq:action1} and expanding out the action in \eqref{eq:action0}
through first order in $\epsilon$ gives
\begin{multline}
  \calS = \int_{\cal V} d^4 x \left\{ [\Omega ]_0 +
    \epsilon \eta^I_a (x^\mu) \left[
      \frac{ \partial \Omega }{ \partial \phi^I_a }
    \right]_0 \right.\\
\left.
{}+ \epsilon \partial_\nu \eta^I_a (x^\mu) \left[
      \frac{ \partial \Omega }{ \partial (\partial_\nu \phi^I_a) }
    \right]_0 + {\cal O}(\epsilon^2) \right\}
\end{multline}
where $[ \cdots ]_0$ indicates that the quantity inside the brackets
is evaluated at $\epsilon = 0$ and we implicitly sum over the $a$
index.
Integrating by parts on the $\partial_\mu \eta^I_a$ terms gives
\begin{multline}
  \calS =\int_{\cal V} d^4 x \left\{ [\Omega ]_0 + \epsilon \,
    \eta_a^I \left[ \frac{ \partial \Omega }{ \partial \phi_a^I } -
      \partial_\mu \frac{ \partial \Omega }{ \partial (\partial_\mu
        \phi^I_a) } \right]_0 \right\} \\
{}+ \epsilon \oint_{\partial
  {\cal V}} \!\!\! d\Sigma_\mu \frac{ \partial \Omega }{ \partial (
  \partial_\mu \phi^I_a) } \eta_a^I  + {\cal O}(\epsilon^2)
\label{eq:action10}
\end{multline}
where $\partial {\cal V}$ is the boundary of the spacetime volume
${\cal V}$ and $d\Sigma_\mu$ is a surface area element on $\partial
{\cal V}$ pointing {\rm out} of ${\cal V}$.

As in discrete mechanics, if the boundary of the space-time volume is
fixed then there should be no contribution to the action's variation.
For notational convenience, we define the current densities $\Pi^\mu_{aI} (x^\mu)$ as
\begin{align}
	\Pi^\mu_{I 1} = \frac{ \pd \Omega }{ \pd ( \pd_\mu \phi_1^I) } ~~, ~~ \Pi^\mu_{I 2} = - \frac{ \pd \Omega }{ \pd ( \pd_\mu \phi_2^I) }
\label{eq:Pimui1and2}
\end{align}
in the $1,2$ basis and 
\begin{align}
	\Pi^\mu_{I +} = \frac{ \pd \Omega }{ \pd ( \pd_\mu \phi_-^I) } ~~, ~~ \Pi^\mu_{I -} = \frac{ \pd \Omega }{ \pd ( \pd_\mu \phi_+^I) }
\label{eq:Pimuipm}
\end{align}
in the $\pm$ basis.
These expressions can be condensed by the introduction of a ``metric'' $c_{ab}$
that can be used to raise and lower the history indices. In the $1,2$ labels
the metric is $c_{ab} = {\rm diag}(1,-1)$ and in the $+,-$ labels is $c_{ab} = {\rm offdiag}(1,1)$
so that (\ref{eq:Pimui1and2}) and (\ref{eq:Pimuipm}) are given succinctly as
\begin{align}
	\Pi^\mu_{aI} = c_{ab} \frac{ \pd \Omega }{ \pd ( \pd_\mu \phi_1^I) } ~~  \Longrightarrow ~~ \Pi^a_I {}^\mu = c^{ab} \Pi^\mu_{bI}  .
\label{eq:Pimuiab11}
\end{align}
In discrete mechanics the equality condition for the variations at the
final time ($\eta_-(t_f) = 0 = \dot{\eta}_-(t_f)$) and the vanishing
of $\eta_a$ at the initial time guarantees that the boundary terms do
not contribute to the variation. In continuum mechanics, we instead
have a  spacetime volume ${\cal V}$ with a boundary $\pd {\cal
  V}$. When describing nonconservative field theories, which are
necessarily ones that can evolve with nonequilibrium or
nonstationary dynamics, one has to solve the equations of motion from
a set of initial data specified at $t = t_i$ to a final time $t =
t_f$. If $V$ denotes a spatial $3$-volume with boundary $\pd V$
then the surface integrals in \eqref{eq:action10} take the following
form, 
\begin{equation}
\begin{split}
  \oint_{\pd {\cal V}} \!\!\! d\Sigma_\mu \frac{ \pd \Omega }{ \pd (
    \pd_\mu \phi^I_a) } \eta_a^I
  = {} & \int_{V} \! d^{3}x \ \pi_{I}^a \eta_a^I \bigg|_{t_i}^{t_f} \\
  & +  \int_{t_i}^{t_f} \! dt \oint_{\pd V} \! dS_i \, \Pi^{ai}_I \eta_a^I
\end{split}
\end{equation}
where 
\begin{align}
  \pi_I^{a} \equiv \Pi_I^{a 0} =
  \frac{ \pd \Omega }{ \pd (\pd_0 \phi_a^I) }
\end{align}
is the momentum canonically conjugate to $\phi_a^I$.
In the $\pm$ coordinates this equals
\begin{align}
  \oint_{\pd {\cal V}} \!\!\! d\Sigma_\mu
  \frac{ \pd \Omega }{ \pd ( \pd_\mu \phi^I_a) } \eta_a^I
 = {} & \int_{V} \! d^{3}x \ \bigg[\pi_{-I} \eta_+^I + \pi_{+I} \eta_-^I \bigg]_{t_i}^{t_f}  \label{eq:continuumboundary1z}  \\
& + \int_{t_{i}}^{t_{f}}\!\!\! dt \oint_{\pd V} \!\!\! dS_i \,
\left(\Pi^{i}_{-I} \eta_+^I+ \Pi^{i}_{+I} \eta_-^I \right) . \nn 
\end{align}
Generalizing Fig.~\ref{fig:histories} to field theories implies that at the initial time the variations individually vanish,
\begin{align}
  \label{eq:eta-initial}
  \eta_-^I (t_i, \bx) = 0 = \eta_+^I (t_i, \bx) 
\end{align}
for $\bx \in V$. 
In addition, the variations at the final time are equal so that the quantity in brackets in (\ref{eq:continuumboundary1z}) vanishes if we take
the continuum version of the equality condition introduced earlier, namely,
\begin{align}
  \label{eq:eta-pi-final}
  \eta_-^I (t_f, \bx) = 0 = \pi_{-}^I (t_f, \bx)
\end{align}
for $\bx \in V$. 
We are thus left with the surface integrals over $\pd V$,
\begin{align}
  \oint_{\pd {\cal V}} \!\!\! d\Sigma_\mu
  \frac{ \pd \Omega }{ \pd ( \pd_\mu \phi^I_a) } \eta_a^I  =
  \int_{t_{i}}^{t_{f}} \!\!\! dt \oint_{\pd V} \!\!\! dS_i \, 
  \left(\Pi^{i}_{-I} \eta_+^I  + \Pi^{i}_{+I} \eta_-^I\right)  .
\label{eq:surface10}
\end{align}
With the following two conditions, the surface integrals for non-dynamical
boundaries will vanish:
\begin{align}
  \label{eq:eta-boundary}
  \eta_1^I (t, \bx_S) &= \eta_2^I (t, \bx_S ) &&\longrightarrow &\eta_-^I
  (t, \bx_S ) &= 0\,, \\
  \label{eq:Pi-boundary}
  \Pi^i_{1I} (t, \bx_S) &= \Pi^i_{2I} (t, \bx_S)
  &&\longrightarrow &\Pi^i_{-I} (t, \bx_S) &= 0\,.
\end{align}
For certain problems, we can better justify these conditions by
analogy with discrete mechanics. For example, if the $\phi_{-}^{I}$
satisfy a linear, homogeneous, second order PDE, then
\eqref{eq:eta-pi-final} uniquely determines
$\phi_{-}^{I}(D^{-}(V_{f}))=0$ in the \emph{past domain of dependence} $D^-$
of $V_{f}$ (the spatial volume $V$ at $t=t_{f}$). The vanishing of
$\phi_{-}^{I}$ would be extended to the entire interior of ${\cal V}$
by giving conditions
\eqref{eq:eta-boundary}-\eqref{eq:Pi-boundary}. Then, just as in
discrete mechanics, we find that the minus variables vanish throughout
the entire solution.

The variation of the action is then given by $[ \pd S / \pd
\epsilon ]_0$, which is stationary when
\begin{align}
  0 = {} &  \left[ \frac{ \pd S }{ \pd \epsilon} \right]_0
\end{align}
and is satisfied for any $\eta_{a}^I(x^\alpha)$ provided that
\begin{equation}
  \pd_\mu \frac{ \pd \Omega }{ \pd (\pd_\mu \phi^I_a) } = \frac{ \pd
    \Omega }{ \pd \phi^I_a }  .
\label{eq:eom1}
\end{equation}

In the physical limit (``PL'') all ``$-$'' variables vanish and all
``$+$'' variables take their physical values and describe the
accessible degrees of freedom. This means that, in the $\pm$ basis,
only the equation with $a=-$ survives because that equation takes
derivatives with respect to the ``$-$'' variables and so only the terms in
$\Omega$ that are perturbatively linear in the ``$-$'' variables will
contribute in the physical limit, giving 
\begin{equation}
  \left[\pd_\mu \frac{ \pd \Omega }{ \pd (\pd_\mu \phi^I_-) } - \frac{ \pd
    \Omega }{ \pd \phi^I_- }\right]_{\rm PL} = 0 ,
\label{eq:eom1b}
\end{equation}
which can be written in terms of $\calL$ and $\calK$ as 
\begin{equation}
  \pd_\mu \frac{ \pd {\cal L}}{ \pd (\pd_\mu \phi^I) } - \frac{ \pd
    {\cal L}}{ \pd \phi^I } = \left[ \frac{ \pd {\cal K}}{ \pd
      \phi^I_- } - \pd_\mu \frac{ \pd {\cal K}}{ \pd (\pd_\mu
      \phi^I_-) }   \right]_{\rm PL}  =: \calQ_I  \, .
\label{eq:eom2}
\end{equation}
All of the results in this section can be extended to fields on a
curved background space-time in a straightforward manner following
standard techniques (see e.g., \cite{MTW}).

Finally, choosing or finding $\calK$ for a specific problem can be
accomplished using the approaches (and possibly others) discussed
in Sec.~\ref{sec:choosingK}.

\subsection{Noether's theorem generalized}\label{sec:ContinuumNoether}

We show here how Noether's theorem is generalized due to
nonconservative forces and interactions. Many of the manipulations are
similar to those encountered for discrete mechanical systems in
Sec.~\ref{sec:discreteNoether}. Consider the transformations,
\begin{align}
	x^\mu & \to x^\mu + \delta x^\mu \label{eq:xtransformation}\\
	\phi^I & \to \phi^I + \delta \phi^I + \delta x^\mu \partial_\mu \phi^I 
\label{eq:phitransformation131z}
\end{align}
with
\begin{align}
	\delta x^\mu & = \epsilon^\alpha \, \left[ \frac{ \partial x^\mu }{ \partial \epsilon^\alpha } \right]_{\epsilon^\alpha = 0} =: \epsilon^\alpha \, \xi_\alpha^\mu \\
	\delta \phi^I & = \epsilon'{}^{a} \, \left[ \frac{ \partial \phi ^I }{ \partial \epsilon'{}^{a} } \right]_{\epsilon'{}^{a} = 0 } =: \epsilon'{}^{a} \, \omega^I_{a} 
\label{eq:omegaI1}
\end{align}
and $\epsilon^\alpha$ and $\epsilon'{}^{a}$ are small parameters associated with (the Lie algebras of) the symmetry groups in question that keep the following {\it conservative} action invariant
\begin{align}
	S = \int_{\cal V} \!\!\! d^4x \, {\cal L} \big( \phi^I (x^\alpha), \partial_\mu \phi^I (x^\alpha), x^\mu \big) .
\end{align}
The index $a$ in (\ref{eq:omegaI1}) should not be confused with the history labels, which will not appear in this section.
The invariance of the action implies, using similar manipulations as
in Sec.~\ref{sec:discreteNoether}, that
\begin{align}
	0 = {} & \!\! \int_\calV \!\!\! d^4x \bigg\{ \epsilon^\alpha \bigg[ \xi^\mu_\alpha \frac{\partial \calL}{\partial x^\mu}+  \xi^\mu_\alpha  \partial_\mu \phi^I \frac{ \partial \calL }{ \partial \phi^I} + \partial_\nu ( \xi^\mu_\alpha  \partial_\mu \phi^I ) \frac{ \partial \calL }{ \partial (\partial_\nu \phi ) }  \nonumber \\
		& {\hskip0.2in}  - \partial_\mu \big( \xi^\mu_\alpha {\cal L} \big) \bigg]  + \epsilon'{}^{a} \bigg[ \omega^I_{a} \frac{ \partial \calL }{ \partial \phi^I } + \partial_\mu \omega^I_{a} \frac{ \partial \calL }{ \partial ( \partial_\mu \phi^I ) } \bigg] \bigg\}  .
\end{align}
Rearranging terms using the product rule for partial derivatives gives
\begin{align}
	0 = {} & \!\! \int _\calV d^4 x \bigg\{ \epsilon^\alpha \bigg[ \partial_\nu \bigg( \xi^\mu_\alpha \partial_\mu \phi^I \frac{ \partial \calL }{ \partial ( \partial_\nu \phi^I ) } - \xi^\nu_\alpha \calL  \bigg) +\xi^\mu_\alpha \frac{\partial \calL}{\partial x^\mu}  \nonumber \\
		& {\hskip0.7in} - \xi^\mu_\alpha \partial_\mu \phi^I \bigg(  \partial_\nu \frac{ \partial \calL }{ \partial (\partial_\nu \phi^I ) }  - \frac{ \partial \calL }{ \partial \phi^I }\bigg) \bigg] \nonumber \\
		&  + \! \epsilon'{}^{a} \! \bigg[  \pd_\nu \! \bigg( \! \omega^I_{a} \frac{ \partial \calL }{ \partial ( \partial_\nu \phi^I) } \bigg) \! - \! \omega^I_{a} \bigg( \! \partial_\mu \frac{ \partial \calL }{ \partial ( \partial_\mu \phi^I ) } -  \frac{ \partial \calL}{ \partial \phi^I } \bigg)   \bigg] \bigg\}
	\label{eq:noetherdS10}
\end{align}
The quantity
\begin{align}
	T_\mu{}^\nu \equiv \partial_\mu \phi^I \frac{ \partial \calL }{ \partial ( \partial_\nu \phi^I ) } - \calL \, \delta^\nu_\mu 
	\label{eq:Tmunu1}
\end{align}
is the canonical stress-energy-momentum (or simply stress) tensor. 
The Noether current density associated with coordinate transformation \eqref{eq:xtransformation} appears in the first divergence term of
\eqref{eq:noetherdS10}, 
given by
\begin{align}
	\xi_\alpha^\mu T_\mu{}^\nu = \xi^\mu_\alpha\partial_\mu \phi^I \frac{ \partial \calL }{ \partial ( \partial_\nu \phi^I ) } - \calL \, \xi^\nu_\alpha 
\end{align}
The first term on the last line of (\ref{eq:noetherdS10}) is the Noether current density 
associated with the transformation in (\ref{eq:phitransformation131z}),
\begin{align}
	J^\nu_{a} = \omega^I_{a} \frac{ \partial \calL }{ \partial ( \partial_\nu \phi^I) }  .
\end{align}
We may now use the Euler-Lagrange equations of motion in (\ref{eq:eom2})
to write (\ref{eq:noetherdS10}) in terms of $T_{\alpha}{}^\nu$, $J^\nu_{a}$, and $\calQ_I$ as
\begin{align}
	0 = {} & \int_{\calV} d^4 x \bigg\{ \epsilon^\alpha \bigg[ \pd_\nu \xi^\mu_\alpha T_\mu{}^\nu + \xi^\mu_\alpha \frac{ \pd \calL }{ \pd x^\mu } - \xi^\mu_\alpha \pd_\mu \phi^I \calQ_I \bigg] \nn \\
		& {\hskip0.4in} + \epsilon'{}^{a} \bigg[  \pd_\nu J^\nu_{a} - \omega^I_{a} \calQ_I \bigg]  \bigg\} .
\end{align}
Finally, since $\epsilon^\alpha$ and $\epsilon'{}^{a}$ are independent then each factor in square brackets must vanish for the whole integral to vanish. The result is
\begin{align}
	\pd_\nu (\xi_\alpha^\mu T_\mu{}^\nu) & = - \xi^\mu_\alpha \frac{ \pd \calL }{ \pd x^\mu } + \xi^\mu_\alpha \pd_\mu \phi^I \calQ_I   ,
\label{eq:divTmunu1} \\
	\pd_\nu J^\nu_{a} & = \omega^I_{a} \calQ_I  .
\label{eq:divJnu1}
\end{align}
We see that a nonzero $\calQ_I$ and explicit $x^\mu$ dependence of the Lagrangian density source (or drain) the system's Noether current. Once $\calK$ is known, through a calculation from integrating out degrees of freedom, coarse-graining, or otherwise specified, one may calculate how the Noether current changes for the accessible degrees of freedom. That is, once the nonconservative action is known one can compute how energy density, angular momentum density, etc., changes (see below). For a closed system having conservative interactions, the nonconservative generalized interactions vanish, $\calQ_I = 0$, and we recover Noether's theorem. 

A more convenient but equivalent form for the
divergences in (\ref{eq:divTmunu1}) and (\ref{eq:divJnu1}) is found using similar
manipulations as performed for discrete systems in
Sec.~\ref{sec:discreteNoether}. We quote the result here, which is
\begin{align}
	\pd_\nu \big(\xi_\alpha^\mu T_\mu{}^\nu + \xi^\mu_\alpha \pd_\mu \phi^I \kappa_I^\nu \big) = {} & - \xi^\mu_\alpha \frac{ \pd \calL }{ \pd x^\mu } + \xi^\mu_\alpha \pd_\mu \phi^I \bigg[ \frac{ \pd \calK }{ \pd \phi^I_- } \bigg]_{\rm PL}   \nn \\
		& + \xi^\mu_\alpha \pd_\nu \pd_\mu \phi^I \kappa_I^\nu ,  \\
	\pd_\nu \big( J_{a}^\nu + \omega^I_{a} \kappa_I^\nu \big) = {} & \omega^I_{a} \bigg[ \frac{ \pd \calK }{ \pd \phi^I_- } \bigg]_{\rm PL} + \pd_\mu \omega^I_{a} \, \kappa^\mu_I
\end{align}
where 
\begin{align}
	\kappa_I^\mu \equiv \bigg[ \frac{ \pd \calK }{ \pd ( \pd_\mu \phi^I_- ) } \bigg]_{\rm PL}
\label{eq:Noetherkappa1}
\end{align}
is the part of the total current density $\Pi^\mu_I$ that is associated with nonconservative interactions.
The left sides are the divergence of a shifted current densities defined by
\begin{align}
	\calT_\mu{}^\nu \equiv {} & T_\mu{}^\nu +\pd_\mu \phi^I \kappa_I^\nu  ,
\label{eq:calTmunu1} \\
	\calJ^\nu_{a} \equiv {} & J_{a}^\nu + \omega^I_{a} \kappa_I^\nu  .
\label{eq:calJnu1}
\end{align}
The contributions that come from $\kappa^\mu_I$ are corrections to the stress-energy
and current density that result from the open system's interaction with the inaccessible
or eliminated degrees of freedom. We can regard $\calT_\alpha{}^\nu$ and $J^\nu_{a}$ as the 
total stress-energy and current density of the accessible degrees of freedom that
include contributions from nonconservative interactions.
Our alternative expressions of Noether's theorem generalized to nonconservative
field theories are thus given by
\begin{align}
	\pd_\nu (\xi_\alpha^\mu \calT_\mu{}^\nu) & = - \xi^\mu_\alpha \frac{ \pd \calL }{ \pd x^\mu } + \xi^\mu_\alpha \pd_\mu \phi^I \bigg[ \frac{ \pd \calK }{ \pd \phi^I_- } \bigg]_{\rm PL} + \xi^\mu_\alpha \pd_\nu \pd_\mu \phi^I \kappa_I^\nu   ,
\label{eq:divTmunu1b}  \\
	\pd_\nu \calJ_{a}^\nu & = \omega^I_{a} \bigg[ \frac{ \pd \calK }{ \pd \phi^I_- } \bigg]_{\rm PL} + \pd_\mu \omega^I_{a} \, \kappa^\mu_I .
\label{eq:divJnu2}
\end{align}
For Lagrangians with space-time translation symmetry, generated by $\xi_\alpha^\mu = \delta_\alpha^\mu$, we find the expression for the divergence of the total stress tensor, 
\begin{align}
	\pd_\nu \calT_\mu{}^\nu & = -\frac{ \pd \calL }{ \pd x^\mu } +\pd_\mu \phi^I \bigg[ \frac{ \pd \calK }{ \pd \phi^I_- } \bigg]_{\rm PL} +  \pd_\nu \pd_\mu \phi^I \kappa_I^\nu   .
\label{eq:divTmunu2}
\end{align}

For various reasons, the canonical stress-energy tensor $T_\alpha{}{}^\nu$ or ${\cal T}_\alpha{}^\nu$ are not necessarily the preferred quantities for calculating the energy, momenta, and fluxes of fields. For example, the canonical stress-energy tensor does not source gravitational fields in almost all theories of gravitation, including general relativity. As another example, the canonical stress-energy tensor is not gauge invariant in electromagnetism because space-dependent gauge transformations do not commute with spatial translations. 
In concluding this subsection, we mention that the standard techniques for building a symmetric stress-energy tensor $\theta_{\mu\nu}$ from $T_{\mu\nu}$ follows in the same way as for conservative field theories. These manipulations also carry through for making a symmetric nonconservative stress-energy tensor $\Theta_{\mu\nu}$ from ${\cal T}_{\mu\nu}$. 
Finally, the divergences of $\Theta_{\mu\nu}$ and $\calT_{\mu\nu}$ (or $\theta_{\mu\nu}$ and $T_{\mu\nu}$) are the same and equal the right hand side of (\ref{eq:divTmunu2}) (or (\ref{eq:divTmunu1})).
For more details, see~\cite{Soper, Goldstein, jackson_classical_1999}, for example.

Equations (\ref{eq:calTmunu1})-(\ref{eq:divTmunu2}) constitute some of the main results of this paper.
As in discrete mechanics, there are several interesting consequences. The first
indicates that the total stress, energy, and momenta of the accessible degrees
of freedom include contributions from the nonconservative current density $\kappa_I^\mu$.
The second is that when $\kappa_I^\mu$ is non-zero that the divergence of the stress-energy tensor
necessarily depends on two derivatives of the accessible field variable, as seen 
in the last term in (\ref{eq:divTmunu2}). Another key point is that (\ref{eq:calTmunu1})-(\ref{eq:divTmunu2}) are computed
{\it directly} from the nonconservative potential density $\calK$. Therefore, once
$\calK$ is known then one can directly calculate how the stress-energy tensor and
Noether current density change with time without having to perform separate
calculations to explicitly compute these quantities.

\subsection{Internal energy and closure conditions}
\label{sec:continuumClosure}

In our previous discussion of nonconservative discrete mechanics we considered systems where the energy contained in the inaccessible degrees of freedom, the ``internal'' energy, could feed back into  the dynamics of the the accessible subsystem. Such systems required a closure condition, in addition to the variational principle that determines the dynamics of the accessible degrees of freedom, in order to close the system of equations and allow for the system to be solved. In this section, we consider the internal energy density and closure conditions for continuum systems.

In continuum mechanics, accessible degrees of freedom are often generated through coarse graining procedures, where the the ``fast'' or ``microscopic'' degrees of freedom are treated as inaccessible, while macroscopically averaged, or ``slow'' quantities become the accessible degrees of freedom. In such systems, the energy contained in the inaccessible degrees of freedom are often included in the internal energy density, which may be related to local thermodynamic parameters like density and entropy through an equation of state. If we had access to the full conservative Lagrangian of the system, ${\cal L}_{\rm full}$, including all microscopic degrees of freedom and their interactions, then, in the absence of external forces, the total energy would be conserved such that the divergence of the full conservative stress-energy tensor would have zero time component
\begin{align}
\pd_\mu [T_{\rm full}]_0{}^\mu = 0 
\end{align}

For some coarse-grained systems we can include an internal energy density in the Lagrangian, ${\cal L}$, that accounts for all of the energy of the microscopic inaccessible degrees of freedom, such that the system is \emph{closed}. The \emph{total} nonconservative stress-energy tensor ${\cal T}_\mu{}^\nu = T_\mu{}^\nu + \tau_\mu{}^\nu$ then includes the contributions from the inaccessible momentum flux as well as the energy in the inaccessible subsystem, allowing us to equate the time component of its divergence with that of the full stress-energy tensor
\begin{align}
\pd_\nu {\cal T}_0{}^\nu = \pd_\nu [T_{\rm full}]_0{}^\nu = 0, \label{eq:fullTclosure}
\end{align}
which we take to be the closure condition for closed continuum systems. 

As in the discrete case, other conditions to close the system of
equations are possible, though these in practice will depend on the
specifics of particular systems. We illustrate through specific
examples in Sec.~\ref{sec:ContinuumExamples}.

\section{Examples in field theory} \label{sec:ContinuumExamples}

To show how to use our new formalism for nonconservative classical
field theories that we have developed in the previous section, we
provide several examples, starting with a simple example of coupled
scalar fields, which serves as an analog of the two-oscillator example
from Sec.~\ref{sec:Review}. One of the most useful properties of
action formulations is the ability to construct actions additively for
various interactions and fields. In this spirit we next explore
several example physical systems starting with a simple perfect fluid, then
developing and adding action terms describing various interactions,
through the nonconservative potential ${\cal K}$. These include heat
diffusion, viscous dissipation, and viscoelasticity.

\subsection{Two coupled scalar fields}
\label{sec:phichi}

Consider two relativistic scalar fields, $\phi(x^\alpha)$ and $\chi(x^\alpha)$, nonlinearly coupled to each other and mutually evolving in a flat spacetime from initial data specified at a given instant of time. The action for this (closed) system is 
\begin{align}
	S [ \phi, \chi] = \! \int \!\! d^4x \bigg\{ \frac{1}{2} \partial_\alpha \phi \partial^\alpha \phi + \frac{1}{2} \partial_\alpha \chi \partial^\alpha \chi + \frac{g}{2} \phi^2 \chi \bigg\}
\end{align}
where $g$ is a coupling constant and $\partial_\alpha = \partial / \partial x^\alpha = ( \partial / \partial t, \partial/ \partial x^i)$ with 
$\pd^\alpha = \eta^{\alpha \beta} \pd_\beta = ( \pd / \pd t, - \pd / \pd x^i)$ and $\eta_{\alpha \beta} = {\rm diag} (1, -1, -1, -1)$ the Minkowski metric in rectangular coordinates.
We choose to integrate out the $\chi$ field at the level of the action. Often such a choice would be motivated by the physics of the problem or the relative scales but our choice is motivated by pedagogy.

Upon doubling both fields and choosing to work with the $\pm$ representation,  the nonconservative action is
\begin{align}
	{\cal S} [ \phi_a, \chi_a ] = {} &  \int d^4 x \bigg\{ \partial_\alpha \phi_- \partial^\alpha \phi_+ + \partial_\alpha \chi_- \partial^\alpha \chi_+ \nonumber \\
		& + \frac{g}{2}  \phi^2_+ \chi_- + g \phi_- \phi_+ \chi_+ + \frac{g}{8} \phi_-^2 \chi_- \bigg\}
\label{eq:phichiS2}
\end{align}
The equations of motion for $\chi_\pm$ are linear,
\begin{align}
	\partial^2  \chi_+  = \frac{ g }{2} \phi_+^2 + \frac{g}{8} \phi_-^2 ~~~~ {\rm and} ~~~~
	\partial^2 \chi_- = g \phi_- \phi_+
\end{align}
where $\pd^2 = \pd_\alpha \pd^\alpha$.
Just as in the discrete example with two harmonic oscillators in
Sec.~\ref{sec:IllustrativeRevisited}, the $\chi_+$ equation is solved
using the retarded Green's function since the initial data is
non-trivial in the physical limit while the $\chi_-$ equation is
solved with the advanced Green's function since the data at the final
time $t_f$ is fixed by the equality condition. Therefore,
\begin{align}
	\chi_+ (x^\alpha) & = \chi^{(h)} (x^\alpha) + \int d^4x' \, G_{\rm ret}(x^\alpha, x'{}^\alpha) \nonumber \\
		& {\hskip0.75in} \times \bigg[ \frac{g}{2} \phi_+^2 (x'{}^\alpha) + \frac{ g}{ 8 } \phi_-^2 (x'{}^\alpha) \bigg]  \\
	\chi_- (x^\alpha) & = \! \int \!\! d^4x' \, G_{\rm adv} (x^\alpha, x'{}^\alpha) \Big[ g \phi_- (x'{}^\alpha) \phi_+ (x'{}^\alpha) \Big]
\end{align}
where $\chi^{(h)} (x^\alpha)$ is a homogeneous solution. Note that $\chi_-$ has no homogeneous contribution because we are solving a second order partial differential equation with trivial final data. Substitution of these solutions back into \eqref{eq:phichiS2} gives the nonconservative effective action for $\phi_\pm$,
\begin{align}
	{\cal S}_{\rm eff} [ \phi_a] = {} & \int d^4x \bigg\{ \partial_\alpha \phi_- \partial^\alpha \phi_+ + g \phi_- \phi_+ \chi^{(h)} \bigg\} \nonumber \\
		& + \frac{g^2}{2} \int d^4x \, d^4x' \, \phi_-(x)  \phi_+ (x) G_{\rm ret}(x,x') \nonumber \\
		& {\hskip0.5in} \times \bigg[ \phi_+^2 (x') + \frac{ 1 }{ 4 } \phi_-^2 (x') \bigg]
\end{align}
from which we read off that 
\begin{align}
	{\cal L} = {} & \int d^4 x \, \frac{1}{2} \partial_\alpha \phi \partial^\alpha \phi \\
	{\cal K}  = {} & \frac{g^2}{2} \int d^4x \, d^4x' \, \phi_-(x)  \phi_+ (x) G_{\rm ret}(x,x') \nonumber \\
		& {\hskip0.2in} \times \bigg[ \phi_+^2 (x') + \frac{ 1 }{ 4 } \phi_-^2 (x') \bigg]
\end{align}
The equations of motion for $\phi$ follow by varying ${\cal S}_{\rm eff}$ with respect to $\phi_-$ or by applying (\ref{eq:eom2}) to the equations directly above. Both calculations give the same result,
\begin{align}
	\partial^2 \phi(x) = {} &  \chi^{\!(h)\!} (x) \phi(x) + \frac{g^2}{2} \phi(x) \!\! \int \! d^4 x' \, G_{\rm ret}(x,x') \phi^2(x')  \nn
\end{align}
which depends on the past nonlinear evolution of $\phi(x)$ 
and is the correct equation one would have found by integrating 
out $\chi$ at the level of the equations of motion.

For further examples in relativistic field theories, see~\cite{Galley:2012qs} for integrating out radiative gravitational perturbations
in the post-Newtonian approximation for the compact binary inspirals due to the emission of gravitational waves. 
See also the interesting work of Kevrekidis~\cite{Kevrekidis:PhysRevA.89.010102} who applies nonconservative field theory with
the collective coordinate~\cite{dauxois2006physics} (or variational~\cite{malomed2002variational, malomed2006soliton}) method to find very accurate and practical approximate equations of motion and solutions for 
 nonlinear wave propagation in dissipative sine-Gordon and $\phi^4$ models.

\subsection{Hydrodynamics}\label{sec:Hydro}

An extensive class of classical field theories can be found in problems in hydrodynamics, 
which are pervasive throughout many disciplines and applications.
Many fluids of theoretical and practical interest are dissipative (e.g., viscous friction) and
involve transport processes (e.g., heat diffusion) that are indicative of fluids in thermodynamical non-equilibrium.
In this section we will develop nonconservative actions for classical hydrodynamics, including irreversible processes. 

Andersson and Comer~\cite{Andersson2013}, have recently begun to develop a relativistic description of hydrodynamic dissipation and heat transport using a
constrained convective variational principle (see e.g.,~\cite{Prix:2002jn}), based on geometric considerations and the interaction of lower dimensional matter-space fields, without requiring a near equilibrium expansion. 

Another recent approach to describe dissipative fluids is with effective field theory 
techniques \cite{Endlich:2012vt, Grozdanov:2013dba} where viscous effects are 
included in the action via a perturbative derivative expansion of fluid elements in their comoving frame 
with respect to a stationary background flow.
While these methods show promise for being  guided by the underlying symmetries of
the problem they seem to be inapplicable to non-smooth background flows (e.g., turbulence and shocks). 

Here we adopt a more pragmatic approach and seek to construct actions, using our nonconservative formalism,
transparently in familiar variables to reproduce well-known results in classical hydrodynamics that are useful for practical applications.

In the Eulerian description (see Appendix \ref{sec:ContinuumMechPrelim}) 
 actions for perfect fluids often require a relatively large number of constraints (via Lagrange multipliers) 
to impose conservation of entropy, mass density, Lin number, etc. (see e.g., \cite{ray1972lagrangian}).
Additionally, one takes the mass and entropy densities as dynamical degrees of freedom, 
which is awkward since these quantities are merely functions that characterize and track some average properties of 
the coarse-grained microscopic variables comprising the fluid element. For example, one does not
vary the action for a free particle with respect to its mass in addition to its position.
While the Eulerian description of fluids will be more familiar to most readers, it is far easier to construct
actions for fluid dynamics without constraints in the Lagrangian description (see e.g., \cite{Eckart1960, Katz1961,
  Seliger1968, Morrison:1998zz}).

In Lagrange coordinates (see Appendix \ref{sec:ContinuumMechPrelim}), a fluid element with label $a^A$ (for material-space indices $A = 1, 2, 3$) traces a path in time with coordinates $q^i (t, a^A)$ (for Eulerian-space indices $i = 1, 2, 3$). 
Therefore, the fluid as a whole is a field over the coordinates $(t, a^A)$ with three scalar component functions
indicating the coordinate of a piece of the fluid at time $t$ and label $a^A$. 
We take this field as the dynamical degree of freedom for a fluid. In doing so, we are implicitly coarse-graining
the large number of microscopic degrees of freedom associated with the
individual molecules/atoms comprising the fluid.\footnote{%
We assume that the fluid's microscopic degrees of freedom become
thermalized at a much faster timescale than that of the fluid
elements' trajectories, which are a set of collective variables that
arise from a coarse-graining procedure. We also implicitly assume that
the coarse-graining procedure has introduced an entropy parameter that
parametrizes the fluid elements' internal energy.}
The field $q^i (t, a^A)$ only captures some of the relevant or accessible degrees of freedom and constitutes a set of collective variables 
that are effectively open because the kinetic energy of a fluid element may be transferred to heat energy, which is a thermodynamic (or collective) accounting of the change in the average velocity of the molecules of the fluid element.
In the examples that follow, we build actions for the dynamics of $q^i (t, a^A)$ and additional thermodynamic quantities
as befits the system in consideration. We also assume a single-species fluid for simplicity, 
though this can be straightforwardly generalized.
See Appendix \ref{sec:ContinuumMechPrelim} for further discussion.

We now discuss constructing actions for fluids that exhibit
nonconservative processes, which may include viscous dissipation and
 heat diffusion. 
Before studying more complicated problems involving heat fluxes and/or viscous dissipation, we begin 
 by reviewing the action for an adiabatic inviscid (e.g., perfect) fluid.  
 We refer the reader to Appendix \ref{sec:ContinuumMechPrelim} 
 for the language and notation we use in the following examples.

\subsubsection{Perfect (Inviscid and Adiabatic) Fluids} \label{sec:PerfectFluid}

If a fluid element is in local thermodynamic equilibrium
we can write the internal energy density (per unit Eulerian coordinate
volume) as $\bar{\varepsilon} = \bar{\varepsilon}(\bar{\rho},\bar{s})$
where $\bar{s}$, the Eulerian entropy density, and
$\bar{\rho}$, the Eulerian mass density, are volume densities of the
extensive thermodynamic variables. The differential of
$\bar{\varepsilon}$ is then
\begin{align}
d\bar{\varepsilon} = \mu \,d\bar{\rho} + T \,d\bar{s}
\end{align}
where the local chemical potential (per unit mass), $\mu \equiv (\pd\bar{\varepsilon}/\pd \bar{\rho})_{\bar{s}}$, and the local temperature $T \equiv (\pd\bar{\varepsilon}/\pd \bar{s})_{\bar{\rho}}$ are the intensive thermodynamic variables.  

For a perfect, isentropic fluid with no viscosity or heat transport a conservative Lagrangian density
can then be written as (we roughly follow the action formulation of~\cite{Morrison:1998zz} for a perfect fluid)
\begin{align}
	{\cal L} =  \frac{1}{2} \rho \dot{\bq}^2 - J \bar{\varepsilon}\left(\frac{\rho}{J}, \frac{s}{J}\right) 
\label{eq:PerfectL}
\end{align}
where the time independent mass density for a particular Lagrangian label $a$ is
$\rho = \rho(a) = J \bar{\rho}$ and we have defined 
\begin{align}
	\dot{\bq}^2 \equiv  \dot{q}^i g_{ij} \dot{q}^j = \dot{q}^i \dot{q}_i \, .
\end{align}
Both $\rho$ and $J\bar\varepsilon$ transform as scalar densities of
weight $+1$, and therefore so does $\calL$.
Since there is no heat generation or
heat transport the entropy density for a particular Lagrangian
coordinate label $a$ is similarly time independent and given by $s =
s(a) = J \bar{s}$. In addition, there are no dissipative or 
nonconservative processes here so $\calK = 0$.

The conservative action is constructed from ${\cal L}$ as
\begin{align}
	S = \int dt \, d^3 a \, {\cal L}  \, .
\end{align}
The equations of motion for a fluid element's path follow from
\eqref{eq:eom2} with $\phi^I (t, x) \to q^i (t,a)$.
A straightforward calculation yields
\begin{align}
	& \rho \frac{ \partial }{ \partial t_a} \left( g_{ij} \dot{q}^j \right) - \frac{1}{2} \bar{\pd}_i g_{jk} \dot{q}^j \dot{q}^k \nonumber \\
	& \qquad - A_i{}^A \pd_A \bar{P} + A_i{}^A e_j^B \pd_A e_B^j \, \bar{P}= 0
\label{perfect_eom_with_i0}
\end{align}
where we have recalled the identity $\partial_A A_i^A = 0$ from (\ref{eq:cofactoridentity1}),
and defined the pressure $\bar{P}$\ 
(also a scalar density of weight $+1$) by\footnote{This expression for pressure arises from the extensivity of the energy $E$, entropy $S$, volume $V$, and particle number $N$ in a thermodynamic system, which generates the relation
$E = TS - \bar{P}V + \mu m_p N$ for a single species fluid where $m_p$ is the mass per particle and $N$ is the particle number. }
\begin{align}
	\bar{P} \equiv \mu \bar{\rho} + T\bar{s} - \bar{\varepsilon} \, .
\label{eq:defP}
\end{align}
In deriving (\ref{perfect_eom_with_i0}) we have been careful to
account for the fact that $g_{ij}$ and $J$ have non-vanishing
gradients of $q$ so that $\pd \calL / \pd q^i$ gives a
contribution. The details of the full calculation are given in
Appendix \ref{app:perfectfluid}.
We recognize from \eqref{eq:acc-down-in-pdta} and
(\ref{eq:covDform1}-\ref{eq:withweight1}) that
the first two and last two terms in (\ref{perfect_eom_with_i0})
combine to give covariant derivatives, $D/Dt_a$ and $\nabla_A$,
respectively,
\begin{align}
	\rho \frac{ D v_i }{ D t_a } - A_i{}^A \nabla_A \bar{P} = 0  
\label{perfect_eom_with_i}
\end{align}
with $v_i = g_{ij}\dot{q}^j$.
Equation (\ref{perfect_eom_with_i}) can be written in the more familiar Eulerian form
\begin{align}
	\bar{\rho} \, \bar{\pd}_t v_i + v^j \bar{\nabla}_j v_i +  \bar{\nabla}_i \bar{P} = 0,
\label{eq:Perfecteom}
\end{align}
using (\ref{eq:MaterialDer11}).
For rectilinear coordinates in flat space such that $g_{ij}(t, q) = \delta_{ij}$ we have
\begin{align}
	\bar{\rho} \, \bar{\pd}_t v_i + v^j \bar{\pd}_j v_i +  \bar{\pd}_i \bar{P} = 0.
\label{eq:PerfecteomFlat}
\end{align}

The continuity equation is expressed in the Lagrange coordinates simply as $\partial_t \rho (a) = 0$.
Applying \eqref{eq:ScalarDensityDot} to this we obtain the continuity equation in Eulerian variables,
\begin{align}
	\bar{\pd}_t \bar{\rho} + \bar{\nabla}_i (v^i \bar{\rho}) = 0.
\label{eq:CovariantContinuity1}
\end{align}

From the Lagrangian density ${\cal L}$ in (\ref{eq:PerfectL}), the canonical stress-energy tensor for the perfect fluid has $a$-coordinate components $\calT_\alpha{}^\beta$ given in \eqref{eq:calTmunu1} by
\begin{equation}
\label{eq:PerfectT}
\begin{aligned}
		\calT_0{}^0  &=  \frac{1}{2}\rho \dot{\bq}^2 + J \bar{\varepsilon} \left( \frac{\rho}{J}, \frac{s}{J} \right) , \\
		\calT_0{}^B &=  J \bar{P} \dot{q}^B, \\
		\calT_A{}^0 &=  \rho \dot{q}_A  ,
 \\
		\calT_A{}^B & = \delta_A{}^B \left[- \frac{1}{2} \rho \dot{\bq}^2 + J \bar{P} + J \bar{\varepsilon} \left( \frac{\rho}{J}, \frac{s}{J} \right) \right] \, ,
\end{aligned}
\end{equation}
where $\dot{q}^A \equiv v^A = e_i^A v^i$. 

The time component of the stress-energy tensor divergence in (\ref{eq:divTmunu2}) 
yields the energy equation in Lagrangian variables for a perfect fluid,
\begin{align}
	\pd_t \left( \frac{1}{2}\rho \dot{\bq}^2 + J \bar{\varepsilon} \left( \frac{\rho}{J}, \frac{s}{J} \right) \right) + \nabla_B \left( \dot{q}^B J \bar{P} \right) = 0,
\end{align}
where we note that ${\cal L}$ does not depend explicitly on time, $\partial \calL / \partial t_a = 0$.
We also remark that, after some manipulation, the energy equation can be written in a manifestly covariant form.
Calculating the derivatives and using the continuity equation, $\partial_t \rho(a) = 0$, gives
$\dot{q}^i$ contracted with the equations of motion in  (\ref{perfect_eom_with_i}).
In Eulerian form 
the equation above is
\begin{align}
	\bar{\pd}_t \left(\frac{1}{2} \bar{\rho} {\bm v}^2 + \bar{\varepsilon}(\bar{\rho}, \bar{s}) \right) + \bar{\nabla}_i \left[ v^i \left(\frac{1}{2} \bar{\rho} \bv^2 + \bar{h}\right)\right] = 0,
\end{align}
where the quantity $\bar{h} \equiv \bar{\varepsilon}(\bar{\rho}, \bar{s}) + \bar{P}$ is 
the fluid's enthalpy density. These expressions agree with those given in 
standard texts (see e.g., \cite{landau1986fluids, Batchelor}). 

The spatial components of the stress-energy tensor divergence in (\ref{eq:divTmunu2}) yield 
\begin{align}
&\pd_t (\rho \dot{q}_i e^i_A) + \pd_A \left( -\frac{1}{2}\rho \dot{\bq}^2 + J\bar{P} + J\bar{\varepsilon} \right) \nonumber \\ &\qquad \qquad \qquad = -\frac{1}{2}\dot{\bq}^2 \pd_A \rho + \pd_A (J \bar{\varepsilon} )
\end{align}
where it is important to note that ${\cal L}$ {\it does} depend explicitly on $a$ through the functions $\rho(a)$ and $J$.
After some manipulation we find that the above equation equals the (manifestly covariant) equations of motion in (\ref{perfect_eom_with_i}) contracted with $e^i_A$.

\subsubsection{Viscous Isentropic Fluids 
	(The ``Cold Stone'' Limit)}\label{sec:ColdStoneFluid}

Next, it is instructive to consider the action for a locally isentropic viscous fluid in an ``open'' system such that all heat generated by viscous dissipation is removed through some external mechanism,\footnote{We refer to this rather artificial, but pedagogically useful, system as the ``Cold Stone'' limit of a viscous fluid, in tribute to the ``Cold Stone Creamery'' chain of ice-cream dispensaries, where viscous (yet delicious) fluids are routinely mixed in thermal contact with a heat sink.} such that $\pd_t s = 0$.  

We can account for the leading effect of viscous dissipation through $\calK$
in a couple of different ways. First, we can appeal to the concept that the ``$-$'' variables
can be interpreted as a kind of virtual displacement and identify $\calK$ 
as the work done on the fluid element by viscous friction when it undergoes a small virtual strain 
of $[u_-]_{AB}$. 
Second, we can 
consider linear combinations of the strain rate tensor ${\bm \gamma}$ and
the strain tensor ${\bm u}$ (see App.~\ref{sec:physical-quantities} for definitions)
  along with constants giving a scalar density with units
of energy per unit volume. In either case, we have\footnote{We choose this as the lowest order
  dissipative term since terms with ${\cal K} \propto q_-^i
  [\dot{q}_+]_i$ will yield diffusion-type processes, as we will see
  in Sec.~\ref{sec:inviscid-fluid-diffusion} when considering heat
  diffusion.}
\begin{align}
	{\cal L} &= \frac{1}{2}\rho \dot{{\bm q}}^2 - J \bar{\varepsilon} \left( \frac{\rho}{J}, \frac{s}{J} \right), \\
	{\cal K} &= - [u_-]_{A B} \, \sigma_+^{A B} 
\label{eq:Kcoldstone43b}
\end{align}
where the viscous stress tensor for an isotropic viscous stress can be modeled by
\begin{align}
	\sigma^{A B} &= \left( \eta_s P^{ABCD} + \eta_b C^{AB}C^{CD}\right) \gamma_{CD},
\label{eq:ViscousStress} \\
	&\equiv {\cal V}^{A BC D}\,\gamma_{C D}
\end{align}
and
\begin{align}
	P^{ABCD} \equiv C^{A(C} C^{D)B} - \frac{1}{3} C^{AB} C^{CD}
\label{eq:projector1}
\end{align}
is the projection tensor that
converts arbitrary rank-$2$ tensors to symmetric and trace-free ones.
The coefficients in (\ref{eq:ViscousStress}) are the dynamic or shear viscosity, $\eta_s$, and the
bulk or volume viscosity, $\eta_b$, which are both scalar
densities (weight +1) and may generally be functions of position
and the local thermodynamic variables. Anisotropic viscous stress
will have a more complicated form for ${\cal V}^{A B C D}$.

In Eulerian coordinates the stress tensor transforms as a tensor density giving
\begin{align}
\bar{\sigma}^{ij} &= \left[\bar{\eta}_s P^{ijk\ell} + \bar{\eta}_b g^{ij} g^{k\ell}\right] \gamma_{k\ell},\\
&\equiv \bar{{\cal V}}^{ijk\ell} \, \gamma_{k\ell}
\end{align}
where $\bar{\eta}_s$ and $\bar{\eta}_b$ are scalar densities that
transform such that the kinematic viscosities in each coordinate
system are the same, $\bar{\eta}_s/\bar{\rho} = \eta_s/\rho$ and
$\bar{\eta}_b/\bar{\rho} = \eta_b/\rho$.
The transformation between the Eulerian and Lagrangian stress tensor components is given by 
\begin{align}
	\sigma^{AB} = J e^A_i e^B_j \bar{\sigma}^{ij}.
\label{eq:sigmatransformation1}
\end{align}

The equations of motion for our ``Cold Stone'' fluid can be found from (\ref{eq:eom2}).
Since ${\cal L}$ is identical to that of the perfect fluid, its
contribution to the equations of motion is the same as in \eqref{perfect_eom_with_i}.
However, the nonvanishing $\calK$ gives rise to a dissipative force on the fluid element.

With 
\begin{align}
	[u_-]_{AB} & \equiv \frac{1}{2}([ C_1]_{AB} - [C_2] _{AB}) \\
		& = \frac{1}{2}\pd_A q^i_1 \, \pd_B q^j_1 \, g_{ij}(q_1) - \frac{1}{2}\pd_A q^i_2 \, \pd_B q^j_2 \, g_{ij}(q_2) \nn
\end{align}
we expand $\calK$ in $q_-$ to find
\begin{align}
	{\cal K}
	=  - \frac{1}{2} q_-^i \big( \sigma^{AB} e^j_A e^k_B \bar{\pd}_i g_{jk} \big)_+ \!\!\! - \pd_A q^i_- \big( \sigma^A{}_B e^B_i  \big)_+  + {\cal O}(-^2).
\end{align}
The non-conservative terms of the equations of motion \eqref{eq:eom2} are given by
\begin{align}
\calQ_i &= \pd_A (\sigma^A{}_B e_i^B) - \frac{1}{2} \sigma^{AB} e_i^C \pd_C C_{AB} - \sigma^A{}_B e^j_A \bar{\pd}_i e^B_j , \nonumber\\
&= \pd_A (\sigma^A{}_B e_i^B) - \sigma^A{}_B \Gamma^B_{AC} e^C_i - \sigma^A{}_B e^j_A \bar{\pd}_j e^B_i  , \nonumber\\
&= e_i^B \nabla_A \sigma^A{}_B 
\end{align}
where $\Gamma^B{}_{AC}$ are the Christoffel connection coefficients in
Lagrangian coordinates, given explicitly in
\eqref{eq:Christoffel-L}.
This gives the the equations of motion for the `Cold Stone' fluid, 
\begin{align}
	\rho \frac{D v_i }{Dt_a}  + A_i{}^A \nabla_A \bar{P} = e^B_i \nabla_A \sigma^A{}_B
\label{eq:ColdStoneEom}
\end{align}
which are equivalent the Navier-Stokes momentum conservation equations. Note that we have already computed the left side of (\ref{eq:ColdStoneEom}) in the previous subsection. In the Eulerian coordinates this becomes
\begin{align}
	\bar{\rho} \, \bar{\pd}_t v_i  + \bar{v}^j \bar{\nabla}_j v_i +  \bar{\nabla}_i \bar{P} = \bar{\nabla}_j \bar{\sigma}^j{}_i \,.
\label{eq:ColdStoneEom2}
\end{align}
For rectilinear Euler coordinates in a flat space these become
\begin{align}
	\bar{\rho} \, \bar{\pd}_t v_i  + \bar{v}^j \bar{\pd}_j v_i +  \bar{\pd}_i \bar{P} = \bar{\pd}_j \bar{\sigma}^j{}_i ,
\end{align}
which is the familiar form of the Navier-Stokes equations of motion for viscous fluid flows.
Note that we used (\ref{eq:C-from-g-e}) and (\ref{eq:sigmatransformation1}) in arriving at this form.

The canonical stress energy tensor when including nonconservative interactions is
given by \eqref{eq:calTmunu1} such that
\begin{align}
	\calT_0{}^0 & = \frac{1}{2} \rho \dot{\bq}^2 + J \bar{\varepsilon} \left( \frac{\rho}{J} , \frac{ s }{ J } \right)   \nn \\
	\calT_0{}^B & =  J \bar{P} \dot{q}^B -  \sigma^B{}_C e^C_i  \nn \\
	\calT_A{}^0 & = \rho \dot{q}_A  \nn \\
	\calT_A{}^B & = \delta_A{}^B \left[ -\frac{1}{2} \rho \dot{\bq}^2 + J \bar{P} + J \bar{\varepsilon}\left( \frac{ \rho}{J}, \frac{ s}{J} \right) \right] -\sigma_A{}^B 
\end{align}
The spatial components of the divergence of the total stress-energy tensor \eqref{eq:divTmunu2} once again yield the equations of motion \eqref{eq:ColdStoneEom}, while the time component gives
\begin{align}
\pd_t \bigg[\frac{1}{2} \rho \dot{\bm q}^2 &+ J \bar{\varepsilon}(\bar{\rho}, \bar{s}) \bigg] +  \nabla_B (\dot{q}^B J \bar{P} - \dot{q}^C \sigma^{B}{}_C) \nonumber\\
&= - \sigma^{A}{}_B \nabla_A \dot{q}^B \,,
\end{align}
which is
the energy equation for our ``Cold Stone'' fluid. In the Euler coordinates this can be written as
\begin{align}
\bar{\pd}_t \left[ \frac{1}{2}\bar{\rho} {\bm v}^2 + \bar{\varepsilon}(\bar{\rho}, \bar{s}) \right]  &+ \bar{\nabla}_i \left\{ v^i \left(\frac{1}{2}\bar{\rho} \bv^2 + \bar{h}  \right) - v^j\, \bar{\sigma}^{i}{}_j\right\}\nonumber\\
 &= - \bar{\sigma}^{ij} \bar{\nabla}_i v_j,
\end{align}
where the first term inside the divergence, $ \bar{\nabla}_i \{ \ldots \}$, we recognize as the energy flux due to the mass transfer of the fluid, while the second term is the energy flux due to viscous shear. The right hand side is simply the rate of energy loss due to viscous dissipation.

\subsubsection{A Viscous Perfectly Insulating Fluid} \label{sec:InsulatingFluid}
We now consider a viscous fluid that has perfectly insulating fluid
elements, such that no heat diffusion is allowed between adjacent
fluid elements. The system is ``closed'' such that viscously
dissipated energy is deposited as heat into the fluid elements, in
contrast with the open system described in the Cold Stone case. To do
this we relax the condition on the entropy field, allowing it to
explicitly depend on both time and the local fluid coordinate, such
that $s = s(t, \ba)$, noting that the second law of thermodynamics
requires that $\pd_t s(t, \ba) \geq 0$.  This acts as an explicit time
dependence in the Lagrangian, which can be viewed as an externally
specified function that does not depend on the dynamical degrees of
freedom $\bq$.

The Euler-Lagrange equations give equations of motion for $q^i$  that are the same as for the Cold Stone case in (\ref{perfect_eom_with_i}). 
We also recover the continuity equation in (\ref{eq:CovariantContinuity1}).
As noted in Sec.~\ref{sec:continuumClosure} the equations of motion
and the continuity equation are no longer enough to
``close'' the system of equations. In general, we need an additional
equation describing the evolution of the entropy $s(t, \ba)$, 
which must be specified in addition to the variational principle.
We shall show that this comes about through a ``closure relation.''

Consider the zeroth component of the divergence of the total stress
tensor, namely, with space-time index $\mu=0$ in \eqref{eq:divTmunu2}, 
\begin{align}
\pd_\nu {\cal T}_0{}^\nu &=
\pd_t \bigg[\frac{1}{2} \rho \dot{\bm q}^2 + J \bar{\varepsilon}(\bar{\rho}, \bar{s}) \bigg] +  \nabla_A \left( \dot{q}^A J \bar{P} - \dot{q}^B \sigma^{A}{}_B \right) \nonumber\\
	&= T \pd_t s - \sigma^A{}_B  \nabla_A \dot{q}^B  \, .
\end{align}
Note that the Lagrangian density has an explicit time dependence through the entropy, $s(t, a)$, such that
\begin{align}
	\frac{\pd \calL }{\pd t} = -T \pd_t s(t, a).
\end{align}

Since this fluid is a \emph{closed} system, (see
Sec.~\ref{sec:continuumClosure}) we expect all energy dissipated from
the accessible degrees of freedom to go into the internal energy such
that the total energy is conserved.  We then apply the closed system
closure condition $\pd_\nu {\cal T}_0{}^\nu = 0$, which gives the
energy and entropy equations
\begin{align}
&\pd_t \bigg( \frac{1}{2} \rho \dot{\bm q}^2 + J\bar{\varepsilon}(\bar{\rho}, \bar{s}) \bigg) +  \nabla_A \left(\dot{q}^A J \bar{P} - \dot{q}^B \sigma^{A}{}_B \right) = 0,\\
&T \pd_t s = \sigma^A{}_B  \nabla_A \dot{q}^B \, ,
\end{align}
which can be written in terms of the Eulerian coordinates
\begin{align}
	&\bar{\pd}_t \left( \frac{1}{2}\bar{\rho} {\bm v}^2 + \bar{\varepsilon}(\bar{\rho}, \bar{s}) \right)  + \bar{\nabla}_i \left[ v^i \left(\frac{1}{2}\bar{\rho} \bv^2 + \bar{h}  \right) - v_j\, \bar{\sigma}^{ij}\right] = 0,  \\
	& \bar{\pd}_t \bar{s} + \bar{\nabla}_i ( v^i \bar{s}) =  \frac{1}{T} \bar{\sigma}^{ij} \bar{\nabla}_i v_j.
\end{align}
Thus we see that entropy is generated by the irreversible viscous dissipation, and that the second Law of Thermodynamics, $\pd_t s \geq 0$, requires that the coefficients $\eta_s$ and $\eta_b$ in the viscous stress tensor be positive. 

We note that the open isentropic system condition, $\pd_t s = 0$, and closed system condition, $\pd_\nu {\cal T}_0{}^\nu = 0 $, are not the only ways to close the system of equations. Other systems can be specified giving different entropy and energy equations. We could imagine, for example, some fixed fraction of the energy going into the internal energy of the fluid while the rest escapes to external inaccessible degrees of freedom, or an isothermal condition on the fluid specifying the energy and entropy equations by a fixed temperature condition. For the simple fluid systems that follow, however, the most useful condition is the closure condition $\pd_\nu {\cal T}_0{}^\nu = 0$, which allows us to recover the expected behavior for the isolated fluid system, where the inaccessible degrees of freedom are only the microscopic coarse-grained degrees of freedom that contribute to the internal energy.

\subsubsection{Inviscid Fluid with Heat Diffusion}
\label{sec:inviscid-fluid-diffusion}
To consider a fluid with heat diffusion we adopt the approach of
Prix~\cite{Prix:2002jn} and Andersson and
Comer~\cite{2010RSPSA.466.1373A} who treat the entropy as an auxiliary
massless fluid, with its own degrees of freedom. Here we will take the
labels of the entropy `fluid' $\alpha^{A_s}(t, \ba)$ to be auxiliary
degrees of freedom, which will have components labeled by the indices
$A_s, B_s, C_s$. This is equivalent to treating the diffusive heat 
flux as a separate degree of freedom. We can then construct a Jacobian matrix
$e^{A_s}_A \equiv (\pd \alpha^{A_s}/\pd a^A)_t$ and determinant ${\cal
  J}_s = \det(e^{As}_A)$. We also have the inverse Jacobian $e^A_{A_s}
= \frac{1}{{\cal J}_s} [{\cal A}_s]_{A_s}{}^A$ , where $[{\cal
  A}]_{A_s}{}^A \equiv (\pd {\cal J}_s/\pd e^{A_s}_A)$ is the cofactor
matrix. Again, this assumes that the Jacobian is non-zero and we are
in regions free from shocks. We can also express the metric in the
${\bm \alpha}$ coordinates that we can use to raise and lower the
$A_s$ indices,
\begin{align}
{\cal C}_{A_s B_s} \equiv e^A_{A_s} e^B_{B_s} C_{AB} = e^A_{A_s} e^B_{B_s} e^i_A e^j_B g_{ij}.
\end{align}

The mass density and velocity of the material are not directly
affected by the perturbations of the entropy fluid label
$\alpha$. However, the entropy density is now given by
\begin{align}
s(t, a) = J \bar{s}(t, q) = \mathcal{J}_{s}\tilde{s}(t, \alpha)
\end{align}
where the $\tilde{s}$ denotes the entropy density in the $(t,  \alpha)$ coordinates. As in the closed system above, we allow the entropy density to be time dependent even in the $\alpha$ coordinates, such that $\tilde{s} = \tilde{s}(t,  \alpha)$. Here, the second law of thermodynamics requires that we have locally $\pd \tilde{s}/\pd t_\alpha \geq 0$. 

The velocity of the entropy fluid is given by
\begin{align}
v_s^i &\equiv \frac{\pd}{\pd t_\alpha} q_s^i(t, \alpha)  \\ 
	&= \frac{\pd}{\pd t_a} q^i (t, a_s(t, \alpha))  + \pd_A q^i \frac{\pd}{\pd t_\alpha} a^{A}_s(t, \alpha)
\end{align}
where $a^A_s(t, \alpha^{A_s})$ and $q_s^i (t, \alpha^{A_s}$) are the material and Eulerian coordinate positions, respectively, of
the entropy fluid element labeled by coordinates $\alpha^{A_s}$. We can utilize the
relative velocity identity \eqref{eq:relvel} to
write
\begin{align}
	\frac{\pd a_s^A}{\pd t _\alpha} = - \left[\frac{\pd \alpha^{A_s}}{\pd a^A} \right]^{-1} \frac{\pd \alpha^{A_s}}{\pd t_a} 
		= - e^A_{A_s} \pd_t \alpha^{A_s}
\end{align}
which gives
\begin{align}
v_s^i = v^i - e^i_A e^A_{A_s} \pd_t \alpha^{A_s}
\end{align}
and defines the relative velocity, $\Delta^i = v_s^i - v^i$ such that
\begin{align}
\Delta^i = - e^i_A e^A_{A_s} \pd_t \alpha^{A_s}.
\end{align}

Consider the Lagrangian and ${\cal K}$ density for a fluid with isotropic heat diffusion given by
\begin{align}
{\cal L} &= \frac{1}{2} \rho \dot{q}^2 - J \bar{\varepsilon} \left( \frac{\rho}{J} , \frac{s}{J} \right) \\
{\cal K} &= -\zeta_+ [\Delta_+]_i q^i_{s-} = - \zeta_+\left[\pd_t \alpha_+\right]_{A_s} \alpha_-^{A_s} + {\cal O}(-^3)
\label{eq:Kheatdiffusion37m}
\end{align}
where $\zeta = J \bar{\zeta}$ is a scalar density of
weight $+1$ that is related to the thermal resistivity (see below) and
may in general depend on time, position and the local thermodynamic
variables. For more general anisotropic heat diffusion, we can
replace the scalar density $\zeta$ with a tensor density.
We also note that $q^i_{s-} \equiv q_{s1}^i - q_{s2}^i =  e^i_A e^A_{A_s}
\alpha^{A_s}_- + {\cal O}(-^3)$, since near the physical limit, the minus
variables are elements of a vector space. The expression $[\pd_t \alpha_+]_{A_s}$
is shorthand for $(\pd \alpha^{B_s}/\pd t_a){\cal C}_{A_s B_s}$, which is geometrically
well defined.

The equations of motion for $q^i$, obtained by varying with respect to $q^i_-$, remains unchanged from that of the perfect fluid
in (\ref{eq:Perfecteom}) and the continuity equation (\ref{eq:CovariantContinuity1}).
 
 To get the equations of motion for the entropy fluid we vary the nonconservative 
 Lagrangian with $\phi^I \to \alpha^{A_s}$ in (\ref{eq:eom2}) and simply find
 \begin{align}
s \, e^A_{A_s}  \nabla_A T = \zeta \,{\cal C}_{A_s B_s}\, \pd_t \alpha^{B_s} \label{eq:alphaEOM}
 \end{align}
 We have used the fact that $T$ is a scalar (and not a scalar density) to write $\pd_A T = \nabla_A T$.

We can define the Eulerian diffusive heat flux density to be
\begin{align}
\bar{{\cal F}}^i \equiv T \bar{s} \Delta^i \label{eq:HeatFluxdef}
\end{align}
which transforms as a vector density of weight $+1$, such that 
\begin{align}
{\cal F}^A &\equiv J e^{A}_{i} \bar{\cal F}^i = - Ts \,e^A_{A_s} \pd_t \alpha^{A_s}, \\ 
\tilde{\cal F}^{A_s} & \equiv \frac{ J }{ {\cal J}_s}   e^{A_s}_A e^A_i \bar{\cal F}^i = -T\tilde{s} \,\pd_t \alpha^{A_s}.  
\end{align}
The equation of motion \eqref{eq:alphaEOM} can then be rearranged to give, 
\begin{align}
{\cal F}_A = -\frac{T{s}^2}{{\zeta}} \nabla_A T
\end{align}
which simply has Eulerian form
\begin{align}
\bar{{\cal F}}_i = -\frac{T\bar{s}^2}{\bar{\zeta}} \bar{\nabla}_i T = - \bar{\kappa} \bar{\nabla}_i T
\end{align}
This is simply Fourier's law of heat conduction where the scalar density $\bar{\kappa} \equiv T \bar{s}^2/\bar{\zeta}$ is the thermal conductivity of the material. The thermal resistivity of the material  is defined to be the inverse of the conductivity $1/\bar{\kappa} = \bar{\zeta}/(T\bar{s}^2)$.

The components of the canonical stress-energy tensor are
\begin{equation}
\begin{aligned}
\calT_0{}^0  &= \frac{1}{2}\rho \dot{\bq}^2 + J \bar{\varepsilon}\\
\calT_0{}^B &=  J \bar{P} \dot{q}^B + \Delta^B \,T s \\
\calT_A{}^0 &= \rho \dot{q}_A ,\\
\calT_A{}^B &= \delta_A{}^B \left[- \frac{1}{2} \rho \dot{\bq}^2 + J \bar{P} + J \bar{\varepsilon} - J T \bar{s} \right]
\end{aligned}
\end{equation}
where, in the last line, we recognize $\bar{P} + \bar{\varepsilon} - T \bar{s}$ as the Gibbs free energy density.
Note that the nonconservative piece, $\tau_\mu{}^\nu$, vanishes so that $\calT_\mu{}^\nu = T_\mu{}^\nu$ since $\kappa_I^\mu = 0$ for this system.
 
We can then write the divergence of the stress-energy tensor as
\begin{align}
\pd_\nu \calT_\mu{}^\nu &= -\frac{\pd \calL}{\pd a^\mu} + \pd_\mu \alpha^{A_s} \left[\frac{\pd {\cal K}}{\pd \alpha_-^{A_s}}\right]_{\rm PL}
\end{align}
Note that the right hand side has no contributions from derivatives with respect to $q^i$ or $e^i_A$.
The spatial components of the divergence give the sums of the equations of motion for the $\bq$ and ${\bm \alpha}$ degrees of freedom, while
the time component of the divergence yields
\begin{align}
&\pd_\nu {T}_0{}^\nu = - \frac{\pd \calL}{\pd t} +  \pd_t \alpha^{A_s} \left[\frac{\pd {\cal K}}{\pd \alpha_-^{A_s}}\right]_{\rm PL}
\end{align}

Combined with the closure condition, $\pd_\nu {\cal T}_0{}^\nu = 0$,
the time component gives the energy equation
\begin{align}
&\pd_t  \bigg( \frac{1}{2}\rho \dot{q}^2 + J \bar{\varepsilon} \bigg) +  \nabla_A (\dot{q}^A \bar{P} + \Delta^A T s )
= 0,
\end{align}
and the entropy equation
\begin{align}
{\cal J}_{s} T \frac{\pd \tilde{s}}{\pd t_\alpha}  = \zeta [\pd_t \alpha]_{As} \pd_t \alpha^{As}.
\end{align}

Noting from the convective derivative \eqref{eq:lie-sum} that 
\begin{align}
	\frac{ {\cal J}_s }{ J } \frac{ \pd \tilde{s} }{ \pd t _\alpha } &= \bar{\pd}_t \bar{s} + {\cal L}_{{\bm v}_s } \bar{s}\\ 
		&=  \bar{\pd}_t \bar{s} + \bar{\nabla}_i (v^i \bar{s} + \Delta^i \bar{s}),
\end{align}
where ${\bm v}_s = {\bm v} + {\bm \Delta}$,
and using the equation of motion \eqref{eq:alphaEOM}
we can write the Eulerian version of the energy equation
\begin{align}
\bar{\pd}_t \left(\frac{1}{2}\bar{\rho} v^2 + \bar{\varepsilon} \right) + \bar{\nabla}_i \left\{v^i\left(\frac{1}{2} \bar{\rho}v^2 + \bar{h} \right) - \bar{\kappa} \bar{\nabla}^i T\right\} = 0
\end{align}
and the entropy equation
\begin{align}
\bar{\pd}_t \bar{s} + \bar{\nabla}_i \left[ (v^i + \Delta^i ) \bar{s} \right] = \frac{\bar{\kappa}}{T^2}  \bar{\nabla}_i T \, \bar{\nabla}^i T
\end{align}
From this we can see immediately that the second law of thermodynamics $\pd \tilde{s}/\pd t_\alpha \geq 0$ requires that $\bar{\kappa} \geq 0$, implying $\zeta \geq 0$. 

The entropy equation can be re-written in a more familiar form relating the entropy gain to the diffusive heat flux using the definition of the heat flux \eqref{eq:HeatFluxdef} and the equation of motion \eqref{eq:alphaEOM}
\begin{align}
\bar{\pd}_t \bar{s} + \bar{\nabla}_i ( v^i \bar{s}) = -\frac{1}{T} \bar{\nabla}_i \bar{{\cal F}}^i= \frac{1}{T}\bar{\nabla}_i ( \bar{\kappa} \bar{\nabla}^i T).
\end{align}

\subsubsection{Navier-Stokes with Heat Flow} \label{sec:Navier-Stokes}
We are now prepared to construct an action to generate the full equations
of irreversible fluid dynamics, including the effects of viscosity, heating, and heat diffusion.
The action for a Navier-Stokes fluid with heat conduction is given in terms of $\calL$ and $\calK$ by
\begin{align}
	{\cal L} &= \frac{1}{2}\rho \dot{\bm q}^2 - J \bar{\varepsilon} ( \bar{\rho}, \bar{s} )
\label{eq:calLNS1} \\
	{\cal K} &=- [u_-]_{AB} \, \sigma_+^{AB} - \zeta_+ [\Delta_+]_i q_{s-}^i
\label{eq:calKNS1}
\end{align}
where, as in the previous section and following \cite{Prix:2002jn, 2010RSPSA.466.1373A}, $q_s^i$ 
are the Eulerian coordinates of an entropy element labeled by $\alpha$
such that $q_s^i = q^i(t, a_s(t, \alpha))$.
 
As before we choose $q^i(t, a)$ and $\alpha^{A_s}(t, a)$ to be the degrees of freedom. 
We also fix the mass density of the material fluid to the flow of the $a$-coordinates, $J\bar{\rho}(t,q) = \rho(a)$, 
but allow the entropy density to have an explicit time dependence $(J/{\cal J}_s) \bar{s} = \tilde{s}(t, \alpha)$ with respect to the $\alpha$-coordinates.  
Notice that $\calK$ in (\ref{eq:calKNS1}) is given by the sum of the nonconservative potentials in 
(\ref{eq:Kcoldstone43b}) and (\ref{eq:Kheatdiffusion37m}). Furthermore, the first contribution in 
(\ref{eq:calKNS1}) is independent of the entropy fluid.

From (\ref{eq:calLNS1}) and (\ref{eq:calKNS1}), we can obtain the equations 
of motion for a viscous fluid with heat diffusion. 
Variation with respect to $q^i(t,a)$ yields the Navier-Stokes momentum equations
given already in (\ref{eq:ColdStoneEom2}),
\begin{align}
	\bar{\rho} \, \bar{\pd}_t v_i  + \bar{v}^j \bar{\nabla}_j v_i +  \bar{\nabla}_i \bar{P} = \bar{\nabla}_j \bar{\sigma}^j{}_i  \, .\nonumber
\end{align}
Likewise, $\pd_t \rho=0$ gives the continuity equation
in (\ref{eq:CovariantContinuity1}),
\begin{align}
	\bar{\pd}_t \bar{\rho} + \bar{\nabla}_i (v^i \bar{\rho}) = 0 \, . \nonumber
\end{align}
Finally, variation with respect to $\alpha^{A_s}(t,a)$ yields the diffusive 
heat flux density in (\ref{eq:HeatFluxdef})
\begin{align}
	\bar{s}\bar{\nabla}_i T= - \bar{\zeta} \Delta_i  ~~ \Longrightarrow  ~~ \bar{{\cal F}}_i = -\bar{\kappa} \bar{\nabla}_i T \nonumber
\end{align}
which is an expression of Fourier's law of heat conduction. Notice that the additivity of the
viscosity and heat diffusion pieces in $\calK$ have allowed us to  recycle previous
calculations for deriving the equations of motion.

The components of the nonconservative stress-energy tensor are found to be
\begin{equation}
\begin{aligned}
\calT_0{}^0  &= \frac{1}{2}\rho \dot{\bq}^2 + J \bar{\varepsilon} \\
\calT_0{}^B &= J \bar{P} \dot{q}^B - \sigma^B{}_C \dot{q}^C  + \Delta^B T s \\
\calT_A{}^0 &= \rho \dot{q}_A \\
\calT_A{}^B &=  \delta_A{}^B \left[- \frac{1}{2} \rho \dot{\bq}^2 + J \bar{P} + J \bar{\varepsilon} - J T \bar{s} \right] - \sigma_A{}^B.
\end{aligned}
\end{equation}
The last two terms in $\calT_0{}^B$ and $\calT_A{}^B$ are
contributions coming from the nonconservative effects of viscosity,
heating, and heat diffusion, and arise because $\kappa_I^\nu$ defined in \eqref{eq:Noetherkappa1} is non-zero.

The divergence of the total stress-energy tensor gives for the $0$th component,
\begin{align}
\pd_\nu {\cal T}_0{}^\nu = {\cal J}_s T \frac{\pd \tilde{s}}{\pd t_\alpha} - \zeta \, [\pd_t \alpha]_{As}\,\pd_t \alpha^{As}  - \sigma^A{}_B \nabla_A \dot{q}^B
\end{align}
Combining this with the closure condition, $\pd_\nu {\cal T}_0{}^\nu = 0$, gives the energy equation, 
 \begin{align}
 \bar{\pd}_t \left(\frac{1}{2}\bar{\rho} \bv^2 + \bar{\varepsilon} \right) &+ \bar{\nabla}_i \bigg\{v^i\left(\frac{1}{2} \bar{\rho} \bv^2 + \bar{h} \right)\nonumber \\
 &\qquad \qquad - v_j\, \bar{\sigma}^{ij} - \bar{\kappa} \bar{\nabla}^i T\bigg\} = 0
 \end{align}
 and the entropy equation
 \begin{align}
 \bar{\pd}_t \bar{s} + \bar{\nabla}_i (v^i \bar{s}) =  \frac{1}{T}\bar{\nabla}_i ( \bar{\kappa} \bar{\nabla}^i T) + \frac{1}{T} \bar{\sigma}^{ij} \bar{\nabla}_i v_j.
 \end{align}

Remarkably, we have been able to generate the entropy evolution equation for a non-equilibrium thermodynamic system by simply using the nonconservative variational principle, and the closed system closure condition. We expect our formalism to apply equally well to other non-equilibrium coarse-grained systems where the (effectively open) accessible degrees of freedom can be described by a nonconservative action that accounts for the energy of the microscopic degrees of freedom that have been coarse-grained out. This allows us capture the thermodynamics of generic non-equilibrium systems using an action principle formulation.

\subsection{Microhydrodynamics and movable boundaries}
\label{sec:microhydro}

We next consider a system where dynamical boundary effects can be expected to dominate. The effect 
of such boundaries are considered in App.~\ref{sec:boundaries} for 
deriving the nonconservative Euler-Lagrange equations of motion and 
generalizing Noether's theorem.

In some applications, a liquid fluid may contain extended objects that 
are very small compared to the typical length scales of the fluid as a 
whole. These particles compose the suspension microstructure of the 
fluid and thus their dynamics is important for many applications, especially 
industrial ones. From a practical point, the particles tend 
to vary between $10^{-3}$ and hundreds of microns in size~\cite{kim1991microhydrodynamics} 
and can have very complicated geometries, which may evolve dynamically 
in time as is the case with microscopic biological swimmers.

If $V$ and $\ell$ are the typical velocity and length scale, respectively, of a small 
particle in a fluid with mass density $\rho$ and (dynamic) viscosity $\eta$ then the 
Reynolds number ${\rm Re}$ is given by
\begin{align}
	{\rm Re} = \frac{ \rho V \ell }{ \eta } 
\label{eq:reynolds1}
\end{align}
At small length scales and low velocities, the effects of viscosity dominate the 
particle's evolution so that ${\rm Re} \ll 1$, which is called the Stokes regime or limit. Therefore, the kinetic energy of the 
fluid's motion generated by the particle will be much smaller than the internal 
energy of any fluid element.

Starting from the action for a viscous fluid with heat flow from
Sec.~\ref{sec:Navier-Stokes}, we recall from (\ref{eq:calLNS1})
and (\ref{eq:calKNS1}) that the conservative Lagrangian and
nonconservative potential densities are 
\begin{align}
	{\cal L} & = \frac{1}{2} \rho \dot{\bq}^2 - J \bar{\varepsilon} (\bar{\rho}, \bar{s} ) , \nonumber
\\
	{\cal K} & = - [ u_- ] _{AB} \sigma_+ ^{AB} - \zeta_+ [ \Delta_+]_i q_{s-}^i .\nonumber
\end{align}
The ratio of the viscous part of ${\cal K}$ to the inertial part of ${\cal L}$ scales 
like ${\rm Re} \ll 1$. Likewise, the inertial energy density of the fluid is much 
smaller than the internal energy density because the particles move slowly in 
the viscous-dominated regime. Following~\cite{kim1991microhydrodynamics}, 
we also assume that the fluid is thermally well equilibrated and the heat flow is 
negligibly small on the length scales relevant to micro-hydrodynamical 
processes. We also assume the heating due to dissipation is much smaller 
than the internal energy. Consequently, we take the entropy density to be 
time-independent and the fluid to be isothermal. Likewise, the change in the 
density under these conditions is negligible so that the liquid fluid is approximately incompressible,
$\bar{\nabla}_i v^i = \nabla_A ( J \dot{q}^A ) = 0$, 
and we may take the shear viscosity $\bar{\eta}_s$ to be constant 
(the bulk viscosity does not contribute because of the fluid's incompressibility).
Therefore, in the limit of small Reynolds number there are only two terms 
contributing to the action $\calS$,
\begin{align}
	{\cal L} & \approx - J \bar{\varepsilon} (\bar{\rho}, \bar{s})   ,
\label{eq:microL2} \\
	{\cal K} & \approx - [ u _- ]_{AB} \sigma_+^{AB}  .
\label{eq:microK2}
\end{align}
The presence of small  particles that can respond to the fluid as well as act 
upon it suggests that we need to account for such effects. For the purposes 
of deriving the fluid equations, we need to keep track of the surface integrals 
that result from integrating by parts when doing the variational principle. This 
is described for general problems in App.~\ref{sec:boundaries}. 

For $N$ small extended objects in the fluid the microhydrodynamical equations 
of motion follow by applying (\ref{eq:eom3}) to (\ref{eq:microL2}) and (\ref{eq:microK2}), 
which yields 
\begin{align}
	& e^B_i \nabla_A \big( \sigma^A{}_B - \delta^A{}_B J\bar{P} \big) \nn \\
	& ~~~~ = -  \sum_{n=1}^N \oint_{\partial V_n} {\hskip-0.1in} dS'_A \, \delta^3 (a - a' )  \big( \sigma^A{}_B e^B_i - J \bar{P} e^A_i \big) 
\label{eq:stokes1}
\end{align}
after a little algebra where $d S'_A$ is the oriented coordinate area element of the $n^{\rm th}$ surface pointing {\it into} the particle's volume. Here, $\delta^3 (a - a')$ is a Dirac delta function that gives a distributional contribution when $a$ is on a point of the surface $\pd V_n$ of the $n^{\rm th}$ particle. 
The surface integrals are taken over the surfaces, parametrized by coordinates 
$a'$ of each of the $N$ particles in the fluid.
We notice that the right side of (\ref{eq:stokes1}) is simply 
the force density exerted by the particles on the fluid~\cite{kim1991microhydrodynamics}
\begin{align}
	f_i & \equiv \sum_{n=1}^N \oint_{\partial V_n} {\hskip-0.1in} dS'_A \,  \delta^3 (a - a' )  \big( \sigma^A{}_B e^B_i - J \bar{P} e^A_i \big)  .
\label{eq:stokesforce1}
\end{align}
To close the system requires the equations of motion for the particles themselves. As this depends on the particular nature of the problem under consideration (e.g., the geometries of the small bodies and whether or not the particles are active or passive swimmers) we
will not consider this here.

In Eulerian variables (\ref{eq:stokes1}) becomes
\begin{align}
	\bar{\eta}_s \bar{\nabla}^2 v_i - \bar{\nabla}_i \bar{P} & = - \! \sum_{n=1}^N \oint_{\partial V_n} \!\!\! dS'_j \, \delta^3 (x - x')  \big( \bar{\sigma}^{j}{}_i - \delta^{j}{}_i \bar{P} \big)  \nn
\end{align}
where we have written (\ref{eq:stokesforce1}) as volume integrals 
over a divergence using Gauss' theorem, transformed to Eulerian 
variables, and reapplied Gauss' theorem to express the right side 
as a surface integral.

The total canonical stress-energy tensor of the fluid in the Stokes regime has components given by
\begin{align}
	{\cal T}_0{}^0 & = J \bar{\varepsilon} \nn \\
	{\cal T}_0{}^B & = J \bar{P} \dot{q}^B  -  \sigma^B{}_C \dot{q}^C \nn  \\
	{\cal T}_A{}^0 & = 0 \nn \\
	{\cal T}_A{}^B & = \delta_A{}^B \big( J \bar{P} + J \bar{\varepsilon} \big)  - \sigma_A{}^B  .
\end{align}
From (\ref{eq:noetherboundariesdiv2}), the time component of the divergence gives the energy equation,
\begin{align}
	& \pd_t ( J \bar{\varepsilon} ) + \nabla_A \big( J \bar{P} \dot{q}^A - \sigma^A{}_B \dot{q}^B \big)  \nn \\
		& ~~~~ = -  \sigma^A{}_B \nabla_A \dot{q}^B \nn \\
	   & ~~~~ ~~~ - \sum_{n=1}^N \oint_{\partial V_n} \!\!\! dS'_A \, \delta^3 (a - a' )  \big( \sigma^A{}_B \dot{q}^B - J \bar{P} \dot{q}^A \big) .
	  \nn
\end{align}
The first term on the right side is the energy lost due to viscosity in the fluid while the second term is the energy lost by the fluid to moving the small particles. In Eulerian variables, the above expression is 
\begin{align}
	& \bar{\partial}_t \bar{\varepsilon} (\bar{\rho}, \bar{s})  + \bar{\nabla}_i \big( v^i \bar{h} - \bar{\sigma}^{ij} v_j \big) \nonumber \\
	& ~~ = - \bar{\eta_s} \bar{\nabla}_i v_j \bar{\nabla}^i v^j \! - \!\! \sum_{n=1}^N \! \oint_{\partial V_n} {\hskip-0.15in} dS'_i \, \delta^3 (x - x' )  \big( \bar{\sigma}^{j}{}_i - \delta^{j}{}_i \bar{P} \big) \dot{q}^i
	\nn
\end{align}
where $\bar{h}$ is the enthalpy density.

\subsection{Visco-elastic fluids} \label{sec:Viscoelastic}

We next consider the dynamics of a simplified isotropic viscoelastic
fluid, utilizing the Maxwell model, analogous to the Maxwell elements
in Sec.~\ref{sec:maxw-elem-spring}. While the Maxwell model is a
particularly simple model of viscoelastic behavior, it is instructive
to consider the action describing its dynamics, as it can be easily
generalized to more sophisticated rheological models.

We can attempt to generalize the discrete Lagrangian \eqref{eq:MaxElL} and nonconservative potential \eqref{eq:MaxElK} for the Maxwell element, by replacing the displacements with their analogous strain tensors. To do so we must first introduce another set of dynamical fields, the six symmetric plastic deformation tensor components, $[C_{\rm pl}]_{AB} =  [C_{\rm pl}]_{BA}$, which keep track of the ``equilibrium deformation'' for which the elastic stress is zero. Thus we can define the elastic strain of the visco-elastic fluid as
\begin{align}
[u_{\rm el}]_{{A}{B}} &\equiv \frac{1}{2}\left(C_{AB} - [C_{\rm pl}]_{{A}{B}}\right)\\
&= \frac{1}{2} \left( e^i_A e^j_B g_{ij}-  [C_{\rm pl}]_{{A}B}\right).
\end{align}

We can define the plastic relative strain tensor
with respect to some fiducial tensor $[C_o]_{AB}$ as, 
\begin{align}
[u_{\rm pl}]_{{A}B} = \frac{1}{2}\left([C_{\rm pl}]_{{A}B} - [C_{\rm o}]_{{A}B}\right),
\end{align}
and the plastic rate of strain tensor as
\begin{align}
[\gamma_{\rm pl}]_{{A}B} \equiv \pd_t{[u_{\rm pl}]}_{{A}B} = \frac{1}{2} \pd_t{[C_{\rm pl}]}_{{A}B}.
\end{align}
The elastic rate of strain tensor can then be defined as
\begin{align}
[\gamma_{\rm el}]_{{A}B} \equiv \pd_t{[u_{\rm el}]}_{{A}B} = \gamma_{{A}B} - [\gamma_{\rm pl}]_{{A}B} \label{eq:ElasticStrainRate}.
\end{align}

The internal energy of the viscoelastic fluid must depend on scalar
combinations of the elastic strain tensor. This can be expanded to
quadratic order in the strain tensor giving, for an isotropic
material,
\begin{align}
\bar{\varepsilon} = \bar{\varepsilon}_o(\bar{\rho}, \bar{s}) &+ \bar{\beta}^{AB} [u_{\rm el}]_{AB} + \frac{1}{2} \bar{\mu}_s P^{ABCD}[u_{\rm el}]_{AB}[u_{\rm el}]_{CD}\nonumber\\
&+ \frac{1}{2}\bar{\mu}_b [u_{\rm el}]^{AB}[u_{\rm el}]_{AB} + {\cal O}([u_{\rm el}]^3)\,,
\end{align}
where $\bar{\mu}_s$ is the isotropic elastic shear modulus\footnote{Our
  convention for $\mu_s$ differs from Landau and
  Lifshitz~\cite{landau1986elasticity} by a factor of 2.} and $\bar{\mu}_b$ is the
isotropic elastic bulk modulus, both of which transform as scalar densities and can be functions of
the local thermodynamic variables.  Since the elastic strain is
defined such that the elastic stress $d\bar{\varepsilon} /
d([u_{\rm{el}}]_{AB}) = 0$ when $[u_{\rm el}]^{AB} = 0$, we must have
$\bar{\beta}^{AB} = 0$. The energy per unit mass
$\bar{\varepsilon}_o(\bar{\rho}, \bar{s})$ is the internal energy of
the fluid without elastic deformation.

Keeping terms in the action only up to quadratic order in the elastic strain, we can then define the elastic stress tensor $\sigma_{\rm el}^{AB} = d\bar{\varepsilon}/d([u_{\rm el}]_{AB})$ as
\begin{align}
\sigma_{\rm el}{}^{AB} &\equiv \bigg[\mu_s P^{ABCD} + \mu_b C^{AB}C^{CD}\bigg] [u_{\rm el}]_{C D},\\
	&\equiv {\cal E}^{ABCD}[u_{\rm el}]_{C D}\,.
\end{align}
We can then propose the Lagrangian and ${\cal K}$ densities,
\begin{align}
{\cal L} &= \frac{1}{2}\rho \dot{\bm q}^2 - J \bar{\varepsilon}_o(\bar{\rho}, \bar{s}) - \frac{1}{2}  [u_{\rm el}]_{AB}[\sigma_{\rm el}]^{AB}\\
{\cal K} &= - [u_{\rm pl - }]_{AB} [\sigma_{{\rm v}+}]^{AB} -\zeta_+ [\Delta_+]_i q^i_{s-}
\end{align}
where the viscous stress is given for an isotropic material as in Equation \eqref{eq:ViscousStress} by
\begin{align}
[\sigma_{\rm v}]^{AB} \equiv  {\cal V}^{ABCD} [\gamma_{\rm pl}]_{CD}.
\end{align}
The form of the viscous interaction that appears in $\calK$ is similar to that which
appears in the action for viscous fluids, except that it only acts on the plastic strain component, rather than
the total fluid strain. 

Again, $\pd_t \rho = 0 $ provides the continuity equation, 
\begin{align}
\bar{\pd}_t \bar{\rho} + \bar{\nabla}_i (v^i \bar{\rho}) = 0.
\end{align}

Performing the variation of the proposed viscoelastic action with respect to $\bq_-$, ${\bm C}_{{\rm pl} -}$, and ${\bm \alpha}_-$ and taking the physical limit, we obtain the equations of motion,
\begin{align}
\rho \frac{Dv_i}{Dt_a}+ A_i{}^A \nabla_A \bar{P} - e_i^B\nabla_A[\sigma_{\rm el}]^{A}{}_B&= 0,\\
[\sigma_{\rm v}]^{AB}  - [\sigma_{\rm el}]^{AB} &= 0, \\
e^A_{As} s \nabla_A T - \zeta (\pd_t \alpha)_{As} &= 0,
\end{align}
or in the Eulerian variables
\begin{align}
\bar{\rho} \, \bar{\pd}_t v_i  + \bar{v}^j \bar{\nabla}_j v_i + \bar{\nabla}_i \bar{P} - \bar{\nabla}_i \left([\bar{\sigma}_{\rm el}]^{i}{}_j \right) &= 0,\label{eq:ViscoElasticEOM1}\\
[\bar{\sigma}_{\rm v}]^{ij}  - [\bar{\sigma}_{\rm el}]^{ij} &= 0, \label{eq:ViscoElasticEOM2}\\
\bar{{\cal F}}_i + \bar{\kappa} \bar{\nabla}_i T &= 0 \label{eq:ViscoElasticEOM3}
\end{align}

Equation \eqref{eq:ViscoElasticEOM1} is the momentum conservation equation for the linear Maxwell model, while Equation \eqref{eq:ViscoElasticEOM3} gives Fourier's law.  Equation \eqref{eq:ViscoElasticEOM2} gives the Maxwell model relation for stress ``in series,'' which can be expanded out using the definitions of $[\sigma_{\rm v}]^{AB}$ and $[\sigma_{\rm el}]^{AB}$ to obtain an evolution equation for the plastic deformation components $[C_{\rm pl}]_{AB}$,
\begin{align}
{\cal V}^{ABCD} \pd_t {[C_{\rm pl}]}_{CD} + \frac{1}{2}{\cal E}^{ABCD}\left([C_{\rm pl}]_{CD} -e^i_Ce^j_Dg_{ij} \right) = 0 
\end{align}

We note that the usual constitutive relation for the Maxwell model can be recovered using Equation \eqref{eq:ViscoElasticEOM2} and the strain rate relation Equation \eqref{eq:ElasticStrainRate},
\begin{align}
\gamma_{AB} &= \pd_t([{\cal E}^{-1}]_{CDAB} [\sigma_{\rm el}]^{C D} ) + [{\cal V}^{-1}]_{CDAB} [\sigma_{\rm v}]^{CD}
\end{align} 
Assuming $\pd_t [{\cal E}^{-1}]_{CDAB} = 0$ we have
\begin{align}
{\cal V}^{ABCD}\, \gamma_{CD} = [\tau_{\rm rel}]^{AB}{}_{CD}\, (\pd_t {\sigma}^{CD}) + \sigma^{AB},
\end{align}
where ${\bm \sigma} \equiv {\bm \sigma}_{\rm v} = {\bm \sigma}_{\rm el}$ is the stress, and $[{\bm \tau}_{\rm rel}]^{AB}{}_{CD} $ is the relaxation time tensor given by
\begin{align}
[\tau_{\rm rel}]^{AB}{}_{C D} \equiv {\cal V}^{ABEF}  [{\cal E}^{-1}]_{CDEF}.
\end{align}

In the Euler coordinates this is given as
\begin{align}
\bar{\cal{V}}^{ijk\ell}\, \bar{\gamma}_{k\ell} = [{\tau}_{\rm rel}]^{ij}{}_{k\ell}\, ({\cal L}_{\pd/\pd t_a}\bar{\sigma}^{k\ell}) + \bar{\sigma}^{ij}, 
\end{align}
where
\begin{align}
[{\tau}_{\rm rel}]^{ij}{}_{k\ell} = \bar{{\cal V}}^{ijmn}  [\bar{{\cal E}}^{-1}]_{k\ell mn}.
\end{align}
We also note that this derivation naturally selects the ``upper-convected'' time derivative in the constitutive relation for the Maxwell model by using the Lie derivative ${\cal L}_{\pd/\pd t_a}$ \eqref{eq:lie-sum} acting on a contravariant rank-2 tensor, in contrast to the usual formulation (see e.g.,~\cite{2009rheo.book.....O}, which relies on empirical considerations to select the upper-convective time derivative as the appropriate choice). 

The total stress tensor is given by
\begin{equation}
\begin{aligned}
\calT_0{}^0  &=  \frac{1}{2}\rho \dot{\bq}^2 + J \bar{\varepsilon}_o + \frac{1}{2}[u_{\rm el}]_{AB}[\sigma_{\rm el}]^{AB} \\
\calT_0{}^B &= J\bar{P}\dot{q}^B - \dot{q}^C \sigma_{\rm el}^{B}{}_C +\Delta^B \,T s  \\
\calT_A{}^0 &= \rho \dot{q}_A\\
\calT_A{}^B &=  
\delta_A{}^B \left[- \frac{1}{2} \rho \dot{\bq}^2 + J \bar{P} + J \bar{\varepsilon} - J T \bar{s}\right] - [\sigma_{\rm el}]_A{}^B.
\end{aligned}
\end{equation}

The closure condition $\pd_\nu {\cal T}_0{}^\nu = 0$ and \eqref{eq:divTmunu1} give the energy equation
\begin{align}
&\bar{\pd}_t \left(\frac{1}{2}\bar{\rho} v^2 + \bar{\varepsilon}+ \frac{1}{2}  [u_{\rm el}]_{ij}\bar{\sigma}^{ij} \right)  \nonumber\\
&\qquad + \bar{\nabla}_i \bigg\{v^i\left(\frac{1}{2} \bar{\rho}v^2 + \bar{h} + \frac{1}{2} [u_{\rm el}]_{ij}\bar{\sigma}^{ij}\right)\bigg\}\nonumber\\
&\qquad -\bar{\nabla}_i \bigg\{ v_j\, \bar{\sigma}^{ij} + \bar{\kappa} \bar{\nabla}^i T\bigg\} = 0
\end{align}
as well as the entropy equation 
\begin{align}
\bar{\pd}_t \bar{s} + \bar{\nabla}_i (v^i \bar{s}) = \frac{1}{T} \bar{\nabla}_i(\bar{\kappa} \bar{\nabla}^i T) + \frac{1}{T} [{\gamma}_{\rm pl}]^{ij} \bar{\sigma}_{ij}.
\end{align}

{
\renewcommand{\arraystretch}{1.5}

\begin{table*}[ht]
\begin{center}
\begin{tabular*}{\textwidth}{| @{\extracolsep{\fill} }>{\raggedright}p{75pt}| c | c | c | c | } \hline
Name  & $\calL$ &$\calK$ & DoF &System Type  \\ \hline \hline
Perfect Fluid \S\ref{sec:PerfectFluid}& $\,\,\frac{1}{2}\rho \dot{\bq}^2 - J \bar{\varepsilon}(\bar{\rho}, \bar{s})\,\,$ & - & $\bq$  & conservative\\ 
Cold Stone Fluid \S\ref{sec:ColdStoneFluid} & $\frac{1}{2}\rho \dot{\bq}^2 - J \bar{\varepsilon}(\bar{\rho}, \bar{s})$ & $- \mathbfcal{V}_+{::}({\bm u}_- {\otimes} {\bm \gamma}_+)$ & $\bq$  & \pbox{80pt}{open isentropic\\ ($ \pd_t s = 0$)} \\
Viscous Insulating Fluid \S\ref{sec:InsulatingFluid}& $\frac{1}{2}\rho \dot{\bq}^2 - J \bar{\varepsilon}(\bar{\rho}, \bar{s})$ & $- \mathbfcal{V}_+{::}({\bm u}_- {\otimes} {\bm \gamma}_+)$& $\bq$  & \pbox{80pt}{closed ($\pd_\nu {\cal T}_0{}^\nu = 0 $)}\\ 
Inviscid  Fluid with Heat Diffusion \S\ref{sec:inviscid-fluid-diffusion}& $\frac{1}{2}\rho \dot{\bq}^2 - J \bar{\varepsilon}(\bar{\rho}, \bar{s})$ & $- \zeta_+ ({\bm \alpha}_- \cdot \pd_t {\bm \alpha}_+) $& $\bq, {\bm \alpha}$  & closed ($\pd_\nu {\cal T}_0{}^\nu=0$)\\  
Navier-Stokes Fluid \S\ref{sec:Navier-Stokes}& $\frac{1}{2}\rho \dot{\bq}^2 - J \bar{\varepsilon}(\bar{\rho}, \bar{s})$ & $-\mathbfcal{V}_+{::}({\bm u}_- {\otimes} {\bm \gamma}_+) - \zeta_+ ({\bm \alpha}_- \cdot \pd_t {\bm \alpha}_+)  $& $\bq, {\bm \alpha}$  & closed ($\pd_\nu {\cal T}_0{}^\nu=0$)\\  
Microhydro-dynamics (Stokes Limit) \S\ref{sec:microhydro}& $- J \bar{\varepsilon}(\bar{\rho}, \bar{s})$ & $-\mathbfcal{V}_+{::}({\bm u}_- {\otimes} {\bm \gamma}_+) $& $\bq$  &  \pbox{80pt}{open isentropic\\ ($ \pd_t s = 0$), \\ movable boundaries}\\  
\hline
Perfect Elastic Material & $\frac{1}{2}\rho \dot{\bq}^2 - J \bar{\varepsilon}_o(\bar{\rho}, \bar{s}) -\frac{1}{2} \mathbfcal{E}{::}({\bm u}{\otimes}{\bm u})  $ &  - & $\bq$  & conservative\\ 
Cold Stone Elastic with Dissipation & $\frac{1}{2}\rho \dot{\bq}^2 - J \bar{\varepsilon}_o(\bar{\rho}, \bar{s}) -\frac{1}{2}  \mathbfcal{E}{::}({\bm u}{\otimes}{\bm u})  $ &  $- \mathbfcal{V}_+{::}({\bm u}_- {\otimes} {\bm \gamma}_+)$ & $\bq$  & \pbox{80pt}{open isentropic\\ ($ \pd_t s = 0$)} \\ 
Insulating Elastic with Dissipation & $\frac{1}{2}\rho \dot{\bq}^2 - J \bar{\varepsilon}_o(\bar{\rho}, \bar{s}) -\frac{1}{2} \mathbfcal{E}{::}({\bm u}{\otimes}{\bm u})$ &  $- \mathbfcal{V}_+{::}({\bm u}_- {\otimes} {\bm \gamma}_+)$ & $\bq$  & closed ($\pd_\mu {\cal T}_0^\mu = 0$)\\ 
Elastic with Heat Diffusion& $\frac{1}{2}\rho \dot{\bq}^2 - J \bar{\varepsilon}_o(\bar{\rho}, \bar{s}) -\frac{1}{2}  \mathbfcal{E}{::}({\bm u}{\otimes}{\bm u})  $ &  $- \zeta_+({\bm \alpha}_- \cdot \pd_t {\bm \alpha}_+) $  & $\bq, {\bm \alpha}$  & closed ($\pd_\mu {\cal T}_0^\mu = 0$) \\  
Elastic with Dissipation \&  Heat Diffusion & $\frac{1}{2}\rho \dot{\bq}^2 - J \bar{\varepsilon}_o(\bar{\rho}, \bar{s}) -\frac{1}{2}  \mathbfcal{E}{::}({\bm u}{\otimes}{\bm u}) $ &  $- \mathbfcal{V}_+{::}({\bm u}_- {\otimes} {\bm \gamma}_+) - \zeta_+ ({\bm \alpha}_- \cdot \pd_t {\bm \alpha}_+) $  & $\bq, {\bm \alpha}$  & closed ($\pd_\mu {\cal T}_0^\mu = 0$) \\  
\hline
Cold Stone Maxwell Fluid & $\frac{1}{2}\rho \dot{\bq}^2 - J \bar{\varepsilon}_o(\bar{\rho}, \bar{s}) -\frac{1}{2}  \mathbfcal{E}{::}({\bm u}_{\rm el}{\otimes}{\bm u}_{\rm el}) $ & $-\mathbfcal{V}_+{::}({\bm u}_{{\rm pl}-} {\otimes} {\bm \gamma}_{{\rm pl}+}) $& $\bq, {\bm C}_{\rm pl}$  &\pbox{80pt}{open isentropic\\ ($ \pd_t s = 0$)} \\ 
Insulating Maxwell Fluid & $\frac{1}{2}\rho \dot{\bq}^2 - J \bar{\varepsilon}_o(\bar{\rho}, \bar{s}) -\frac{1}{2}  \mathbfcal{E}{::}({\bm u}_{\rm el}{\otimes}{\bm u}_{\rm el}) $ & $-\mathbfcal{V}_+{::}({\bm u}_{{\rm pl}-} {\otimes} {\bm \gamma}_{{\rm pl}+})$& $\bq, {\bm C}_{\rm pl}$  & closed ($\pd_\nu {\cal T}_0{}^\nu=0$)\\  
Viscoelastic Maxwell Fluid \S\ref{sec:Viscoelastic}& $\frac{1}{2}\rho \dot{\bq}^2 - J \bar{\varepsilon}_o(\bar{\rho}, \bar{s}) -\frac{1}{2}  \mathbfcal{E}{::}({\bm u}_{\rm el}{\otimes}{\bm u}_{\rm el}) $ & $-\mathbfcal{V}_+{::}({\bm u}_{{\rm pl}-} {\otimes} {\bm \gamma}_{{\rm pl}+}) - \zeta_+ ({\bm \alpha}_- \cdot \pd_t {\bm \alpha}_+)  $& $\bq, {\bm C}_{\rm pl}, {\bm \alpha}$  & closed ($\pd_\nu {\cal T}_0{}^\nu=0$)\\
\hline
\end{tabular*}
\caption{The Lagrangian densities (${\cal L}$) and nonconservative potential densities (${\cal K}$) used to construct the nonconservative actions in example continuum systems. Here we use the notation $\mathbfcal{T}{::}({\bm X}{\otimes}{\bm Y}) \equiv {\cal T}^{ABCD} X_{AB} Y_{CD}$ for compactness. ``DoF'' indicates the accessible degrees of freedom (i.e., the generalized coordinates) considered. ``System Type'' denotes whether the accessible degrees of freedom are open, conservative, closed, etc. 
This table demonstrates the modularity of building nonconservative actions for different physical systems.}
\vspace{-1em}
\label{tab:1}
\end{center}
\end{table*}

}

\section{Discussion and outlook}\label{sec:Discussion}

In this paper we have further developed the variational principle for
nonconservative discrete systems (Sec.~\ref{sec:Review}) introduced
in~\cite{Galley:2012hx} and generalized it to include classical field
theories (Sec.~\ref{sec:ContinuumMechanics}). This variational
principle was developed to be consistent with the specification of
initial data for a system
thereby allowing one to accommodate nonconservative interactions
such as time-irreversible processes, at the level of the action.
To accomplish this, the degrees of freedom are doubled, and the
nonconservative action is defined as an integral forward in time along
one set of generalized coordinate paths, and backwards in time along
the second set of paths, with their variations satisfying the
\emph{equality condition} such that the two copies of the coordinates
(and their generalized momenta) are equal, but unspecified at the
final time.  After extremizing the nonconservative action,
\begin{equation*}
  \calS [ \bq_1, \bq_2] = \int_{t_i}^{t_f} \!\!\! dt \, \Lambda
  (\bq_1, \bq_2, \dot{\bq}_1, \dot{\bq}_2, t) \,,
\end{equation*}
the equations of motion are
recovered by applying the physical limit, which equates the two
paths. The nonconservative Lagrangian $\Lambda$ can be separated into
a conservative piece $L(q_{1})-L(q_{2})$ plus a \emph{nonconservative
  potential}, $K(q_{1},q_{2})$, which couples the paths together, and can be used
to model or derive general nonconservative forces acting on the
system. The nonconservative potential may arise from an open system
interaction with inaccessible degrees of freedom or from integrating
out or coarse-graining a subset of the degrees of freedom in a
conservative system.
The equations of motion for the system include the effects of the 
nonconservative potential and are found to satisfy the new Euler-Lagrange 
equations \eqref{eq:nonconsEL4},
\begin{align}
	\frac{ d}{ dt} \frac{ \partial L }{ \partial \dot{q}^I } - \frac{ \partial L }{ \partial q^I } = \left[ \frac{ \partial K }{ \partial q^I_-} - \frac{ d }{ dt } \frac{ \partial K }{ \partial \dot{q}^I_-} \right]_{PL}\nonumber
\end{align}
for discrete mechanics, and \eqref{eq:eom2},
\begin{align}
  \pd_\mu \frac{ \pd {\cal L}}{ \pd (\pd_\mu \phi^I) } - \frac{ \pd
    {\cal L}}{ \pd \phi^I } = \left[ \frac{ \pd {\cal K}}{ \pd
      \phi^I_- } - \pd_\mu \frac{ \pd {\cal K}}{ \pd (\pd_\mu
      \phi^I_-) }   \right]_{\rm PL} \nonumber
\end{align}
for classical field theories.

We have also generalized Noether's theorem
(Sec.~\ref{sec:discreteNoether} \& Sec.~\ref{sec:ContinuumNoether}) to
include the influence of nonconservative interactions on the Noether
currents generated by continuous symmetries of the conservative action
$S = \int dt L$. The energy function of discrete systems and the
canonical stress-energy tensors of continuum systems can be derived
through the Noether currents generated by shift symmetry in time and
space coordinates. By applying the generalized Noether's theorem [see
\eqref{eq:Echarge3}, \eqref{eq:Jcharge3}, 
\eqref{eq:divTmunu2}, and \eqref{eq:divJnu2}] we show how the Noether currents
receive contributions from nonconservative interactions.
For example, the total energy $\calE$ in (\ref{eq:Echarge2}) evolves according to \eqref{eq:Echarge3} as
\begin{align}
\frac{ d \calE}{dt} = - \frac{ \pd L }{ \pd t}
+ \dot{q}^I \bigg[ \frac{ \pd K }{ \pd q_-^I } \bigg]_{\rm PL}
+ \ddot{q}^I \left[\frac{\pd K}{\pd \dot{q}_-^I}\right]_{\rm PL} \,, \nn
\end{align}
where 
\begin{align}
	{\cal E} = E + \dot{q}^I \left[ \frac{ \pd K }{ \pd\dot{q}_-^I } \right]_{\rm PL} \nn
\end{align}
is the total energy of the accessible degrees of freedom and includes 
nonconservative contributions and $E$ is the usual energy function derived from the conservative Lagrangian $L$. For
continuum systems and classical field theories, the canonical stress-energy tensor of the 
accessible degrees of freedom evolves according to \eqref{eq:divTmunu2},
\begin{align}
	\pd_\nu \calT_\mu{}^\nu = - \frac{\pd \calL}{\pd x^\mu}
  + \pd_\mu \phi^I \bigg[ \frac{ \pd \calK}{ \pd \phi_-^I} \bigg]_{\rm PL}  {\hskip-0.12in}
  + \pd_\nu \pd_\mu \phi^I \left[ \frac{\pd \calK}{\pd(\pd_\nu
      \phi_-^I)}\right]_{\rm PL} {\hskip-0.1in} ,\nn
\end{align}
where the total canonical stress-energy tensor (including
nonconservative contributions) is given by
\begin{align}
\calT_\mu{}^\nu \equiv T_\mu{}^\nu +
\pd_\mu \phi^I \left[ \frac{\pd \calK}{\pd(\pd_\nu \phi_-^I)}\right]_{\rm PL}, \nn
\end{align}
and $T_\mu{}^\nu$ is the usual canonical stress tensor derived from the conservative Lagrangian density $\calL$.
More generally, we have showed how the total Noether currents of the accessible degrees of freedom in Secs.~\ref{sec:discreteNoether} and \ref{sec:ContinuumNoether}, which include a contribution from nonconservative interactions, evolve in time due to $K$ or $\calK$, respectively.

This new variational principle can be used to capture the dynamics of an ``accessible'' subset of the total degrees of freedom for a system. These dynamics can include interactions with the ``inaccessible'' degrees of freedom, such as those that have been integrated out, or coarse grained away. This naturally describes ``open system'' conditions, where energy is removed from the accessible degrees of freedom and no longer affects the dynamics. If an internal energy function describing the energy deposited into coarse-grained (inaccessible) degrees of freedom  is included in the Lagrangian, then a ``closed system'' condition, $d{\cal E}/dt = 0$, or $\pd_\nu {\cal T}_0{}^\nu = 0$,  can be used to close the system of equations. This closure condition can be used to generate the energy and entropy equations for isolated systems when combined with the new results of Noether's theorem above. 

This formalism has allowed us to develop nonconservative actions that
include \emph{irreversible} processes such as viscous damping of a
free particle or harmonic oscillator
(Sec.~\ref{sec:ForcedDampedOscillator}), radiation reaction on an
accelerated charge (Sec.~\ref{sec:emrr}), or viscous dissipation and heat
conduction in a non-equilibrium fluid (Sec.~\ref{sec:Hydro}). In the
second half of this paper we have focused on providing examples for
continuum mechanics. We have utilized the advantages of an action
formulation to constructively build actions by cumulatively including
various physical components and interactions. We first considered the
standard perfect conservative fluid, then gradually included
irreversible thermodynamical processes such as dissipation and heat
diffusion (see Table 1), which culminated in developing an
unconstrained action that generates the Navier-Stokes equations of
motion (Sec.~\ref{sec:Navier-Stokes}), including heat transport, with
dissipation and heat diffusion consistent with the second law of
thermodynamics. Applying the appropriate limits at the action level we
easily recover an action for the Stokes regime, and we explored the
interactions due to the movable boundary terms typically encountered
in micro-hydrodynamic systems with particles in suspension
(Sec.~\ref{sec:microhydro}).

We have also developed, for the first time, an unconstrained action
formulation for the Maxwell model of viscoelasticity
(Sec.~\ref{sec:Viscoelastic}). While this is a particularly simple
rheological model, our approach can be easily generalized to produce
more sophisticated models of viscoelastic materials. The
nonconservative action formulation naturally reproduces the well known
``Upper-Convected Maxwell Model'' without needing to make an arbitrary
choice of time derivative, as typically must be done when constructing
the model at the equation of motion level (see
e.g.,~\cite{2009rheo.book.....O}).

In Table \ref{tab:1} we outline the Lagrangian and nonconservative potential densities for actions of various continuum systems, including several examples of elastic materials that were not discussed in the text. This table demonstrates, in particular, the modularity of the action approach for including different physical processes and interactions. 

While we have developed the theory of nonconservative classical Lagrangian mechanics from
requirements to preserve the causal evolution of
degrees of freedom that can lose or gain energy dynamically, one may wonder if there is a fundamental 
connection where the nonconservative action ${\cal S}$ in (\ref{eq:newaction1}) 
can be derived from a more complete quantum theory. We discuss
this in Appendix \ref{sec:quantum} and show that (\ref{eq:newaction1}) is
the action that results by taking the classical limit of the reduced density matrix
for quantum degrees of freedom that couple to variables that have
been integrated out. The reduced density matrix contains doubled degrees of freedom because one
computes true expectation values of operators that evolve from 
a given initial state or density matrix, which is crucial for 
correctly describing the nonequilibrium dynamical behavior of the quantum system.
That the nonconservative action in (\ref{eq:newaction1}) comes from the 
classical limit of a more complete quantum framework is reassuring and strongly indicative
that the nonconservative mechanics developed in this paper is
well-rooted in a ``first principles'' foundation.

We have shown that nonconservative action principles for discrete and
continuum systems allow non-equilibrium processes, such as viscous
dissipation and heat diffusion, to be modeled. We expect the formalism 
we have presented above to be applicable for generic non-equilibrium 
systems, allowing non-equilibrium thermodynamic processes to be described 
by an unconstrained variational principle. This remarkable result arises when 
applying the formalism we have developed for effectively open systems, to 
(coarse-grained) systems that can be closed by including the (internal) energy 
of the inaccessible degrees of freedom. 

Recently Kevrekidis \cite{Kevrekidis:PhysRevA.89.010102} applied the nonconservative 
formalism of \cite{Galley:2012hx} to study nonlinear waves in nonconservative or open systems, 
using the collective coordinate method. This method utilizes the projection of higher-dimensional
dynamics to a lower-dimensional set of degrees of freedom, and thus is well suited to the formalism
we have discussed above. Using the nonconservative action formalism Kevrekidis studied two 
${\cal PT}$-symmetric example field theories (dissipative variants of sine-Gordon and a $\phi^4$ model), showing that
the collective nonconservative variational formulation predicted the behavior of nonlinear waves in these
theories remarkably well compared to the full numerical evolution of the fields.
This approach seems to be a promising and practical avenue for generating accurate approximate solutions
using the nonconservative action for nonlinear field theories.

One obvious application of this formalism
which we have not discussed is non-ideal magneto-hydrodynamics, by
adding resistivity and ambipolar diffusion to the ideal MHD
action~\cite{2014arXiv1407.3884K}. We will address this in forthcoming
work~\cite{MHDfuture}.

Additionally, we have not considered interesting examples from statistical theory 
of fluctuations as it applies to hydrodynamics \cite{PhysRev.91.1505, LL:hydroflucs}.
It would be interesting to apply our formalism to study these problems and explore connections
with previous work in this field (see e.g., \cite{Martin:1973zz, DeDominicis:1977fw, Eyink:1996, Eyink:1998, Kovtun:2014hpa})
where doubled variables also seem to arise for computing statistical correlation functions but we leave this for future work.

It is in principle simple to construct Lorentz invariant actions for
nonconservative relativistic theories.  This will be particularly
useful for dissipative systems as there currently exists no consistent
stable theory for dissipation and heat diffusion in relativistic fluids (see
e.g.,~\cite{Andersson2013, 1979AnPhy.118..341I}.  Prix~\cite{Prix:2002jn}
and Andersson and Comer~\cite{2010RSPSA.466.1373A} showed that causal
heat flow can be induced by the addition of ``entrainment''
interaction between the massless entropy ``fluid'' and the background
flow. This entrainment adds an effective mass to the entropy fluid,
and introduces a timescale that corresponds to the local thermal
equilibration time of a fluid element.

While we have focused on nonconservative Lagrangian mechanics, it is straightforward to apply the formalism to Hamiltonian mechanics. In a future work we will extend the discrete Hamiltonian mechanics from~\cite{Galley:2012hx} to classical field theories, including a description of the phase space structure of the doubled degrees of freedom. This formalism may provide new variational integration techniques for numerically evolving nonconservative initial value problems~\cite{GalleyFuture1}.

Our approach to nonconservative variational principles has been general and is not limited to the few examples we have presented in this paper. We anticipate future applications in many fields including optimal control theory, variational calculus, and non-equilibrium statistical mechanics.

\acknowledgments

We thank A.~Cumming, G.~Holder, M.~Sutton, and  G.~Eyink for useful discussions
as well as I.~Rothstein and A.~Leibovich for discussions during the initial development
of this framework.
We especially thank Nils Andersson for discussions about nonequilibrium fluid 
dynamics. 
C.R.G.~was supported in part by NSF grants CAREER PHY-0956189, PHY-1068881, 
and PHY-1005655 to the California Institute of Technology.
D.T.~was supported in part by the Lorne Trottier Chair in Astrophysics and Cosmology 
as well as the Canadian Institute for Advanced Research.
L.C.S.~acknowledges that support for this work was
provided by the NASA through
Einstein Postdoctoral Fellowship Award Number PF2-130101 issued by the
Chandra X-ray Observatory Center, which is operated by the Smithsonian
Astrophysical Observatory for and on behalf of the National
Aeronautics Space Administration under contract NAS8-03060.

\appendix

\section{Higher derivative nonconservative mechanics}
\label{sec:higherders}

Some systems may naturally involve nonconservative forces and interactions that 
depend on the acceleration of the coordinates and/or higher time derivatives (e.g., see Sec.~\ref{sec:emrr}). We discuss how 
the principle of stationary nonconservative action works
for such systems. Details of the omitted calculations are similar to those  
presented in Sec.~\ref{sec:NonconsDiscLag} and will not be shown here.

For a discrete nonconservative system depending on the first $M$ time derivatives
of the $N$ generalized coordinates $\{ q^I \}_{n=1}^N$, the nonconservative Lagrangian is 
\begin{align}
	\Lambda = \Lambda \big( \bq_a, \dot{\bq}_a, \ddot{\bq}_a, \ldots, \bq^{(M)}_a , t \big)
\end{align}
where the $(M)$ superscript indicates the $M^{\rm th}$ time derivative of $q_a^I$. We will also
use $[q_a^I]^{(M)}$ for an individual element of $\bq_a^{(M)}$. The nonconservative action $\calS [ \bq_a]$ is 
stationary under the $2(M+1)$ variations 
\begin{align}
	\bigg\{  \bq_a^{(m)} (t, \epsilon) = \bq_a^{(m)} (t, 0) + \epsilon \, {\bm \eta}_a^{(m)} (t) \bigg\}_{m=0}^M
\end{align}
if the Euler-Lagrange equations of motion are satisfied
\begin{align}
	\frac{ d \pi^a_I }{ dt } = \frac{ \pd \Lambda }{ \pd q^I_a }
\label{eq:ELeom531zz}
\end{align}
where 
\begin{align}
	\pi^a_I \big( \bq_b,  \ldots, \bq_b^{(M)}, t \big)  \equiv {} & - \sum_{m=1}^M (-1)^{m} \frac{ d^{m-1} }{ dt ^{m-1} } \frac{ \pd \Lambda }{ \pd [q^I_a]^{(m)} } 
\label{eq:higherderpia1} \\ 
	= {} & \frac{ \pd \Lambda }{ \pd \dot{q}^I_a } - \frac{ d }{ dt } \frac{ \pd \Lambda }{ \pd \ddot{q}^I_a } + \frac{d^2}{dt^2 } \frac{ \pd \Lambda }{ \pd \dddot{q}^I_a} -  \cdots \nn
\end{align}
is the total canonical momentum associated with the higher-derivative dependent nonconservative Lagrangian $\Lambda$.
Recall that the history index $a$ can be raised and lowered with the ``metric'' $c_{ab}$ discussed before (\ref{eq:Pimuiab11}).
The sequence of derivatives in (\ref{eq:higherderpia1}) commonly appears in these calculations and are equivalently expressed by the functional derivative. For some functional $F$ of $x(t)$,
\begin{align}
	F[x] = \int_{t_i}^{t_f} \!\!\! dt \, f \big( x, \ldots, x^{(M)}, t \big)  ,
\end{align}
the functional derivative is
\begin{align}
	\frac{ \delta F}{ \delta x(t) } \equiv \frac{ \pd f}{ \pd x } - \frac{ d }{ dt} \frac{ \pd f}{ \pd \dot{x} } + \frac{ d^2 }{ dt^2 } \frac{ \pd f}{ \pd \ddot{x} } - \cdots  = \sum_{m=0}^M \frac{ d ^m}{ dt^m } \frac{ \pd f}{ \pd x^{(m)} }   
\end{align}
Therefore,
\begin{align}
	\pi^a_I \big( \bq_b,  \ldots, \bq_b^{(M)}, t \big) = \frac{ \delta \calS }{ \delta \dot{q}_a^I (t) } 
\end{align}
As in Sec.~\ref{sec:NonconsDiscLag}, the physical limit of (\ref{eq:ELeom531zz}) gives a single set of non-trivial equations of motion, 
which are expressed in terms of $L$ and $K$ by
\begin{align}
	\frac{ d p_I }{ dt } - \frac{ \pd L }{ \pd q^I } = \bigg[ \frac{ \pd K }{ \pd q^I_- } \bigg]_{\rm PL} - \frac{ d\kappa_I }{ dt }
\end{align}
where $p_I$ and $\kappa_I$ are the conservative and nonconservative parts of the total conjugate momentum $\pi_I = [\pi_{+I}]_{\rm PL} = p_I + \kappa_I$, respectively,
\begin{align}
	p_I \big( \bq,  \ldots, \bq^{(M)}, t \big)  \equiv {} & \frac{ \delta S }{ \delta \dot{q}^I(t) } 
\label{eq:higherderp1} \\
	\kappa_I \big( \bq,  \ldots, \bq^{(M)}, t \big)  \equiv {} & \bigg[ \frac{ \delta \calS_K }{ \delta \dot{q}^I_-(t) } \bigg]_{\rm PL}
\label{eq:higherderkappa1}
\end{align}
where $\calS_K$ is the part of the nonconservative action $\calS$ that involves $K$.

Ensuring that we have a well-defined variational principle requires more conditions on the variations and the conjugate momenta 
than in Sec.~\ref{sec:NonconsDiscLag} because the problem depends on higher time derivatives of
the generalized coordinates. We find that $2N M$ conditions are needed to fix the variations at the initial time
\begin{align}
	\bigg\{ {\bm \eta}^{(m)}_- (t_i) = 0 , ~ {\bm \eta}^{(m)}_+(t_i) = 0 \bigg\}_{m=0}^{M-1}
\end{align}
At the final time, the equality condition
is generalized so that $N M$ conditions are imposed on the ${\bm \eta}_-$ variations,
\begin{align}
	\bigg\{ {\bm \eta}_- ^{(m)} (t_f) = 0 \bigg\} _{m=0}^{M-1}
\label{eq:higherderequality1}
\end{align}
and another $N M$ on the total canonical momenta ${\bm \pi}_-$,
\begin{align}
	\bigg\{ \frac{ \delta \calS }{ \delta q_-^{(m)} (t) } \bigg|_{t_f}  = 0\bigg\}_{m=1}^M
\label{eq:higherderequality2}
\end{align}
The $2NM$ conditions in (\ref{eq:higherderequality1}) and (\ref{eq:higherderequality2}) together form the {\it equality condition} for problems that depend on higher time derivatives. For $M=1$ this agrees with the equality condition in (\ref{eq:eqcond2}).

The generalization of Noether's theorem in Sec.~\ref{sec:discreteNoether} likewise is modified to accommodate higher time derivatives of $\bq$. We find that
\begin{align}
	\frac{ d \calE }{ dt } = - \frac{ \pd L}{ \pd t } + \dot{q}^I \bigg[ \frac{ \pd K }{ \pd q^I_- } \bigg]_{\rm PL} + \ddot{q}^I \kappa_I
\end{align}
when $t \to t + \delta t$ and $q^I \to q^I + \epsilon^a \omega^I_a + \dot{q}^I \delta t$, respectively, 
where $\kappa_I$ is given in (\ref{eq:higherderkappa1}) and $\calE = E + \kappa_I \dot{q}^I$ with
\begin{align}
	E \equiv
		 \sum_{m=1}^M [q^I]^{(m)} \frac{ \delta S }{ \delta [q^I]^{(m)} } - L
\end{align}
being the conservative energy function for the system. Also, the nonconservative Noether current $\calJ_a = J_a + \kappa_I \omega^I_a$ satisfies
\begin{align}
	\frac{ d \calJ_a }{ dt } = \omega_a^I \bigg[ \frac{ \pd K}{ \pd q^I_- } \bigg]_{\rm PL} + \dot{\omega}_a^I \kappa_I
\end{align}
where
\begin{align}
	J_a \equiv {} & \sum_{m=1}^M [\omega_a^I]^{(m-1)} \frac{ \delta S }{ \delta [q^I]^{(m)} } 
\end{align}
is the familiar conservative Noether current but for higher time derivative systems.

The corresponding results for higher-derivative nonconservative classical field theories are generalized
in an obvious way following similar manipulations. As such, we do not give their results here.

\section{Dynamical boundary contributions}
\label{sec:boundaries}

In some problems, the volume or region occupied by a field can change with time.
Hence, when the action is varied to get the equations of motion there will generally
be a contribution arising from the boundary terms in (\ref{eq:action10}) to the field's dynamics.
Such a situation occurs,
for example, with a particle in a fluid where the outer boundary of
the fluid is fixed (e.g., the walls of a pipe) but the interior of the particle is not penetrated by the fluid. 
The particle can move in response to traction forces and torques that develop across its
surface. The resulting motion couples back into the fluid effecting the overall flow. In this Appendix, 
we show how these surface contributions affect the Euler-Lagrange equations of motion, Noether's 
theorem, and stress-energy conservation.

\subsection{Euler-Lagrange equations of motion}

For movable boundaries the variation of the integral is given by
\eqref{eq:action10}. With ${\cal V} = [t_i, t_f] \times V$ we can use
the equality condition at the final time and the fixed initial data to
eliminate the boundary contributions at $t_i$ and $t_f$.
This leaves the surface integrals in \eqref{eq:surface10},
\begin{align}
  \oint_{\pd {\cal V}} \!\!\! d\Sigma_\mu \frac{ \pd \Omega }{ \pd (
    \pd_\mu \phi^I_a) } \eta_a^I
    = {} & \int_{t_i}^{t_f} \!\!\! dt \oint_{\pd V} \!\!\! dS_i \, \Pi^{ai}_{I} \eta_a^I
\label{eq:surface20}
\end{align}
where $dS_i$ is the (oriented) coordinate surface area element on the boundary $\pd V$.
Because the interactions between the field and the surface are now
dynamical we are no longer free to set $\eta_-^I$ and $\Pi
_{-I}^i$ to zero on $\pd V$ as we did in
Sec.~\ref{sec:continuumLagrange} because a variation on the
surface translates into a force. Therefore, the action in
\eqref{eq:action10} is
\begin{align}
  S = {} & \int_{\cal V} d^4 x \left\{ [\Omega ]_0 + \epsilon \,
    \eta_a^I \left[ \frac{ \pd \Omega }{ \pd \phi_a^I } - \pd_\mu
      \frac{ \pd \Omega }{ \pd (\pd_\mu \phi^I_a) } \right]_0 \right\}
  \nonumber\\
 & {}+ \epsilon \int_{t_i}^{t_f} \!\!\! dt  \oint_{\pd V} \!\!\! dS_i \, \Pi^{ai}_{I} \eta_a^I + {\cal O}(\epsilon^2)
\end{align}
The surface integral can be written as a volume integral by noting that
\begin{align}
  \oint_{\pd V} {\hskip-0.1in} dS_i \, \frac{ \pd \Omega }{ \pd (\pd_i
    \phi^I_a) } \eta_a^I
     = \! \int_{V} {\hskip-0.05in} d^3x \, \eta_a^I (x^\alpha) \! \oint_{\pd V}
    {\hskip-0.1in} dS'_i \, \delta^3( x - x') \Pi^{ai}_{I}
     \nonumber
\end{align}
where $x \in V$ is in the spatial volume and $x' \in \pd V$
is on the spatial boundary. Then, the action becomes
\begin{align}
  S = {} & \int_{\cal V} d^4 x \left\{ [\Omega ]_0 + \epsilon \,
    \eta_a^I \left[ \frac{ \pd \Omega }{ \pd \phi_a^I } - \pd_\mu
      \frac{ \pd \Omega }{ \pd (\pd_\mu \phi^I_a) }
    \right.\right.\nonumber\\
&\left.\left. {}+ \oint_{\pd V}
      \!\!\! dS'_i \, \delta^3( x - x') \Pi^{ai}_{I} \right]_0  \right\} + {\cal O}(\epsilon^2)
\end{align}
which is stationary when $[ \pd S  /\pd \epsilon ]_0$
vanishes yielding
\begin{align}
    \pd_\mu \frac{ \pd \Omega }{ \pd (\pd_\mu \phi^I_a) } = \frac{ \pd
      \Omega }{ \pd \phi_a^I } +  \oint_{\pd V} \!\!\! dS'_i \,
    \delta^3( x - x') \Pi^{ai}_{I}
\end{align}
where $dS'_i$ points {\it out} of the spatial volume $V$.

In the physical limit, only the equation for $a=-$ survives giving
\begin{align}
	\pd_\mu \frac{ \pd {\cal L}}{ \pd (\pd_\mu \phi^I) } - \frac{ \pd {\cal L}}{ \pd \phi^I } = {} & {\cal Q}_I  + \oint_{\pd V} \!\!\! dS'_i \, \delta^3(x - x') \Pi^i_I
\label{eq:eom3}
\end{align}
where we recall that
\begin{align}
	\Pi^\mu_I (x^\alpha) \equiv \big[ \Pi^\mu_{+I} \big]_{\rm PL} = \frac{ \pd {\cal L} }{ \pd (\pd_\mu \phi^I) } + \left[ \frac{ \pd {\cal K} }{ \pd (\pd_\mu \phi^I_-) } \right]_{\rm PL} 
\end{align}
is the canonical current density and $\calQ_I$ is given in (\ref{eq:eom2}).
The surface integral term in (\ref{eq:eom3}) accounts for the full 
nonconservative momentum flux through the boundary of the 
spatial volume. The generalization to $N$ movable boundaries 
requires summing over $N$ surface integrals.

\subsection{Noether's theorem generalized}

For dynamical boundaries, we use the equations of motion from
\eqref{eq:eom3} to write the divergence of the stress-energy tensor in \eqref{eq:divTmunu1} and
Noether current in (\ref{eq:divJnu1}) as
\begin{align}
	\pd_\nu ( \xi^\mu_\alpha T_\mu{}^\nu ) = {} & - \xi^\mu_\alpha \frac{ \pd \calL }{ \pd x^\mu } + \xi^\mu_\alpha \pd_\mu \phi^I \calQ_I \nn \\
		& + \xi^\mu_\alpha \pd_\mu \phi^I \oint_{\pd V} \!\!\! dS'_i \, \delta^3 (x - x') \Pi^i_I \\
	\pd_\nu J^\nu_a = {} & \omega^I_a \calQ_I +  \omega^I_a \oint_{\pd V} \!\!\! dS'_i \, \delta^3 (x - x') \Pi^i_I
\end{align}
Using the known expression for ${\cal Q}_I$ in terms of ${\cal K}$ from (\ref{eq:eom2}) allows us to rewrite the above expressions in terms of divergences of $\calT_\alpha{}^\nu$ and $\calJ^\nu_a$ as in (\ref{eq:divTmunu2}) and (\ref{eq:divJnu2}), respectively,
\begin{align}
	\pd_\nu ( \xi^\mu_\alpha \calT_\mu{}^\nu ) = {} & - \xi^\mu_\alpha \frac{ \pd \calL }{ \pd x^\mu }  + \xi^\mu_\alpha \pd_\mu \phi^I \bigg[ \frac{ \pd \calK }{ \pd \phi_-^I } \bigg]_{\rm PL} {\hskip-0.12in} + \xi^\mu_\alpha \pd_\nu \pd_\mu \phi^I \kappa^\nu_I \nn \\
		& + \xi^\mu_\alpha \pd_\mu \phi^I \oint_{\pd V} \!\!\! dS'_i \, \delta^3 (x - x') \Pi^i_I 
\label{eq:noetherboundariesdiv2}  \\
	\pd_\nu \calJ^\nu_a = {} &  \omega^I_a \bigg[ \frac{ \pd \calK}{ \pd \phi^I_- } \bigg]_{\rm PL} {\hskip-0.05in} + \kappa^\nu_I \pd_\nu \omega^I_a \nn \\
		 & +  \omega^I_a \oint_{\pd V} \!\!\! dS'_i \, \delta^3 (x - x') \Pi^i_I
\end{align}
and $\kappa_I^\mu$ is given in (\ref{eq:Noetherkappa1}). 

When ${\cal K}=0$ it follows that $\calT_\alpha{}^\nu = T_\alpha{}^\nu$ and ${\cal J}^\nu_a = J^\nu_a$, which are not
necessarily conserved because of momentum fluxes that result from interactions with the surface,
\begin{align}
	\pd_\nu ( \xi^\mu_\alpha T_\mu{}^\nu ) = {} & - \xi^\mu_\alpha \frac{ \pd \calL }{ \pd x^\mu }   + \xi^\mu_\alpha \pd_\mu \phi^I \oint_{\pd V} \!\!\! dS'_i \, \delta^3 (x - x') P^i_I \nn \\
	\pd_\nu J^\nu_a = {} &  \omega^I_a \oint_{\pd V} \!\!\! dS'_i \, \delta^3 (x - x') P^i_I  \nn
\end{align}
as well as any explicit coordinate dependence in $\calL$. Here, $P^\mu_I (x^\alpha) = \partial {\cal L} /  \partial (\partial_\mu \phi^I )$.

\section{Preliminaries of fluids and materials}
\label{sec:ContinuumMechPrelim}

\begin{figure*}[htb]
  \centering
  \includegraphics[width=\textwidth]{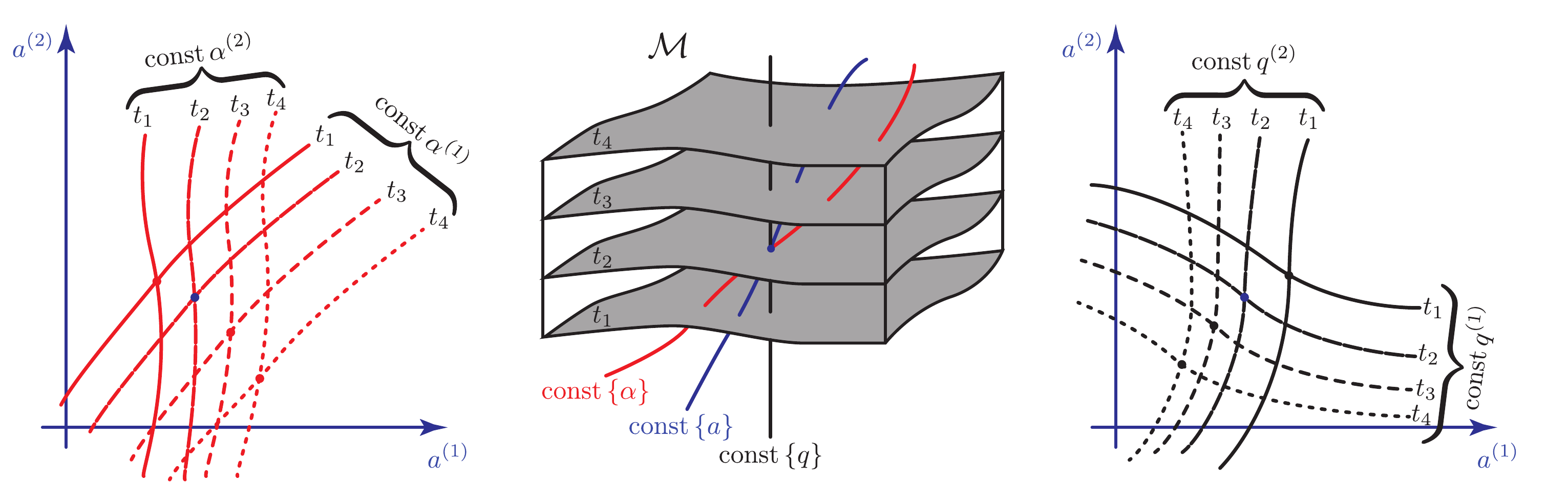}
  \caption{We take a manifold ${\cal M}$ which has natural foliations
    by the surfaces of constant time (middle figure). World lines
    describing paths with fixed coordinates $a$ (e.g., for a given fluid element) and
    $q$ (e.g., for an Eulerian observer) are shown. In the Lagrange picture, this manifests as the
    flow of the Eulerian ($q$) coordinates (right figure)
    relative to the fixed material ($a$) coordinates.
    When another species of fluid is included, such as for describing the
    flow of entropy \cite{Prix:2002jn, 2010RSPSA.466.1373A}, 
    there is an additional fixed coordinate $\alpha$
    that traces a path in the manifold (middle) and flows relative to the
    fixed material ($a$) coordinates (left).
\vspace{-1em}
    }  
\label{fig:coordinatemap}
\end{figure*}

In this appendix we provide mathematical preliminaries that will be 
useful for the hydrodynamics examples given in Sec.~\ref{sec:Hydro}. 

In Sec.~\ref{sec:phichi} we saw an example of a theory which is
formulated in a manifestly Lorentz-covariant, four-dimensional
fashion. In contrast, the Eulerian and Lagrangian descriptions of the
physics of fluids and materials is usually presented in a
three-dimensional form, with time being a parameter. However, this
hides the geometric, coordinate-independent physics. Instead, we will
take a four-dimensional view and treat all quantities as functions,
vectors, tensors, and tensor densities on a manifold with a natural
time function. For background, see any standard differential geometry
text~\cite{spivak1970comprehensive,schutz1980geometrical,wald}.

\subsection{Euler and Lagrange coordinates}

Our geometric setting is a four-dimensional
Lorentzian manifold $\mathcal{M}$, which we declare to have a natural
time function $t$. We only consider coordinate systems
$x^{\mu}=(t,x^{1},x^{2},x^{3})$ which
have $t$ as one of the four coordinate functions,
e.g.,~$x^{0}=t$. Remember that a coordinate system is a collection of
functions. The superscript $\mu$ runs over 0--3 and merely labels the set of
four functions; it is {\it not} a vector index on these functions. We will use Greek letters
to denote spacetime indices that run over 0--3 and Latin
letters for spatial indices over 1--3.

To connect with the usual Eulerian and Lagrangian descriptions of
continuum mechanics, we will introduce (at least) two coordinate
systems below. The Eulerian and Lagrangian descriptions of continuum
mechanics both provide useful insight into the evolution of continuum
flows. While the Eulerian description of fluids is more popular, it is
far easier to construct actions for fluid dynamics without constraints
in the Lagrangian description (see e.g.,~\cite{Eckart1960, Katz1961,
  Seliger1968, Morrison:1998zz}). We will generally adopt the
Lagrangian description when describing the actions and performing the
variations in Sec.~\ref{sec:Hydro}, but also provide the Eulerian form for
the equations of motion.

In the more widely used Eulerian description, material flows past the
``lab frame'' Eulerian coordinates. The Lagrangian formulation instead
describes the evolution from the viewpoint of individual fluid
elements or parcels
tracing out paths in the ``lab frame''.

We denote the Eulerian coordinate system by $q^{\alpha}$, again with $q^{0}=t$.
We will use lower-case Latin
indices to label the spatial Eulerian coordinate functions, $q^{i}$,
and for spatial components of tensors in Eulerian coordinates. Next,
consider a fluid or material flowing in this space, and at some
initial or fiducial time, label the fluid elements with spatial coordinates
$a^{A}$. Now promote these labels to functions on the whole manifold
so that they are constant on each \emph{world-line}, or fluid
parcel's trajectory through spacetime (see
Fig.~\ref{fig:coordinatemap}). That is, each choice of fixed
$(a^{1},a^{2},a^{3})$ labels the trajectory of an individual fluid
parcel through spacetime. Now taking $a^{0}=t$, these four functions
$a^{\alpha}$ are a valid coordinate system. We will use capital
Latin indices from the beginning of the alphabet to label the spatial
Lagrangian coordinate functions, $a^{A}$, and for spatial components
of tensors in Lagrangian coordinates. The simplest and most common
choice for $a$ comes from taking $a^{\alpha}(t=0)=q^{\alpha}(t=0)$.

\subsection{Coordinate transformations}

Each coordinate system comes equipped with its own coordinate basis
vectors ($\frac{\pd}{\pd q^{\alpha}}$ and $\frac{\pd}{\pd
  a^{\alpha}}$) and coordinate basis one-forms ($dq^{\alpha}$ and
$da^{\alpha}$). It is with respect to these bases that vector and
tensor components are evaluated in these coordinate systems.
The spatial bases $\frac{\pd}{\pd q^{i}}$ and $\frac{\pd}{\pd a^A}$ are notationally
unambiguous, and we will abbreviate them as  $\bar{\pd}_{i}$ and $\pd_{A}$, respectively.
However, we must pay special attention to $\frac{\pd}{\pd t}$.

In an arbitrary coordinate system, the meaning of $\frac{\pd}{\pd
  x^{i}}$ is ``partial derivative in the direction of $x^{i}$ while
holding all other $x^{j\ne i}$ constant.'' However, writing $\pd/\pd
t$ lacks the information of which three coordinates are being held
constant, since the coordinate function $t$ is common to all
coordinate systems. Therefore, we introduce the following
notation. We define
\begin{equation}
  \frac{\pd}{\pd t_{x}} \equiv \frac{\pd}{\pd x^{0}} = \frac{\pd}{\pd t}\bigg|_{x^{i}=\text{const}}
\end{equation}
for any coordinate system $x$, either $q$ or $a$ (or others, in the
case of multifluids). These 4-vectors are the tangents to the
world-lines of constant $(x^{1},x^{2},x^{3})$. For convenience we also
introduce $\pd_{t}=\pd/\pd t_{a}$ and $\bar{\pd}_{t}=\pd/\pd t_{q}$.

To see how to transform coordinate components of tensors between the
Lagrangian and Eulerian coordinate system (and importantly, how to
relate $\pd_{t}$ to $\bar{\pd}_{t}$), we must introduce the Jacobian
matrix between the two coordinate systems. The Jacobian matrix is
given by { \renewcommand{\arraystretch}{1.5}
\begin{align}
  \frac{\pd q^{\alpha}}{\pd a^{\beta}} &=
  \begin{bmatrix}
    \frac{\pd t}{\pd t_{a}} & \frac{\pd t}{\pd a^{1}} & \frac{\pd t}{\pd a^{2}} & \frac{\pd t}{\pd a^{3}} \\
    \frac{\pd q^{1}}{\pd t_{a}} & \frac{\pd q^{1}}{\pd a^{1}} & \frac{\pd q^{1}}{\pd a^{2}} & \frac{\pd q^{1}}{\pd a^{3}} \\
    \frac{\pd q^{2}}{\pd t_{a}} & \frac{\pd q^{2}}{\pd a^{1}} & \frac{\pd q^{2}}{\pd a^{2}} & \frac{\pd q^{2}}{\pd a^{3}} \\
    \frac{\pd q^{3}}{\pd t_{a}} & \frac{\pd q^{3}}{\pd a^{1}} & \frac{\pd q^{3}}{\pd a^{2}} & \frac{\pd q^{3}}{\pd a^{3}}
  \end{bmatrix}
\label{eq:Jacobian-block}
  &=
  \begin{bmatrix}
    1 & \mathbf{0} \\
    \mathbf{V}^{\bq}_{\ba} & \mathbf{E}^{\bq}_{\ba}
  \end{bmatrix}
\end{align}
}%
where we have defined $\mathbf{V}^{\bq}_{\ba}$ (labeled with the two
coordinate systems) as the purely spatial
``vector'' which is the part of the relative velocity of the $q$
coordinates, as measured by an observer at constant $a$, with
components $(\mathbf{V}^{\bq}_{\ba})^{i}=\pd q^{i}/\pd t_{a}$. We have
also defined the purely spatial Jacobian matrix $\mathbf{E}^{\bq}_{\ba}$
(also labeled with the two coordinate systems).
From the form of the full $4\times4$
Jacobian matrix we can easily see that purely spatial vectors (ones
which are tangent to surfaces of $t={}$const) may be
transformed between coordinate systems with just the $\mathbf{E}_{\ba}^{\bq}$
matrix. We write the components of $\mathbf{E}_{\ba}^{\bq}$ as
\begin{equation}
  e^{i}_{A} = \frac{\pd q^{i}}{\pd a^{A}} = \pd_{A}q^{i}\,,
 \label{eq:eiAdef1}
\end{equation}
which is sometimes called the deformation gradient in continuum
mechanics.\footnote{%
The symbol $e^{i}_{A}$ is sometimes used to denote an orthonormal
frame field, but for our usage, it is literally just a Jacobian matrix.
}
This Jacobian transforms \emph{purely spatial} vectors ${\bm v}$ from the
Lagrangian coordinate basis to the Eulerian one, and the spatial
components of one-forms ${\bm w}$ in the opposite sense:
\begin{align}
\label{eq:jac-trans-1}
  v^{i} &= e^{i}_{A} v^{A}\,,& w_{A} &= e^{i}_{A}w_{i}\,.
\end{align}
A special case of this is the spatial gradient of a scalar function,
\begin{align}
  \pd_{A} f &= e^{i}_{A} \bar{\pd}_{i} f\,.
\end{align}

The determinant of the spatial Jacobian matrix is sometimes also
called the Jacobian, and it is given by
\begin{align}
	J = \det \, \frac{ \pd q^i }{ \pd a^A } = \det \, e^i_A = \frac{1}{3!} \epsilon_{ijk} \epsilon^{ABC} e^i_A e^j_B e^k_C \,.
\label{eq:Jacobian47}
\end{align}
It is also the determinant of the full $4\times 4$ Jacobian matrix.

When it exists, it is straightforward to see that the inverse matrix,
$\pd a^{\alpha}/\pd q^{\beta}$ has the same block form as
\eqref{eq:Jacobian-block}, but exchanging $a\leftrightarrow q$. It is
also easy to verify that
\begin{align}
  \label{eq:inv-jac-rels}
  \mathbf{E}_{\bq}^{\ba}&=(\mathbf{E}_{\ba}^{\bq})^{-1}\,, &
  \mathbf{V}_{\bq}^{\ba} &=
  -\mathbf{E}_{\bq}^{\ba}\mathbf{V}_{\ba}^{\bq} \,.
\end{align}
We write the components of $\mathbf{E}_{\bq}^{\ba}$ as
\begin{align}
  e^{A}_{i} = \frac{\pd a^{A}}{\pd q^{i}}\,.
\end{align}
Since $e^{i}_A$
is generally not symmetric, neither is its inverse. Naturally the
determinant of the inverse is the multiplicative inverse of $J$,
i.e.~$\det e_{i}^{A} = 1/J$.
The relative velocity identities of \eqref{eq:inv-jac-rels} in components read
\begin{align}
e^{A}{}_{i} &= \left[ e^{i}{}_{A} \right]^{-1} , &
  \frac{\pd a^{A}(t,q)}{\pd t_q} &=
  -e^{A}{}_{i}
	 \frac{\pd q^{i}(t,a)}{\pd t_a} \label{eq:relvel}
\end{align}
With the inverse Jacobian matrix, we have
two more transformation rules to add to \eqref{eq:jac-trans-1}:
\begin{align}
\label{eq:jac-trans-2}
  v^{A} &= e^{A}_{i} v^{i}\,,& w_{i} &= e^{A}_{i}w_{A}\,.
\end{align}

Using Cramer's rule it is typical to
write the inverse matrix in terms of the transposed matrix of
cofactors,
\begin{align}
e_{i}^{A} = \frac{1}{J} A_{i}{}^A\,,
\end{align}
where the cofactor matrix $A_i{}^A$ is defined to be
\begin{align}
  A_i{}^A & \equiv \frac{ \pd J}{ \pd (\pd_A q^i ) } = \frac{ \pd J}{
    \pd e^i_A } 
     = \frac{1}{2!} \epsilon_{ijk} \epsilon^{ABC} e_B{}^j e_C{}^k.
    \label{eq:cofactor1}
\end{align}

From the definition of an inverse matrix, we immediately get two
identities involving the cofactor matrix $A_{i}{}^{A}$: one via the
left inverse and one via the right inverse. These two identities are
\begin{align}
e^{i}_{A} e_{j}^{A} = \delta^{i}{}_{j}
\quad\Longrightarrow\quad
(\pd_{A}q^{i}) A_{j}{}^{A} = J \delta^{i}{}_{j}, \label{eq:CofactorIdentity1}
\end{align}
and
\begin{align}
e_{i}^{A}  e^{i}_{B} = \delta^{A}{}_{B}
\quad\Longrightarrow\quad
A_{i}{}^{A} (\pd_{B} q^{i}) = J \delta^{A}{}_{B} \label{eq:CofactorIdentity2}\,.
\end{align}

We find a third identity from the symmetry of partial
derivatives. Notice in the definition \eqref{eq:cofactor1} that
both derivative indices $B, C$ are contracted with the completely anti-symmetric Levi-Civita
tensor, which has the free upper index $A$. Taking the divergence
on that free index gives
\begin{align}
  \pd_A A_i{}^A
    &= \frac{1}{2!} \epsilon_{ijk} \epsilon^{ABC}
    \big( \pd_A \pd_B q^j \pd_C q^k + \pd_B q^j \pd_A \pd_C q^k \big) \, .
\end{align}
This establishes that $A_{i}{}^{A}$ is divergence-free on its second
(upper) index,
\begin{align}
  \pd_A A_i{}^A &= 0\,.
\label{eq:cofactoridentity1}
\end{align}
This result only holds for the second index, since in general
the cofactor matrix is not symmetric.

With the conventional choice of coordinates [$a^{\mu}(t=0)=q^{\mu}(t=0)$],
the volume of the fluid element at initial time $t=0$ is $d^3q(t=0) =
d^3a$, while at a later time its volume is $d^3q = J d^3a$. The
Jacobian thus accounts for the volume changes that result from
dynamical evolution. Mass conservation then requires the density of a
fluid element to be related to its \emph{initial} density $\rho(a)$ by the
ratio of the volume elements at the two times,
\begin{align}
\bar{\rho}(t,q) = \frac{\rho({ a})} {J}\,, \label{eq:Lagcont}
\end{align}
where the mass element is 
$dM = \bar{\rho} \ d^3q = \rho\ d^3a$.
A quantity which acquires factors of Jacobian determinants when
changing coordinate systems is called a density (e.g.,~scalar density
or tensor density). Specifically, a tensor density $\calT$ of weight
$+w$ transforms from coordinate system $q^{i}$ to coordinate system
$a^{A}$ as
\begin{align}
  \calT_{AB\cdots} &= \left|\det\frac{\pd a^{C}}{\pd q^{k}}\right|^{+w}
  e^{i}_{A}e^{j}_{B}\cdots\calT_{ij\cdots} \\
  &= J^{-w} e^{i}_{A}e^{j}_{B}\cdots\calT_{ij\cdots} \,,
\end{align}
since $J=\det \pd q^{i}/\pd a^{A}$ from \eqref{eq:Jacobian47}.
A simple way to turn a tensor into a tensor density is to multiply it
by a factor of $J^{-w}$. Thus we say that $\bar\rho$ is a scalar
density of weight $+1$.
We adopt the overbar on some tensors to indicate tensor densities of
non-zero weight $w$, which have gained a factor of $1/J^w$ when
transformed to the Eulerian coordinates.

\subsection{Metric and derivatives}

In order to measure the lengths of spatial vectors, we must introduce
a spatial metric ${\bm{ds}}^{2}$. In Eulerian coordinates, it has
components $g_{ij}$ in ${\bm{ds}}^{2}=g_{ij}dq^{i}dq^{j}$, with
inverse metric $g^{ij}$ such that $g^{ik}g_{kj}=\delta^{i}_{j}$. The
metric and inverse are used to ``raise'' and ``lower'' contravariant
and covariant tensor indices. When
expressed in Lagrangian coordinates, it is traditionally represented
as $C_{AB}$ and called the Cauchy-Green deformation tensor~\cite{2009rheo.book.....O}
\begin{equation}
  {\bm{ds}}^{2} = g_{ij}dq^{i}dq^{j} = C_{AB}da^{A}da^{B}\,,
\end{equation}
or, in terms of the Jacobian matrix and its inverse, the deformation
tensor and its inverse are
\begin{align}
\label{eq:C-from-g-e}
  C_{AB} &= e^{i}_{A}g_{ij}e^{j}_{B}\,, &   C^{AB} &= e^{A}_{i}g^{ij}e^{B}_{j}\,.
\end{align}

With the conventional choice of coordinates [$a^{\mu}(t=0)=q^{\mu}(t=0)$],
at time $t=0$, the spatial Jacobian matrix is instantaneously equal to
the identity, $e^{i}_{A}(t=0)=\delta^{i}_{A}$, so we instantaneously
have $C_{AB}(t=0)=g_{AB}$. The metric determinant in different
coordinate systems are related through the Jacobian:
\begin{equation}
  \left.\begin{aligned}
      d^{3}q&= J d^{3}a \\
      \sqrt{g}d^{3}q &= \sqrt{C} d^{3}a
    \end{aligned}
  \right\} \Longrightarrow
  J = \sqrt{\frac{C}{g}}
\end{equation}
where $g\equiv\det g_{ij}$ and $C\equiv\det C_{AB}$.

We now turn to how $t$-components of vectors are related in the
Eulerian and Lagrangian coordinates. Using the full $4\times 4$
Jacobian, we have
\begin{align}
  \frac{\pd}{\pd t_{a}} &= \frac{\pd q^{\alpha}}{\pd a^{0}}
  \frac{\pd}{\pd q^{\alpha}}
  = \frac{\pd}{\pd t_{q}} + \frac{\pd q^{i}}{\pd t_{a}}
  \frac{\pd}{\pd q^{i}}
\end{align}
which would be written in the more traditional fluid mechanics
literature as
\begin{equation}
\label{eq:pdta-pdtq}
\frac{\pd}{\pd t_{a}} = \frac{\pd}{\pd t_{q}} +
\mathbf{V}^{\bq}_{\ba}\cdot\cd_{q}
\end{equation}
where we have defined the ``vector''
$\cd_{q} = (\pd/\pd q^{1},\pd/\pd q^{2},\pd/\pd q^{3})$. In fact, the
combination $\mathbf{V}^{\bq}_{\ba}\cdot\cd_{q}$ is a true tangent vector
so we define the symbol
\begin{equation}
\label{eq:vqa-def}
\mathbf{v}^{\bq}_{\ba}\equiv\mathbf{V}^{\bq}_{\ba} \cdot \cd_{q}
\equiv \pd/\pd t_{a}-\pd/\pd t_{q}
= \pd_{t}-\bar{\pd}_{t}
\end{equation}
and with Eulerian coordinate components
$(\mathbf{v}^{\bq}_{\ba})^{i}=\pd q^{i}/\pd t_{a}$. This vector is
easily seen to be purely spatial, since $\mathbf{v}^{\bq}_{\ba}(t) =
\pd t/\pd t_{a}-\pd t/\pd t_{q} = 0$, and so $\mathbf{v}^{\bq}_{\ba}$
is tangent to surfaces of $t={}$const.\footnote{%
  The notation $\bv(s)$ means the action of vector $\bv$ on the scalar
  $s$, i.e.~to take a directional derivative of $s$ on the
  manifold in the direction of $\bv$.}

Equation~\ref{eq:pdta-pdtq} [or \eqref{eq:vqa-def}] relates three
tangent vectors, and therefore holds as
an identity when acting on scalar functions. This is the usual
``material derivative'' but only when the vector acts on a manifold scalar
function. If we want to take derivatives of tensors on this manifold,
then we will have to use the covariant ``parameter derivative'' along
these tangent vector fields. When acting on a scalar field, the
parameter derivative along curves of constant $a$ agrees with
the material derivative. In general, to evaluate the parameter
derivative we need a covariant derivative (or connection), $\cd$,
and the parameter derivative is given by
\begin{equation}
  \frac{D}{D\lambda}{\bm T} = \cd_{\bm u} {\bm T}
  = u^{\alpha}\cd_{\alpha} {\bm T}
\end{equation}
where $\bm u$ is the 4-vector which is tangent to the curve with
parameter $\lambda$, with components $u^{\alpha} =
dx^{\alpha}(\lambda)/d\lambda$ where the world-line coordinates are
$x^{\alpha}(\lambda)$. When
acting on a scalar, any covariant derivative agrees with the usual
partial derivative, $\cd_{\alpha} f = \pd_{\alpha} f$. When acting on
vectors and one-forms, we need the non-tensorial connection
coefficients $\Gamma^{\alpha}_{\beta\gamma}$
to relate the covariant
and partial derivatives:
\begin{align}
  \cd_{\alpha}A^{\beta} &= \pd_{\alpha}A^{\beta} + \Gamma^{\beta}_{\alpha\gamma} A^{\gamma} \\
  \cd_{\alpha}B_{\beta} &= \pd_{\alpha}B_{\beta} - \Gamma^{\gamma}_{\alpha\beta} B_{\gamma}\,.
  \label{eq:covDform1}
\end{align}
For a tensor density of weight $+w$, we also need to add the single
correction term,
\begin{align}
	-w\Gamma^{\gamma}_{\gamma\alpha}T^{\ldots}_{\ldots} \, .
\label{eq:withweight1}
\end{align}

In this work,
we fix the spatial part of the connection coefficients by demanding
that the covariant derivative is compatible with the spatial metric,
$\bar{\cd}_{i}g_{jk}=0$, so that $g_{ij}\bar{\cd}_{k}W^{j} = \bar{\cd}_{k}W_{i}$. This means
that our connection coefficients are the usual Christoffel
coefficients when the indices are all spatial~\cite{wald}.
In Eulerian coordinates,
the Christoffel coefficients of $\bar{\cd}_{i}$ are\footnote{%
We use an overbar to match the notation for covariant
$(\bar{\nabla}_i$) and partial $(\bar{\pd}_i)$ in Eulerian
coordinates and should not be taken to imply that the connection
coefficients are tensor densities.}
\begin{equation}
  \label{eq:Christoffel-E}
  \bar{\Gamma}^{i}{}_{jk} = \frac{1}{2} g^{il} \left(\bar{\pd}_{j}g_{kl}+\bar{\pd}_{k}g_{jl}-\bar{\pd}_{l}g_{jk} \right)
\,,
\end{equation}
while in Lagrangian coordinates, they are
\begin{equation}
  \label{eq:Christoffel-L}
  \Gamma^{A}{}_{BC} = \frac{1}{2} C^{AD} \left(\pd_{B}C_{CD}+\pd_{C}C_{BD}-\pd_{D}C_{BC} \right)
\,.
\end{equation}
For example, if we are working in Cartesian coordinates and
$g_{ij}=\delta_{ij}$, then $\bar{\Gamma}^{i}_{jk}=0$.

There is still freedom to fix time components of $\Gamma$.\footnote{%
  Thiffeault noted~\cite{2001JPhA...34.5875T} that other covariant
  time derivatives, ${\cal D}$, can be defined by using a non-tensorial rank-2
  quantity $\alpha^{i}{}_{j}$ . In our 4-dimensional language, we see
  that the 3-dimensional quantity $\alpha^{i}{}_{j}$ is simply the
  choice of components of $\Gamma^{i}_{0j}$. Thiffeault separates out
  an arbitrary tensorial part ${\cal H}^{i}{}_{j}$ which may be freely
  specified, and modifies the derivative presented here through
  \begin{equation}
  \begin{split}
    ({\cal D}T)^{i_{1}i_{2}\ldots}_{j_{1}j_{2}\ldots} \equiv
    (\cd_{\bf u}{\bf T})^{i_{1}i_{2}\ldots}_{j_{1}j_{2}\ldots}
   & {}+ {\cal H}^{i_{1}}{}_{k} T^{ki_{2}\ldots}_{j_{1}j_{2}\ldots} +\cdots\\
   & {}- {\cal H}^{k}{}_{j_{1}} T^{i_{1}i_{2}\ldots}_{kj_{2}\ldots} -\cdots
  \end{split}
  \end{equation}
  with $+{\cal H}$ correction terms for contravariant indices and
  $-{\cal H}$ correction terms for covariant indices.
  For example, in Lagrangian coordinates, choosing ${\cal H}_{{A} {B}} =
  {\gamma}_{AB}$ (the rate-of-strain tensor, defined below) yields the
  Jaumann corotational derivative.
}
Consider
for example the acceleration vector of the world lines of constant
$a$,
\begin{align}
  \dot{\bm v}^{i} &\equiv
  \left(\frac{D}{Dt_{a}}{\bm v} \right)^{i} =
  (\cd_{\pd/\pd t_{a}} {\bm v})^{i} \\
  &=
  v^{0}\nabla_{\pd/\pd t_{q}} v^{i}+ v^{j} \nabla_{j} v^{i}\\
  &= \frac{\pd}{\pd t_{q}} v^{i} + \Gamma^{i}_{0\alpha}v^{\alpha}
  + v^{j} \nabla_{j} v^{i}\,
\end{align}
where we are parametrizing the world-line by the parameter $t$, so the
tangent vector has $v^{0}=1$ (in both Eulerian and Lagrangian
coordinates).
If we choose to set $\Gamma^{i}_{0\alpha}=0$,\footnote{%
If we formulate the theory in 4 dimensions with a 4-metric that
satisfies $g_{00}=-1$, $g_{0i}=0$, and $\bar{\pd}_{t}g_{ij}$=0, then
the choice $\Gamma^{\mu}_{0\alpha}=0$ in the $q$ coordinates is
consistent with metric-compatibility with the 4-metric.
}
then this acceleration
vector agrees with the usual notion in the fluid mechanics literature:
that the acceleration vector is the ``material derivative'' of the
velocity vector,
\begin{align}
  \dot{v}^{i} &= \frac{\pd}{\pd t_{q}} v^{i} + v^{j} \bar{\cd}_{j} v^{i}\,, \label{eq:MaterialDer11}
\\
\intertext{or, in Cartesian coordinates with a flat spatial metric,}
  \dot{v}^{i} &= \frac{\pd}{\pd t_{q}} v^{i} + v^{j} \bar{\pd}_{j} v^{i}\,.
\end{align}

Often times we will want to be able to express time derivatives of
components of tensors in both the Eulerian and Lagrangian coordinates. For
example, we might ask: what is the relation between
\begin{align}
  \pd_{t} \left( T^{A}{}_{B} \right)&&\text{and}&&\bar{\pd}_{t}\left(T^{i}{}_{j}\right)?
\end{align}
This is most easily answered with the \emph{Lie derivative}
$\calL_{\bv}$ along some vector field $\bv$. The Lie derivative satisfies
\begin{align}
  \calL_{a\bv+b\bw}{\bm T} =   a\calL_{\bv}{\bm T}+b\calL_{\bw}{\bm T}\,,
\label{eq:Lielinear1}
\end{align}
where $a$ and $b$ are constants.
In some arbitrary coordinate system, the
components of the Lie derivative of some tensor ${\bm T}$ along a
vector field $\bv$ is
\begin{align}
(\calL_{\bv}{\bm T})^{\alpha_{1}\alpha_{2}\ldots}{}_{\beta_{1}\beta_{2}\ldots}
={}& v^{\lambda} \pd_{\lambda} (T^{\alpha_{1}\alpha_{2}\ldots}{}_{\beta_{1}\beta_{2}\ldots})\\
&{}-(\pd_{\lambda} v^{\alpha_{1}}) T^{\lambda\alpha_{2}\ldots}{}_{\beta_{1}\beta_{2}\ldots}
-\ldots\nn\\
&{}+(\pd_{\beta_{1}} v^{\lambda}) T^{\alpha_{1}\alpha_{2}\ldots}{}_{\lambda\beta_{2}\ldots}
+\ldots\nn\\
&{}+w(\pd_{\lambda} v^{\lambda}) T^{\alpha_{1}\alpha_{2}\ldots}{}_{\beta_{1}\beta_{2}\ldots}\nn
\end{align}
where there is a $-\pd_{\lambda}v^{\alpha_{i}}$ correction term for
each contravariant index, a $+\pd_{\beta_{j}}v^{\lambda}$ correction term
for each covariant index, and $w$ is the weight of the tensor
density.

The Lie derivative is easy to evaluate
when in a coordinate system where $\bv$ is a coordinate basis vector:
the components $v^{\alpha}$ will be constant [e.g.,~$(1,0,0,0)$] and so
all the ``correction'' terms will vanish. This is convenient when
taking the Lie derivative along $\pd_{t}$ or $\bar{\pd}_{t}$, which are
easiest to evaluate in their respective coordinate
systems. Specifically,
\begin{align}
(\calL_{\bar{\pd}_{t}}T)^{i\ldots}{}_{j\ldots} &=
\frac{\pd}{\pd t_{q}}\left(
  T^{i\ldots}{}_{j\ldots}\right) \\
(\calL_{\pd_{t}}T)^{A\ldots}{}_{B\ldots} &=
\frac{\pd}{\pd t_{a}}\left(
  T^{A\ldots}{}_{B\ldots}\right)
\end{align}

Using the above properties we can relate partial derivatives along
$\pd_{t}$ to those along $\bar{\pd}_{t}$. Using
$\pd_{t}=\bar{\pd}_{t}+\bv_{\ba}^{\bq}$ from \eqref{eq:vqa-def}, we have from \eqref{eq:Lielinear1}
\begin{equation}
  \label{eq:lie-sum}
  \mathcal{L}_{\pd_{t}} {\bm T}=
  \mathcal{L}_{\bar{\pd}_{t}} {\bm T}
  + \mathcal{L}_{\bv^{\bq}_{\ba}} {\bm T}\,,
\end{equation}
which is true as an operator equation acting on any type of tensor.
The LHS is best evaluated in the $a$ coordinate system, while the
first term on the RHS is best evaluated in the $q$ coordinate
system. This expression can be used to give the so-called ``convective 
derivative'' relating time derivatives in different coordinate systems.
Taking components gives, for example,
\begin{align}
e_{A}^{i}e^{B}_{j}
  \frac{\pd}{\pd t_{a}}\left(
  T^{A}{}_{B}\right)
={}&
  \frac{\pd}{\pd t_{q}}\left( T^{i}{}_{j}\right) +
\left(\calL_{\bv_{\ba}^{\bq}}{\bm T}\right)^{i}{}_{j}\\
={}&
  \frac{\pd}{\pd t_{q}}\left( T^{i}{}_{j}\right)
+v^{k} \bar{\pd}_{k} \left(T^{i}{}_{j}\right) \\
&{}- \left(\bar{\pd}_{k} v^{i} \right)T^{k}{}_{j}
+ \left(\bar{\pd}_{j} v^{k} \right)T^{i}{}_{k} \,. \nn
\end{align}
Another useful example which we will often encounter is the Lie
derivative of a scalar density $\sigma$ of weight $w=+1$, such as the
mass density $\rho$, entropy density, $s$, or energy density
$\varepsilon$. We evaluate \eqref{eq:lie-sum} on $\sigma$,
evaluating the LHS in the $a$ coordinates and the RHS in the $q$
coordinates (then transforming the resulting scalar density back to
$a$ coordinates by multiplying by $J$):
\begin{align}
\pd_{t}\sigma &= J\left[\bar{\pd}_{t} \bar\sigma + v^{i}\bar{\pd}_{i} \bar\sigma +
  \bar\sigma \bar\pd_{i}v^{i}\right] \\
\label{eq:ScalarDensityDot0}
&= J\left[ \bar{\pd}_{t}\bar\sigma + \bar\pd_{i}(\bar\sigma v^{i}) \right]\,.
\end{align}
It is easy to show that $\bar{\pd}_i ( \bar{\sigma} v^i ) = \bar{\nabla}_i ( \bar{\sigma} v^i)$
so that (\ref{eq:ScalarDensityDot0}) can be written as
\begin{align}
	\pd_{t}\sigma = J\left[ \bar{\pd}_{t}\bar\sigma + \bar\nabla_{i}(\bar\sigma v^{i}) \right]\,.
\label{eq:ScalarDensityDot}
\end{align}

When acting on a contravariant rank-2 tensor, the derivative
$\calL_{\pd/\pd t_{a}}$, expanded in $q$-coordinates,
corresponds to the usual ``upper convected derivative'' used in
rheology (see e.g.,~\cite{1950RSPSA.200..523O,2009rheo.book.....O}). When acting on
a covariant rank-2 tensor, it corresponds to the ``lower convected
derivative.''

\subsection{Physical quantities}
\label{sec:physical-quantities}

Now we may define some of the physical quantities necessary to
describe fluids.
We have already seen the spatial velocity of a fluid element of
constant $a$, as measured in $q$ coordinates, is $\bv_{\ba}^{\bq}$,
which are tangent to the fluid worldlines of constant $a$. They are
evaluated in coordinates via the parameter derivative of the
coordinate functions. Since the coordinates are just scalars, the
parameter derivatives agree with partial derivatives:
\begin{align}
  v^{i}=
	\dot{q}^i & = \bigg( \frac{ D }{ Dt_a} q \bigg)^i
		= \big( \nabla_{\pd/\pd t_a} q \big)^i
		= \frac{ \pd }{ \pd t_a} q^i
\end{align}
Also mentioned above, the spatial acceleration vector comes from the
covariant parameter derivative along the world-lines, $\dot{\bv}\equiv
D\bv/Dt_{a}=\cd_{\bv}\bv$. Further, we will continue to use a
covariant derivative which is metric-compatible with the spatial
metric, $\cd_{i}g_{jk}=0$, and fix (some) time components of the
$q$-coordinate connection via $\Gamma^{i}_{0\alpha}=0$, so that the
acceleration is given in components by
\begin{equation}
  \left(\frac{D\bv}{Dt_{a}}\right)^{i}=
  \dot{v}^{i} = \frac{\pd}{\pd t_{q}} v^{i} + v^{j} \bar{\cd}_{j} v^{i}
\,.
\end{equation}
In flat space and Cartesian coordinates,
this coincides with the ``material derivative'' of velocity,
\begin{equation}
  \left(\frac{D\bv}{Dt_{a}}\right)^{i}
  = \frac{\pd}{\pd t_{q}} v^{i} + v^{j} \bar{\pd}_{j} v^{i}\,.
\end{equation}

A useful form of the acceleration which we often encounter is given by
\begin{align}
  \left(\frac{D\bv}{Dt_{a}}\right)_{i}&=
  \dot{v}_{i} = \frac{\pd}{\pd t_{q}} v_{i} + v^{j} \bar{\cd}_{j} v_{i}
\\ &
     = \frac{\pd}{\pd t_{a}} v_{i} - v^{j}\bar{\pd}_{j}v_{i} + v^{j} \bar{\cd}_{j} v_{i}
\\
  \label{eq:acc-down-in-pdta}
  \frac{Dv_{i}}{Dt_{a}}&
     =\frac{\pd}{\pd t_{a}} v_{i}  - v^{j}\bar{\Gamma}^{k}_{ji}v_{k} \,.
\end{align}

It is traditional to define the relative
strain tensor $u_{AB}$, which is (half) the difference between the metric
at any time and a reference tensor $[C_o]_{AB}$
\begin{equation}
   2u_{AB} \equiv C_{AB} - [C_o]_{AB} \,. \label{eq:StrainTensor}
\end{equation}
The reference tensor $[C_o]_{AB}$ is symmetric and constant in the Lagrange coordinates. For
viscous fluids it can be arbitrarily defined to be match the metric at some particular 
time $t=0$. For an elastic material it is useful to set $[C_o]_{AB}$  to be 
the `background' material deformation for which the elastic stress is zero.

Another quantity we will encounter in the equations of motion is the
rate of change of the metric (or Cauchy-Green deformation tensor)
along the fluid world-lines. This is known as the rate of strain
tensor,
\begin{align}
\label{eq:gamma-1}
  \gamma_{AB} &\equiv \frac{1}{2} (\calL_{\pd/\pd t_{a}} {\bm g})_{AB}
  = (\calL_{\pd/\pd t_{a}} {\bm u})_{AB} \\
\label{eq:gamma-2}
&= \frac{1}{2}\frac{\pd}{\pd t_{a}} (C_{AB}) =\frac{\pd}{\pd t_{a}}( u_{AB})\,.
\end{align}

For component calculations, it is useful to expand via
\eqref{eq:C-from-g-e}, $C_{AB}=e^{i}_{A}g_{ij}e^{j}_{B}$, giving
\begin{equation}
\begin{split}
  \gamma_{AB} = \frac{1}{2} \bigg[
&\frac{\pd\dot{q}^i}{\pd a^{A}}\frac{\pd q^j}{\pd a^{B}}g_{ij}+ \frac{\pd\dot{q}^i}{\pd a^{B}}\frac{\pd q^j}{\pd a^{A}}g_{ij} \\
& {}+  \frac{\pd q^i}{\pd a^{B}}\frac{\pd q^j}{\pd a^{A}}\pd_t g_{ij} \bigg] .
\end{split}
\end{equation}

We can also express the rate of strain tensor in Eulerian coordinates,
by using the Lie derivative identity in \eqref{eq:lie-sum}. This
gives
\begin{align}
  \gamma_{ij}&=\frac{1}{2} (\calL_{\pd_t} {\bm g})_{ij} =
  \frac{1}{2} (\calL_{\bar{\pd}_t} {\bm g})_{ij}+
  \frac{1}{2} (\calL_{\bv} {\bm g})_{ij} \\
  \label{eq:gamma-ij}
 &= \frac{1}{2}\left[
    v^{k}\bar{\pd}_{k} g_{ij}
    + (\bar{\pd}_{i}v^{k})g_{kj}
    + (\bar{\pd}_{j}v^{k})g_{ik}
    \right]\,.
\end{align}
The term $\bar{\pd}_{t} g_{ij}$ arising from
$(\calL_{\bar{\pd}_{t}}{\bm g})_{ij}$ vanishes since we are considering
stationary metrics.
The astute reader will note that \eqref{eq:gamma-ij} is valid
even for curved metrics and curvilinear coordinates, since throughout
(\ref{eq:gamma-1}-\ref{eq:gamma-ij}) we have only used a Lie
derivative identity and the stationarity of the metric in $q$
coordinates.

\section{Deriving Euler's equation in coordinate covariant form}
\label{app:perfectfluid}

In this appendix, we give some details of the calculation for deriving
the Euler-Lagrange equations for an adiabatic and inviscid (i.e., perfect) 
fluid. We do this to help fill in the gaps not provided in Sec.~\ref{sec:PerfectFluid}.

The Lagrangian density for a perfect fluid is given in (\ref{eq:PerfectL}) and the Euler-Lagrange
equations of motion in (\ref{eq:eom2}). With $\calK = 0$ we have that
\begin{align}
	\frac{ \partial }{ \partial t_a } \frac{ \partial \calL }{ \partial \dot{q}^k } + \partial _A \frac{ \partial \calL }{ \partial e_A^k } - \frac{ \pd \calL }{ \pd q^k } = 0
\end{align}
upon recalling from (\ref{eq:eiAdef1}) that $e_A^k = \pd_A q^k$.
Computing each term above in turn yields
\begin{align}
	\frac{ \partial }{ \partial t_a } \frac{ \partial \calL }{ \partial \dot{q}^k } = {} & \rho(a) \frac{ \pd }{ \pd t_a } \left( g_{ki} \dot{q}^i \right) 
\end{align}
and
\begin{align}
	\partial _A \frac{ \partial \calL }{ \partial e_A^k } = {} & \partial_A \left( - \frac{ \pd J }{ \pd e_A^k } \bar{\varepsilon} - J \frac{ \pd \bar{\varepsilon} }{ \pd J } \frac{ \pd J }{ \pd e_A^k } \right)  \\
		= {} & - A_k{}^A \pd_A \left( \bar{\varepsilon} + J \frac{ \partial \bar{\varepsilon}}{ \pd J } \right)
\end{align}
upon using the identity $\pd_A A_k{}^A = 0$ in (\ref{eq:cofactoridentity1}). The term $\pd \calL / \pd q^k$ has two contributions that are vital for ensuring that the fluid equations of motion are covariantly represented in any coordinate system. The first comes from the fact that the space-time metric $g_{ij} = g_{ij}(t, q(t,a))$ is evaluated on the coordinates of the fluid element's trajectory. The second comes from the fact that $\pd J / \pd q^k$ is generally non-zero in curved spaces and/or in curvilinear coordinates, such as the Lagrangian coordinates $a^A$. Therefore,
\begin{align}
	\frac{ \pd \calL}{ \pd q^k } = {} & \frac{1}{2} \rho \, \bar{\pd}_k g_{ij} \, \dot{q}^i \dot{q}^j - \bar{\pd}_k \left( J \bar{\varepsilon} \right) \, .
\label{eq:dLdq41c}
\end{align}
It is straightforward to show using the definition of the Christoffel
connection coefficients [given in \eqref{eq:Christoffel-E}]
that the first term in (\ref{eq:dLdq41c}) can be written as
\begin{align}
	\frac{1}{2} \bar{\pd}_k g_{ij} \, \dot{q}^i \dot{q}^j = g_{im} \bar{\Gamma}^m_{jk} \dot{q}^i \dot{q}^j = \dot{q}^j \bar{\Gamma}^m_{jk} \left( g_{m i } \dot{q}^i  \right)  \, .
\end{align}
Likewise, using (\ref{eq:Jacobian47}), Jacobi's formula, and \eqref{eq:Christoffel-L} one can show that
\begin{align}
	\bar{\pd}_k J = e_k^A \pd_A J = J e_k^A e^B_i \pd_A e_B^i 
		= A_k{}^A \Gamma^B_{BA} 
\,.
\end{align}
Then, the second term in (\ref{eq:dLdq41c}) is
\begin{align}
	- \bar{\pd}_k \left( J \bar{\varepsilon} \right)
		& = - A_k{}^A \Gamma^B_{BA} \left( \bar{\varepsilon} + J \frac{ \pd \bar{\varepsilon} }{ \pd J} \right)  \, .
\end{align}

Combining these results all together gives for the Euler-Lagrange equations the following expression
\begin{align}
	& \rho \frac{ \pd }{ \pd t_a } \left( g_{ki} \dot{q}^i \right) - \rho \dot{q}^j \bar{\Gamma}^m_{jk} \left( g_{m i } \dot{q}^i  \right)  \nonumber \\
		& {\hskip0.2in} - A_k{}^A \pd_A \left( \bar{\varepsilon} + J \frac{ \partial \bar{\varepsilon}}{ \pd J } \right)  + A_k{}^A \Gamma^B_{BA} \left( \bar{\varepsilon} + J \frac{ \pd \bar{\varepsilon} }{ \pd J} \right) = 0 \, .  \nonumber
\end{align}
Written in this way we see that the first line is proportional to the
covariant parameter derivative of $g_{ki}\dot{q}^i = \dot{q}_k$ from
\eqref{eq:acc-down-in-pdta}, and that the second line is
proportional to the covariant spatial derivative of the pressure,
\begin{align}
	- \bar{P} = \bar{\varepsilon} + J \frac{ \pd \bar{\varepsilon} }{ \pd J }  = \bar{\varepsilon} - \mu \bar{\rho} - T \bar{s}  \, .
\end{align}
In general curvilinear coordinates and a possibly curved space we thus find
\begin{align}
	\rho \frac{D \dot{q}_k}{ Dt_a}  + A_k{}^A \nabla_A \bar{P} = 0  \, .
\end{align}
Since only covariant derivatives that are compatible with the metric appear in this expression we can freely raise the spatial index ``$k$'' by contracting both sides with the metric $g^{ik}$ to give,
\begin{align}
	\rho \frac{ D \dot{q}^i }{ Dt_a}  + g^{ij} A_j {}^A \nabla_A \bar{P } = 0  \, .
\end{align}
If one is working in rectangular coordinates in a flat space then the first term in (\ref{eq:dLdq41c}) for $\partial \calL / \partial q^k$ vanishes while the second one generally remains except for steady flows.

\section{The classical limit of nonequilibrium quantum theory}
\label{sec:quantum}

In this appendix, we show how the nonconservative action in \eqref{eq:newaction1} 
can be derived from the classical limit of quantum 
mechanical systems in nonequilibrium (see e.g., \cite{CalzettaHu, Weiss}). This result
provides a fundamental justification for the manipulations and assumptions
we have made in extending classical Lagrangian and Hamiltonian 
mechanics and field theories to nonconservative systems and dissipative problems.
For presentation purposes, we focus 
on discrete quantum mechanical systems but the steps are 
similar for nonequilibrium quantum field theories.

Consider a closed (i.e., unitary) quantum system with accessible 
degrees of freedom $q(t)$ and variables 
$\bQ(t) = \{Q^I(t) \}_{I=1}^N$ that we will eliminate. 
The action is 
\begin{align}
	S[ q, \bQ] = S_q[q] + S_Q[ \bQ] + S_{\rm int} [q, \bQ]
\label{eq:claction1}
\end{align}
where the last term accounts for mutual interactions.

The density matrix $\hat{\rho}(t)$ at the initial time $t_i$ is given
by $\hat{\rho}_i = \hat{\rho}(t_i)$. At some arbitrary future final time $t_f$
the density matrix will have unitarily evolved to $\hat{\rho}_f = \hat{\rho}(t_f)$ via 
the time-evolution operator $\hat{U} (t_f, t_i)$ so that
\begin{align}
	\hat{\rho}_f = \hat{U} (t_f, t_i ) \hat{\rho}_i \hat{U}^\dagger (t_f, t_i) \, ,
\label{eq:densitymx1}
\end{align}
which has matrix elements given by
\begin{align}
	\langle q_f, \bQ_f | \hat{\rho}_f | q'_f, \bQ'_f \rangle \, .
\label{eq:densitymx2}
\end{align}
The {\it reduced} density matrix for $\hat{q}(t)$ 
is defined as the trace  
taken over states at the final time of \eqref{eq:densitymx2},
\begin{align}
	\rho_{\rm red}(q_f, q'_f) & \equiv \! \int \!\! d^NQ_f \langle q_f, \bQ_f | \hat{\rho}_f | q'_f, \bQ'_f = \bQ_f \rangle   \, .
\label{eq:densitymx4}
\end{align}
The reduced density matrix is the quantity that one computes expectation 
values of an operator $\hat{\cal O}$ through tracing over $q$ in the usual manner,
$\langle \hat{\cal O}  \rangle = {\rm Tr} ( \hat{\rho}_{\rm red} \hat{\cal O} )$.
Assuming that the initial states $\{ |q_i, \bQ_i \rangle \}$ are complete it follows
that we can put two insertions of the identity operator,
\begin{align}
	\hat{\mathbbold{1}} = \int dq_i \int d^N Q_i \, | q_i, \bQ_i \rangle \langle q_i, \bQ_i |
\end{align}
into \eqref{eq:densitymx4}
which yields
\begin{align}
	\rho_{\rm red}(q_f, q'_f) = {} & \int dq_i dq'_i \int d^N Q_i d^NQ'_i d^NQ_f  \nonumber \\
		& \times \langle q_f, \bQ_f | \hat{U}(t_f, t_i) | q_i, \bQ_i \rangle \nonumber \\
		& \times \langle q_i, \bQ_i | \hat{\rho}_i | q'_i, \bQ'_i \rangle \nonumber \\
		& \times \langle q'_i, \bQ'_i | \hat{U}^\dagger(t_f, t_i) | q'_f , \bQ'_f \rangle
\end{align}
upon using \eqref{eq:densitymx1} to express $\rho_f$ in terms of $\rho_i$. 
The matrix element in the second line
has a well known path integral representation in terms of the 
classical (conservative) action in \eqref{eq:claction1},
\begin{align}
	\langle q_f, \bQ_f | \hat{U}(t_f, t_i) | q_i, \bQ_i \rangle = \int_{q_i}^{q_f} \!\!\! {\cal D} q(t) \int_{\bQ_i}^{\bQ_f} \!\!\! {\cal D} \bQ(t) \, e^{i S[q, \bQ] / \hbar}  \,. \nn
\end{align}
Assuming for simplicity that the initial states are factorized, 
$|q_i, \bQ_i \rangle = |q_i \rangle \otimes | \bQ_i \rangle$ so that 
\begin{align}
	\langle q_i, \bQ_i | \hat{\rho}_i | q'_i, \bQ'_i \rangle = \rho_i^{(q)} (q_i, q'_i) \rho_i^{(\bQ)} (\bQ_i, \bQ'_i) \, ,
\end{align}
we find that the reduced density matrix can be written as
\begin{align}
	\rho_{\rm red}(q_f, q'_f) = {} & \int dq_i dq'_i \int_{q_i}^{q_f} {\hskip-0.1in} {\cal D}q(t) \int_{q'_i}^{q'_f} {\hskip-0.1in}  {\cal D} q'(t) \, 
	\rho_i^{(q)} (q_i, q'_i)
	\nonumber \\
		& \times  e^{i ( S_q[q] - S_q[q'] + S_{\rm infl} [q, q'] ) / \hbar } 
\label{eq:densitymx3}
\end{align}
where $S_{\rm infl}$ is called the influence action and determines 
the influence on the accessible variables $q$ 
of all the contributions from the integrated out $\bQ$ quantum 
degrees of freedom \cite{FeynmanVernon:AnnPhys24},
\begin{align}
	e^{i S_{\rm infl} [ q, q'] /\hbar }\equiv {} & \int d^NQ_i d^NQ'_i d^N Q_f \int_{\bQ_i}^{\bQ_f} {\hskip-0.1in} {\cal D} \bQ(t) \int_{\bQ'_i}^{\bQ_f} {\hskip-0.1in} {\cal D} \bQ'(t) \,  \nonumber  \\
		& \times \rho_i^{(\bQ)} (\bQ_i, \bQ'_i) \label{eq:inflaction1}  \\
		& \times e^{ i ( S_Q [\bQ] - S_Q[\bQ'] + S_{\rm int} [q, \bQ] - S_{\rm int}[q', \bQ'] ) / \hbar }  \nonumber
\end{align}
Notice that the insertion of two identity operators into the reduced 
density matrix and their corresponding path integral representations 
leads quite naturally to the appearance of a doubled set of degrees 
of freedom for the whole problem. That this happens is a result of the 
density matrix at the final time $\hat{\rho}_f$ being evolved by two 
unitary operators in \eqref{eq:densitymx1} that, in turn, is due to the 
structure of evolving a system from some initial state in a causal manner. 
This formulation of quantum theory is sometimes referred to as the 
``in-in'' formalism \cite{Schwinger:JMathPhys2, Keldysh:JEPT20, Jordan:PRD33, CalzettaHu:PRD35} 
because one keeps track of how operators of nonequilibrium quantum 
systems evolve from the ``in'' state to the ``in'' state (i.e., expectation values) 
as opposed to the ``out'' state (i.e., matrix elements and amplitudes).

In many systems where the path integrals can be computed, 
$S_{\rm infl}$ contains an imaginary part,
\begin{align}
	\big| e^{i S_{\rm infl}[q, q'] / \hbar}  \big| = e^{- {\rm Im} S_{\rm infl} [q, q'] / \hbar}
\label{eq:inflfunctional1}
\end{align}
that drives the accessible variables to decohere towards classicality. 
For example, if we take the $\{Q^I(t) \}_{I=1}^N$ to be harmonic oscillators
coupled linearly to $q$,
\begin{align}
	& S_Q [ \bQ] + S_{\rm int}[q, \bQ] \nonumber \\
	& ~~ =  \int_{t_i}^{t_f} \!\!\! dt \, \bigg\{ \frac{1}{2} M \dot{\bQ}^2 - \frac{1}{2} M \Omega \bQ^2 + \sum_{I=1}^N \lambda q Q^I(t) \bigg\},
\label{eq:qmosc1}
\end{align}
then it follows that \cite{CalzettaRouraVerdaguer:PhysicaA319, CalzettaHu}
\begin{align}
	{\rm Im} \, S_{\rm infl} [ q, q'] = {} & \frac{\lambda^2}{4} \int_{t_i}^{t_f} \!\!\! dt \, dt' \, q_-(t) q_-(t') \nonumber \\
		& {\hskip0.15in} \times \sum_{I=1}^N \big\langle \big\{ \hat{Q}^I(t), \hat{Q}^{I}(t') \big\} \big\rangle  
	\label{eq:imagInflAction1}
\end{align}
where the expectation value is taken in the (zero-mean, Gaussian) 
initial state $| \bQ_i \rangle$ of the $\{Q^I \}_{I=1}^N$. Here, 
$q_- (t) = q(t) - q'(t)$ is the quantum version of the ``$-$'' variable
that appears in the nonconservative classical mechanics formulation of
Secs.~\ref{sec:Review} and \ref{sec:ContinuumMechanics}. 
For a large class of initial states, the imaginary part of the influence
action is positive so that the norm of (\ref{eq:inflfunctional1})
decays super exponentially when $q_-$ increases thereby suppressing 
quantum fluctuations and inciting decoherence. As a result, the phase in 
the path integrand of (\ref{eq:densitymx3}) becomes peaked around 
configurations where $q_-$ is small, which causes the reduced density 
matrix to be (nearly) diagonal (see e.g., \cite{FeynmanVernon:AnnPhys24, CalzettaRouraVerdaguer:PhysicaA319, CalzettaHu}).

We can now take the $\hbar \to 0$ limit of the reduced density matrix
in (\ref{eq:densitymx3}) to find --- using the stationary phase approximation ---
that the dominant contribution to the reduced density matrix comes from\footnote{
The ${\cal O}(\hbar^{1/2})$ corrections come from quantum fluctuations 
that manifest as classical stochastic forces $\xi$ coupled to $q_-$ 
through $\int dt \, q_-(t) \xi(t)$. Following \cite{FeynmanVernon:AnnPhys24, CalzettaRouraVerdaguer:PhysicaA319}, 
for the oscillator example in \eqref{eq:qmosc1} and \eqref{eq:imagInflAction1}, 
the statistical two-point correlations, 
$\langle \xi(t) \xi(t') \rangle_\xi \propto \hbar \sum_{I=1}^N \langle \{ \hat{Q}^I (t), \hat{Q}^I (t') \} \rangle$, 
imply that $\xi  \sim \hbar^{1/2}$.
} 
\begin{align}
	{\cal S}[ q, q'] \equiv S_q[ q ] - S_q [q'] + {\rm Re} S_{\rm infl} [ q, q'] + {\cal O}(\hbar^{1/2})
\end{align}
When written out in terms of Lagrangians we find a familiar 
expression, namely (\ref{eq:newaction1}), 
\begin{align}
	{\cal S} [ q, q'] = \int_{t_i}^{t_f} \!\!\! dt \, \Big[ L(q, \dot{q}) - L(q', \dot{q}' ) + K (q, q', \dot{q}, \dot{q}' ) \Big]  \nonumber
\end{align}
where the real part of the influence action is the time integral of the 
nonconservative potential $K$.
Therefore, the classical limit of the quantum theory in non-equilibrium
is described by exactly the nonconservative action written down in 
\cite{Galley:2012hx} and \eqref{eq:newaction1}. Of course, in cases
where $S_{\rm infl}$ cannot be computed explicitly there still exists 
a formal expression for it given in \eqref{eq:inflaction1} and thus a relationship with $K$. 
The result of this appendix can also be found by using the more abstract ``in-in'' generating functional 
language to derive the nonconservative action ${\cal S}$ from the coarse-grained effective action
 (see e.g., \cite{CalzettaHu}) via a loop expansion in powers of $\hbar$.

This appendix has established that nonconservative classical mechanics is derivable from
a more complete fundamental quantum theory, provide there is 
sufficient decoherence that a classical limit can be reached.
See also recent work in \cite{kuwahara2013classical} who quantize the nonconservative Hamiltonian of \cite{Galley:2012hx} and find a relation with Thermo Field Dynamics. In addition, see
 work in \cite{Polonyi:2014rpa} for further connections between ${\cal S}$ and  the ``in-in''
formulation of non-equilibrium quantum field theories.

\bibliography{continuum-mechanics}

\begin{thebibliography}{74}%
\makeatletter
\providecommand \@ifxundefined [1]{%
 \@ifx{#1\undefined}
}%
\providecommand \@ifnum [1]{%
 \ifnum #1\expandafter \@firstoftwo
 \else \expandafter \@secondoftwo
 \fi
}%
\providecommand \@ifx [1]{%
 \ifx #1\expandafter \@firstoftwo
 \else \expandafter \@secondoftwo
 \fi
}%
\providecommand \natexlab [1]{#1}%
\providecommand \enquote  [1]{``#1''}%
\providecommand \bibnamefont  [1]{#1}%
\providecommand \bibfnamefont [1]{#1}%
\providecommand \citenamefont [1]{#1}%
\providecommand \href@noop [0]{\@secondoftwo}%
\providecommand \href [0]{\begingroup \@sanitize@url \@href}%
\providecommand \@href[1]{\@@startlink{#1}\@@href}%
\providecommand \@@href[1]{\endgroup#1\@@endlink}%
\providecommand \@sanitize@url [0]{\catcode `\\12\catcode `\$12\catcode
  `\&12\catcode `\#12\catcode `\^12\catcode `\_12\catcode `\%12\relax}%
\providecommand \@@startlink[1]{}%
\providecommand \@@endlink[0]{}%
\providecommand \url  [0]{\begingroup\@sanitize@url \@url }%
\providecommand \@url [1]{\endgroup\@href {#1}{\urlprefix }}%
\providecommand \urlprefix  [0]{URL }%
\providecommand \Eprint [0]{\href }%
\providecommand \doibase [0]{http://dx.doi.org/}%
\providecommand \selectlanguage [0]{\@gobble}%
\providecommand \bibinfo  [0]{\@secondoftwo}%
\providecommand \bibfield  [0]{\@secondoftwo}%
\providecommand \translation [1]{[#1]}%
\providecommand \BibitemOpen [0]{}%
\providecommand \bibitemStop [0]{}%
\providecommand \bibitemNoStop [0]{.\EOS\space}%
\providecommand \EOS [0]{\spacefactor3000\relax}%
\providecommand \BibitemShut  [1]{\csname bibitem#1\endcsname}%
\let\auto@bib@innerbib\@empty
\bibitem [{\citenamefont {Hamilton}(1834)}]{hamilton1834general}%
  \BibitemOpen
  \bibfield  {author} {\bibinfo {author} {\bibfnamefont {W.~R.}\ \bibnamefont
  {Hamilton}},\ }\href@noop {} {\bibfield  {journal} {\bibinfo  {journal}
  {Phil. Trans. Roy. Soc. Lon}\ ,\ \bibinfo {pages} {95}} (\bibinfo {year}
  {1834})}\BibitemShut {NoStop}%
\bibitem [{\citenamefont {{Goldstein}}\ \emph {et~al.}(2002)\citenamefont
  {{Goldstein}}, \citenamefont {{Poole}},\ and\ \citenamefont
  {{Safko}}}]{Goldstein}%
  \BibitemOpen
  \bibfield  {author} {\bibinfo {author} {\bibfnamefont {H.}~\bibnamefont
  {{Goldstein}}}, \bibinfo {author} {\bibfnamefont {C.}~\bibnamefont
  {{Poole}}}, \ and\ \bibinfo {author} {\bibfnamefont {J.}~\bibnamefont
  {{Safko}}},\ }\href@noop {} {\emph {\bibinfo {title} {{Classical
  mechanics}}}},\ \bibinfo {edition} {3rd}\ ed.\ (\bibinfo  {publisher}
  {Addison-Wesley},\ \bibinfo {year} {2002})\BibitemShut {NoStop}%
\bibitem [{\citenamefont {{Noether}}(1918)}]{Noether1918}%
  \BibitemOpen
  \bibfield  {author} {\bibinfo {author} {\bibfnamefont {E.}~\bibnamefont
  {{Noether}}},\ }\href@noop {} {\bibfield  {journal} {\bibinfo  {journal}
  {Nachr.~D.~Koenig.~Gesellsch.~D.~Wiss.~Zu Goettingen, Math-phys.~Klasse 1918:
  p.~235-237}\ ,\ \bibinfo {pages} {235}} (\bibinfo {year} {1918})}\BibitemShut
  {NoStop}%
\bibitem [{\citenamefont {Dirac}(1950)}]{dirac1950generalized}%
  \BibitemOpen
  \bibfield  {author} {\bibinfo {author} {\bibfnamefont {P.~A.~M.}\
  \bibnamefont {Dirac}},\ }\href@noop {} {\bibfield  {journal} {\bibinfo
  {journal} {Can. J. Math}\ }\textbf {\bibinfo {volume} {2}},\ \bibinfo {pages}
  {129} (\bibinfo {year} {1950})}\BibitemShut {NoStop}%
\bibitem [{\citenamefont {Calzetta}\ and\ \citenamefont
  {Hu}(2008)}]{CalzettaHu}%
  \BibitemOpen
  \bibfield  {author} {\bibinfo {author} {\bibfnamefont {E.~A.}\ \bibnamefont
  {Calzetta}}\ and\ \bibinfo {author} {\bibfnamefont {B.-L.~B.}\ \bibnamefont
  {Hu}},\ }\href@noop {} {\emph {\bibinfo {title} {Nonequilibrium Quantum Field
  Theory}}}\ (\bibinfo  {publisher} {Cambridge University Press},\ \bibinfo
  {year} {2008})\BibitemShut {NoStop}%
\bibitem [{\citenamefont {DeGroot}\ and\ \citenamefont
  {Mazur}(1984)}]{degroot1984non}%
  \BibitemOpen
  \bibfield  {author} {\bibinfo {author} {\bibfnamefont {S.}~\bibnamefont
  {DeGroot}}\ and\ \bibinfo {author} {\bibfnamefont {P.}~\bibnamefont
  {Mazur}},\ }\href {http://books.google.com/books?id=HFAIv43rlGkC} {\emph
  {\bibinfo {title} {Non-equilibrium Thermodynamics}}},\ Dover Books on Physics
  Series\ (\bibinfo  {publisher} {Dover Publications},\ \bibinfo {year}
  {1984})\BibitemShut {NoStop}%
\bibitem [{\citenamefont {Rayleigh}(1896)}]{rayleigh1896theory}%
  \BibitemOpen
  \bibfield  {author} {\bibinfo {author} {\bibfnamefont {J.~W. S.~B.}\
  \bibnamefont {Rayleigh}},\ }\href@noop {} {\emph {\bibinfo {title} {The
  theory of sound}}},\ Vol.~\bibinfo {volume} {2}\ (\bibinfo  {publisher}
  {Macmillan},\ \bibinfo {year} {1896})\BibitemShut {NoStop}%
\bibitem [{\citenamefont {{Virga}}(2014)}]{2014arXiv1406.6906V}%
  \BibitemOpen
  \bibfield  {author} {\bibinfo {author} {\bibfnamefont {E.~G.}\ \bibnamefont
  {{Virga}}},\ }\href@noop {} {\bibfield  {journal} {\bibinfo  {journal} {ArXiv
  e-prints}\ } (\bibinfo {year} {2014})},\ \Eprint
  {http://arxiv.org/abs/1406.6906} {arXiv:1406.6906 [math-ph]} \BibitemShut
  {NoStop}%
\bibitem [{\citenamefont {Bauer}(1931)}]{bauer1931dissipative}%
  \BibitemOpen
  \bibfield  {author} {\bibinfo {author} {\bibfnamefont {P.~S.}\ \bibnamefont
  {Bauer}},\ }\href@noop {} {\bibfield  {journal} {\bibinfo  {journal}
  {Proceedings of the National Academy of Sciences of the United States of
  America}\ }\textbf {\bibinfo {volume} {17}},\ \bibinfo {pages} {311}
  (\bibinfo {year} {1931})}\BibitemShut {NoStop}%
\bibitem [{\citenamefont {Bateman}(1931)}]{bateman1931dissipative}%
  \BibitemOpen
  \bibfield  {author} {\bibinfo {author} {\bibfnamefont {H.}~\bibnamefont
  {Bateman}},\ }\href@noop {} {\bibfield  {journal} {\bibinfo  {journal}
  {Physical Review}\ }\textbf {\bibinfo {volume} {38}},\ \bibinfo {pages} {815}
  (\bibinfo {year} {1931})}\BibitemShut {NoStop}%
\bibitem [{\citenamefont {Dreisigmeyer}\ and\ \citenamefont
  {Young}(2003)}]{DreisigmeyerYoung:2003}%
  \BibitemOpen
  \bibfield  {author} {\bibinfo {author} {\bibfnamefont {D.~W.}\ \bibnamefont
  {Dreisigmeyer}}\ and\ \bibinfo {author} {\bibfnamefont {P.~M.}\ \bibnamefont
  {Young}},\ }\href {http://stacks.iop.org/0305-4470/36/i=30/a=307} {\bibfield
  {journal} {\bibinfo  {journal} {Journal of Physics A: Mathematical and
  General}\ }\textbf {\bibinfo {volume} {36}},\ \bibinfo {pages} {8297}
  (\bibinfo {year} {2003})}\BibitemShut {NoStop}%
\bibitem [{\citenamefont {{Staruszkiewicz}}(1970)}]{1970AnP...480..362S}%
  \BibitemOpen
  \bibfield  {author} {\bibinfo {author} {\bibfnamefont {A.}~\bibnamefont
  {{Staruszkiewicz}}},\ }\href {\doibase 10.1002/andp.19704800404} {\bibfield
  {journal} {\bibinfo  {journal} {Annalen der Physik}\ }\textbf {\bibinfo
  {volume} {480}},\ \bibinfo {pages} {362} (\bibinfo {year}
  {1970})}\BibitemShut {NoStop}%
\bibitem [{\citenamefont {{Jaranowski}}\ and\ \citenamefont
  {{Sch{\"a}fer}}(1997)}]{1997PhRvD..55.4712J}%
  \BibitemOpen
  \bibfield  {author} {\bibinfo {author} {\bibfnamefont {P.}~\bibnamefont
  {{Jaranowski}}}\ and\ \bibinfo {author} {\bibfnamefont {G.}~\bibnamefont
  {{Sch{\"a}fer}}},\ }\href {\doibase 10.1103/PhysRevD.55.4712} {\bibfield
  {journal} {\bibinfo  {journal} {\prd}\ }\textbf {\bibinfo {volume} {55}},\
  \bibinfo {pages} {4712} (\bibinfo {year} {1997})}\BibitemShut {NoStop}%
\bibitem [{\citenamefont {Galley}(2013)}]{Galley:2012hx}%
  \BibitemOpen
  \bibfield  {author} {\bibinfo {author} {\bibfnamefont {C.~R.}\ \bibnamefont
  {Galley}},\ }\href {\doibase 10.1103/PhysRevLett.110.174301} {\bibfield
  {journal} {\bibinfo  {journal} {Phys. Rev. Lett.}\ }\textbf {\bibinfo
  {volume} {110}},\ \bibinfo {pages} {174301} (\bibinfo {year} {2013})},\
  \Eprint {http://arxiv.org/abs/1210.2745} {arXiv:1210.2745 [gr-qc]}
  \BibitemShut {NoStop}%
\bibitem [{\citenamefont {Machta}\ \emph {et~al.}(2013)\citenamefont {Machta},
  \citenamefont {Chachra}, \citenamefont {Transtrum},\ and\ \citenamefont
  {Sethna}}]{machta2013parameter}%
  \BibitemOpen
  \bibfield  {author} {\bibinfo {author} {\bibfnamefont {B.~B.}\ \bibnamefont
  {Machta}}, \bibinfo {author} {\bibfnamefont {R.}~\bibnamefont {Chachra}},
  \bibinfo {author} {\bibfnamefont {M.~K.}\ \bibnamefont {Transtrum}}, \ and\
  \bibinfo {author} {\bibfnamefont {J.~P.}\ \bibnamefont {Sethna}},\
  }\href@noop {} {\bibfield  {journal} {\bibinfo  {journal} {Science}\ }\textbf
  {\bibinfo {volume} {342}},\ \bibinfo {pages} {604} (\bibinfo {year}
  {2013})}\BibitemShut {NoStop}%
\bibitem [{\citenamefont {Burgess}(2007)}]{Burgess:EFT}%
  \BibitemOpen
  \bibfield  {author} {\bibinfo {author} {\bibfnamefont {C.~P.}\ \bibnamefont
  {Burgess}},\ }\href@noop {} {\bibfield  {journal} {\bibinfo  {journal} {Ann.
  Rev. Nucl. Part. Sci.}\ }\textbf {\bibinfo {volume} {57}},\ \bibinfo {pages}
  {329} (\bibinfo {year} {2007})},\ \Eprint
  {http://arxiv.org/abs/hep-th/0701053} {hep-th/0701053} \BibitemShut {NoStop}%
\bibitem [{\citenamefont {Manohar}(1997)}]{Manohar:1996cq}%
  \BibitemOpen
  \bibfield  {author} {\bibinfo {author} {\bibfnamefont {A.~V.}\ \bibnamefont
  {Manohar}},\ }\href {\doibase 10.1007/BFb0104294} {\bibfield  {journal}
  {\bibinfo  {journal} {Lect.Notes Phys.}\ }\textbf {\bibinfo {volume} {479}},\
  \bibinfo {pages} {311} (\bibinfo {year} {1997})},\ \Eprint
  {http://arxiv.org/abs/hep-ph/9606222} {arXiv:hep-ph/9606222 [hep-ph]}
  \BibitemShut {NoStop}%
\bibitem [{\citenamefont {Rothstein}(2002)}]{Rothstein:EFT1}%
  \BibitemOpen
  \bibfield  {author} {\bibinfo {author} {\bibfnamefont {I.~Z.}\ \bibnamefont
  {Rothstein}},\ }in\ \href@noop {} {\emph {\bibinfo {booktitle} {Prepared for
  Theoretical Advanced Study Institute in Elementary Particle Physics (TASI
  2002): Particle Physics and Cosmology: The Quest for Physics Beyond the
  Standard Model(s), Boulder, Colorado}}}\ (\bibinfo {year} {2002})\ \Eprint
  {http://arxiv.org/abs/hep-ph/0308266} {hep-ph/0308266} \BibitemShut {NoStop}%
\bibitem [{\citenamefont {Goldberger}(2007)}]{Goldberger:LesHouches}%
  \BibitemOpen
  \bibfield  {author} {\bibinfo {author} {\bibfnamefont {W.}~\bibnamefont
  {Goldberger}},\ }in\ \href@noop {} {\emph {\bibinfo {booktitle} {Paricle
  Physics and Cosmology: The Fabric of Spacetime, Volume LXXXVI: Lecture notes
  of the Les Houches Summer School}}},\ \bibinfo {editor} {edited by\ \bibinfo
  {editor} {\bibfnamefont {F.}~\bibnamefont {Bernardeau}}, \bibinfo {editor}
  {\bibfnamefont {C.}~\bibnamefont {Grojean}}, \ and\ \bibinfo {editor}
  {\bibfnamefont {J.}~\bibnamefont {Dalibard}}}\ (\bibinfo  {publisher}
  {Elsevier Science},\ \bibinfo {year} {2007})\ \Eprint
  {http://arxiv.org/abs/hep-ph/0701129} {hep-ph/0701129} \BibitemShut {NoStop}%
\bibitem [{\citenamefont {{Galley}}\ \emph {et~al.}()\citenamefont {{Galley}},
  \citenamefont {{Tsang}},\ and\ \citenamefont {{Stein}}}]{GalleyFuture1}%
  \BibitemOpen
  \bibfield  {author} {\bibinfo {author} {\bibfnamefont {C.~R.}\ \bibnamefont
  {{Galley}}}, \bibinfo {author} {\bibfnamefont {D.}~\bibnamefont {{Tsang}}}, \
  and\ \bibinfo {author} {\bibfnamefont {L.~C.}\ \bibnamefont {{Stein}}},\
  }\href@noop {} {}\bibinfo {note} {(in preparation)}\BibitemShut {NoStop}%
\bibitem [{\citenamefont {Wheeler}\ and\ \citenamefont
  {Feynman}(1949)}]{WheelerFeynman}%
  \BibitemOpen
  \bibfield  {author} {\bibinfo {author} {\bibfnamefont {J.~A.}\ \bibnamefont
  {Wheeler}}\ and\ \bibinfo {author} {\bibfnamefont {R.~P.}\ \bibnamefont
  {Feynman}},\ }\href@noop {} {\bibfield  {journal} {\bibinfo  {journal} {Rev.
  Mod. Phys.}\ }\textbf {\bibinfo {volume} {21}},\ \bibinfo {pages} {425}
  (\bibinfo {year} {1949})}\BibitemShut {NoStop}%
\bibitem [{\citenamefont {Weiss}(1999)}]{Weiss}%
  \BibitemOpen
  \bibfield  {author} {\bibinfo {author} {\bibfnamefont {U.}~\bibnamefont
  {Weiss}},\ }\href@noop {} {\emph {\bibinfo {title} {Quantum Dissipative
  Systems}}},\ \bibinfo {edition} {2nd}\ ed.\ (\bibinfo  {publisher} {World
  Scientific},\ \bibinfo {year} {1999})\BibitemShut {NoStop}%
\bibitem [{\citenamefont {Scheck}(1996)}]{Scheck}%
  \BibitemOpen
  \bibfield  {author} {\bibinfo {author} {\bibfnamefont {F.}~\bibnamefont
  {Scheck}},\ }\href@noop {} {\emph {\bibinfo {title} {Mechanics: From Newton's
  Laws to Deterministic Chaos}}},\ \bibinfo {edition} {3rd}\ ed.\ (\bibinfo
  {publisher} {Springer, Berlin},\ \bibinfo {year} {1996})\BibitemShut
  {NoStop}%
\bibitem [{\citenamefont {Goldberger}\ and\ \citenamefont
  {Rothstein}(2006)}]{Goldberger:2005cd}%
  \BibitemOpen
  \bibfield  {author} {\bibinfo {author} {\bibfnamefont {W.~D.}\ \bibnamefont
  {Goldberger}}\ and\ \bibinfo {author} {\bibfnamefont {I.~Z.}\ \bibnamefont
  {Rothstein}},\ }\href {\doibase 10.1103/PhysRevD.73.104030} {\bibfield
  {journal} {\bibinfo  {journal} {Phys.~Rev.}\ }\textbf {\bibinfo {volume}
  {D73}},\ \bibinfo {pages} {104030} (\bibinfo {year} {2006})},\ \Eprint
  {http://arxiv.org/abs/hep-th/0511133} {arXiv:hep-th/0511133 [hep-th]}
  \BibitemShut {NoStop}%
\bibitem [{\citenamefont {Galley}\ \emph {et~al.}(2010)\citenamefont {Galley},
  \citenamefont {Leibovich},\ and\ \citenamefont {Rothstein}}]{Galley:2010es}%
  \BibitemOpen
  \bibfield  {author} {\bibinfo {author} {\bibfnamefont {C.~R.}\ \bibnamefont
  {Galley}}, \bibinfo {author} {\bibfnamefont {A.~K.}\ \bibnamefont
  {Leibovich}}, \ and\ \bibinfo {author} {\bibfnamefont {I.~Z.}\ \bibnamefont
  {Rothstein}},\ }\href {\doibase 10.1103/PhysRevLett.105.094802} {\bibfield
  {journal} {\bibinfo  {journal} {Phys. Rev. Lett.}\ }\textbf {\bibinfo
  {volume} {105}},\ \bibinfo {pages} {094802} (\bibinfo {year} {2010})},\
  \Eprint {http://arxiv.org/abs/1005.2617} {arXiv:1005.2617 [gr-qc]}
  \BibitemShut {NoStop}%
\bibitem [{\citenamefont {Jackson}(1999)}]{jackson_classical_1999}%
  \BibitemOpen
  \bibfield  {author} {\bibinfo {author} {\bibfnamefont {J.~D.}\ \bibnamefont
  {Jackson}},\ }\href {http://cdsweb.cern.ch/record/490457} {\emph {\bibinfo
  {title} {Classical electrodynamics}}},\ \bibinfo {edition} {3rd}\ ed.\
  (\bibinfo  {publisher} {Wiley},\ \bibinfo {address} {New York, {NY}},\
  \bibinfo {year} {1999})\BibitemShut {NoStop}%
\bibitem [{\citenamefont {Landau}\ and\ \citenamefont
  {Lifshitz}(1986{\natexlab{a}})}]{landau1986fieldtheory}%
  \BibitemOpen
  \bibfield  {author} {\bibinfo {author} {\bibfnamefont {L.}~\bibnamefont
  {Landau}}\ and\ \bibinfo {author} {\bibfnamefont {E.}~\bibnamefont
  {Lifshitz}},\ }\href@noop {} {\emph {\bibinfo {title} {Classical Theory of
  Fields}}},\ Course of Theoretical Physics\ (\bibinfo  {publisher}
  {Butterworth-Heinemann},\ \bibinfo {year} {1986})\BibitemShut {NoStop}%
\bibitem [{\citenamefont {Koga}(2004)}]{Koga:PRE70}%
  \BibitemOpen
  \bibfield  {author} {\bibinfo {author} {\bibfnamefont {J.}~\bibnamefont
  {Koga}},\ }\href {\doibase 10.1103/PhysRevE.70.046502} {\bibfield  {journal}
  {\bibinfo  {journal} {Phys.~Rev.~E}\ }\textbf {\bibinfo {volume} {70}},\
  \bibinfo {pages} {046502} (\bibinfo {year} {2004})}\BibitemShut {NoStop}%
\bibitem [{\citenamefont {{Ober-Bl{\"o}baum}}\ \emph
  {et~al.}(2013)\citenamefont {{Ober-Bl{\"o}baum}}, \citenamefont {{Tao}},
  \citenamefont {{Cheng}}, \citenamefont {{Owhadi}},\ and\ \citenamefont
  {{Marsden}}}]{OberBlobaum2013498}%
  \BibitemOpen
  \bibfield  {author} {\bibinfo {author} {\bibfnamefont {S.}~\bibnamefont
  {{Ober-Bl{\"o}baum}}}, \bibinfo {author} {\bibfnamefont {M.}~\bibnamefont
  {{Tao}}}, \bibinfo {author} {\bibfnamefont {M.}~\bibnamefont {{Cheng}}},
  \bibinfo {author} {\bibfnamefont {H.}~\bibnamefont {{Owhadi}}}, \ and\
  \bibinfo {author} {\bibfnamefont {J.~E.}\ \bibnamefont {{Marsden}}},\ }\href
  {\doibase 10.1016/j.jcp.2013.02.006} {\bibfield  {journal} {\bibinfo
  {journal} {Journal of Computational Physics}\ }\textbf {\bibinfo {volume}
  {242}},\ \bibinfo {pages} {498 } (\bibinfo {year} {2013})},\ \Eprint
  {http://arxiv.org/abs/1103.1859} {arXiv:1103.1859 [math.NA]} \BibitemShut
  {NoStop}%
\bibitem [{\citenamefont {Misner}\ \emph {et~al.}(1973)\citenamefont {Misner},
  \citenamefont {Thorne},\ and\ \citenamefont {Wheeler}}]{MTW}%
  \BibitemOpen
  \bibfield  {author} {\bibinfo {author} {\bibfnamefont {C.~W.}\ \bibnamefont
  {Misner}}, \bibinfo {author} {\bibfnamefont {K.~S.}\ \bibnamefont {Thorne}},
  \ and\ \bibinfo {author} {\bibfnamefont {J.~A.}\ \bibnamefont {Wheeler}},\
  }\href@noop {} {\emph {\bibinfo {title} {Gravitation}}}\ (\bibinfo
  {publisher} {Freeman, San Francisco},\ \bibinfo {year} {1973})\BibitemShut
  {NoStop}%
\bibitem [{\citenamefont {Soper}(2008)}]{Soper}%
  \BibitemOpen
  \bibfield  {author} {\bibinfo {author} {\bibfnamefont {D.~E.}\ \bibnamefont
  {Soper}},\ }\href@noop {} {\emph {\bibinfo {title} {Classical Field
  Theory}}}\ (\bibinfo  {publisher} {Dover Publications, New York},\ \bibinfo
  {year} {2008})\BibitemShut {NoStop}%
\bibitem [{\citenamefont {Galley}\ and\ \citenamefont
  {Leibovich}(2012)}]{Galley:2012qs}%
  \BibitemOpen
  \bibfield  {author} {\bibinfo {author} {\bibfnamefont {C.~R.}\ \bibnamefont
  {Galley}}\ and\ \bibinfo {author} {\bibfnamefont {A.~K.}\ \bibnamefont
  {Leibovich}},\ }\href {\doibase 10.1103/PhysRevD.86.044029} {\bibfield
  {journal} {\bibinfo  {journal} {Phys.~Rev.}\ }\textbf {\bibinfo {volume}
  {D86}},\ \bibinfo {pages} {044029} (\bibinfo {year} {2012})},\ \Eprint
  {http://arxiv.org/abs/1205.3842} {arXiv:1205.3842 [gr-qc]} \BibitemShut
  {NoStop}%
\bibitem [{\citenamefont {Kevrekidis}(2014)}]{Kevrekidis:PhysRevA.89.010102}%
  \BibitemOpen
  \bibfield  {author} {\bibinfo {author} {\bibfnamefont {P.~G.}\ \bibnamefont
  {Kevrekidis}},\ }\href {\doibase 10.1103/PhysRevA.89.010102} {\bibfield
  {journal} {\bibinfo  {journal} {Phys.~Rev.~A}\ }\textbf {\bibinfo {volume}
  {89}},\ \bibinfo {pages} {010102} (\bibinfo {year} {2014})}\BibitemShut
  {NoStop}%
\bibitem [{\citenamefont {Dauxois}\ and\ \citenamefont
  {Peyrard}(2006)}]{dauxois2006physics}%
  \BibitemOpen
  \bibfield  {author} {\bibinfo {author} {\bibfnamefont {T.}~\bibnamefont
  {Dauxois}}\ and\ \bibinfo {author} {\bibfnamefont {M.}~\bibnamefont
  {Peyrard}},\ }\href@noop {} {\emph {\bibinfo {title} {Physics of solitons}}}\
  (\bibinfo  {publisher} {Cambridge University Press},\ \bibinfo {year}
  {2006})\BibitemShut {NoStop}%
\bibitem [{\citenamefont {Malomed}(2002)}]{malomed2002variational}%
  \BibitemOpen
  \bibfield  {author} {\bibinfo {author} {\bibfnamefont {B.~A.}\ \bibnamefont
  {Malomed}},\ }\href@noop {} {\bibfield  {journal} {\bibinfo  {journal}
  {Progress in Optics}\ }\textbf {\bibinfo {volume} {43}},\ \bibinfo {pages}
  {71} (\bibinfo {year} {2002})}\BibitemShut {NoStop}%
\bibitem [{\citenamefont {Malomed}(2006)}]{malomed2006soliton}%
  \BibitemOpen
  \bibfield  {author} {\bibinfo {author} {\bibfnamefont {B.~A.}\ \bibnamefont
  {Malomed}},\ }\href@noop {} {\emph {\bibinfo {title} {Soliton management in
  periodic systems}}}\ (\bibinfo  {publisher} {Springer},\ \bibinfo {year}
  {2006})\BibitemShut {NoStop}%
\bibitem [{\citenamefont {{Andersson}}\ and\ \citenamefont
  {{Comer}}(2013)}]{Andersson2013}%
  \BibitemOpen
  \bibfield  {author} {\bibinfo {author} {\bibfnamefont {N.}~\bibnamefont
  {{Andersson}}}\ and\ \bibinfo {author} {\bibfnamefont {G.~L.}\ \bibnamefont
  {{Comer}}},\ }\href@noop {} {\bibfield  {journal} {\bibinfo  {journal} {ArXiv
  e-prints}\ } (\bibinfo {year} {2013})},\ \Eprint
  {http://arxiv.org/abs/1306.3345} {arXiv:1306.3345 [gr-qc]} \BibitemShut
  {NoStop}%
\bibitem [{\citenamefont {Prix}(2004)}]{Prix:2002jn}%
  \BibitemOpen
  \bibfield  {author} {\bibinfo {author} {\bibfnamefont {R.}~\bibnamefont
  {Prix}},\ }\href {\doibase 10.1103/PhysRevD.69.043001} {\bibfield  {journal}
  {\bibinfo  {journal} {Phys.~Rev.}\ }\textbf {\bibinfo {volume} {D69}},\
  \bibinfo {pages} {043001} (\bibinfo {year} {2004})},\ \Eprint
  {http://arxiv.org/abs/physics/0209024} {arXiv:physics/0209024 [physics]}
  \BibitemShut {NoStop}%
\bibitem [{\citenamefont {Endlich}\ \emph {et~al.}(2013)\citenamefont
  {Endlich}, \citenamefont {Nicolis}, \citenamefont {Porto},\ and\
  \citenamefont {Wang}}]{Endlich:2012vt}%
  \BibitemOpen
  \bibfield  {author} {\bibinfo {author} {\bibfnamefont {S.}~\bibnamefont
  {Endlich}}, \bibinfo {author} {\bibfnamefont {A.}~\bibnamefont {Nicolis}},
  \bibinfo {author} {\bibfnamefont {R.~A.}\ \bibnamefont {Porto}}, \ and\
  \bibinfo {author} {\bibfnamefont {J.}~\bibnamefont {Wang}},\ }\href {\doibase
  10.1103/PhysRevD.88.105001} {\bibfield  {journal} {\bibinfo  {journal}
  {Phys.~Rev.}\ }\textbf {\bibinfo {volume} {D88}},\ \bibinfo {pages} {105001}
  (\bibinfo {year} {2013})},\ \Eprint {http://arxiv.org/abs/1211.6461}
  {arXiv:1211.6461 [hep-th]} \BibitemShut {NoStop}%
\bibitem [{\citenamefont {Grozdanov}\ and\ \citenamefont
  {Polonyi}(2013)}]{Grozdanov:2013dba}%
  \BibitemOpen
  \bibfield  {author} {\bibinfo {author} {\bibfnamefont {S.}~\bibnamefont
  {Grozdanov}}\ and\ \bibinfo {author} {\bibfnamefont {J.}~\bibnamefont
  {Polonyi}},\ }\href@noop {} {\  (\bibinfo {year} {2013})},\ \Eprint
  {http://arxiv.org/abs/1305.3670} {arXiv:1305.3670 [hep-th]} \BibitemShut
  {NoStop}%
\bibitem [{\citenamefont {Ray}(1972)}]{ray1972lagrangian}%
  \BibitemOpen
  \bibfield  {author} {\bibinfo {author} {\bibfnamefont {J.~R.}\ \bibnamefont
  {Ray}},\ }\href@noop {} {\bibfield  {journal} {\bibinfo  {journal}
  {J.~Math.~Phys.}\ }\textbf {\bibinfo {volume} {13}},\ \bibinfo {pages} {1451}
  (\bibinfo {year} {1972})}\BibitemShut {NoStop}%
\bibitem [{\citenamefont {{Eckart}}(1960)}]{Eckart1960}%
  \BibitemOpen
  \bibfield  {author} {\bibinfo {author} {\bibfnamefont {C.}~\bibnamefont
  {{Eckart}}},\ }\href {\doibase 10.1063/1.1706053} {\bibfield  {journal}
  {\bibinfo  {journal} {Physics of Fluids}\ }\textbf {\bibinfo {volume} {3}},\
  \bibinfo {pages} {421} (\bibinfo {year} {1960})}\BibitemShut {NoStop}%
\bibitem [{\citenamefont {{Katz}}(1961)}]{Katz1961}%
  \BibitemOpen
  \bibfield  {author} {\bibinfo {author} {\bibfnamefont {S.}~\bibnamefont
  {{Katz}}},\ }\href {\doibase 10.1063/1.1706330} {\bibfield  {journal}
  {\bibinfo  {journal} {Physics of Fluids}\ }\textbf {\bibinfo {volume} {4}},\
  \bibinfo {pages} {345} (\bibinfo {year} {1961})}\BibitemShut {NoStop}%
\bibitem [{\citenamefont {{Seliger}}\ and\ \citenamefont
  {{Whitham}}(1968)}]{Seliger1968}%
  \BibitemOpen
  \bibfield  {author} {\bibinfo {author} {\bibfnamefont {R.~L.}\ \bibnamefont
  {{Seliger}}}\ and\ \bibinfo {author} {\bibfnamefont {G.~B.}\ \bibnamefont
  {{Whitham}}},\ }\href {\doibase 10.1098/rspa.1968.0103} {\bibfield  {journal}
  {\bibinfo  {journal} {Royal Society of London Proceedings Series A}\ }\textbf
  {\bibinfo {volume} {305}},\ \bibinfo {pages} {1} (\bibinfo {year}
  {1968})}\BibitemShut {NoStop}%
\bibitem [{\citenamefont {Morrison}(1998)}]{Morrison:1998zz}%
  \BibitemOpen
  \bibfield  {author} {\bibinfo {author} {\bibfnamefont {P.}~\bibnamefont
  {Morrison}},\ }\href {\doibase 10.1103/RevModPhys.70.467} {\bibfield
  {journal} {\bibinfo  {journal} {Rev. Mod. Phys.}\ }\textbf {\bibinfo {volume}
  {70}},\ \bibinfo {pages} {467} (\bibinfo {year} {1998})}\BibitemShut
  {NoStop}%
\bibitem [{\citenamefont {Landau}\ and\ \citenamefont
  {Lifshitz}(1986{\natexlab{b}})}]{landau1986fluids}%
  \BibitemOpen
  \bibfield  {author} {\bibinfo {author} {\bibfnamefont {L.}~\bibnamefont
  {Landau}}\ and\ \bibinfo {author} {\bibfnamefont {E.}~\bibnamefont
  {Lifshitz}},\ }\href@noop {} {\emph {\bibinfo {title} {Fluid Mechanics}}},\
  Course of Theoretical Physics\ (\bibinfo  {publisher}
  {Butterworth-Heinemann},\ \bibinfo {year} {1986})\BibitemShut {NoStop}%
\bibitem [{\citenamefont {Batchelor}(1967)}]{Batchelor}%
  \BibitemOpen
  \bibfield  {author} {\bibinfo {author} {\bibfnamefont {G.~K.}\ \bibnamefont
  {Batchelor}},\ }\href@noop {} {\emph {\bibinfo {title} {An Introduction to
  Fluid Dynamics}}}\ (\bibinfo  {publisher} {Cambridge University Press,
  Cambridge},\ \bibinfo {year} {1967})\BibitemShut {NoStop}%
\bibitem [{\citenamefont {{Andersson}}\ and\ \citenamefont
  {{Comer}}(2010)}]{2010RSPSA.466.1373A}%
  \BibitemOpen
  \bibfield  {author} {\bibinfo {author} {\bibfnamefont {N.}~\bibnamefont
  {{Andersson}}}\ and\ \bibinfo {author} {\bibfnamefont {G.~L.}\ \bibnamefont
  {{Comer}}},\ }\href {\doibase 10.1098/rspa.2009.0423} {\bibfield  {journal}
  {\bibinfo  {journal} {Royal Society of London Proceedings Series A}\ }\textbf
  {\bibinfo {volume} {466}},\ \bibinfo {pages} {1373} (\bibinfo {year}
  {2010})},\ \Eprint {http://arxiv.org/abs/0908.1707} {arXiv:0908.1707
  [physics.flu-dyn]} \BibitemShut {NoStop}%
\bibitem [{\citenamefont {Kim}\ and\ \citenamefont
  {Karrila}(1991)}]{kim1991microhydrodynamics}%
  \BibitemOpen
  \bibfield  {author} {\bibinfo {author} {\bibfnamefont {S.}~\bibnamefont
  {Kim}}\ and\ \bibinfo {author} {\bibfnamefont {S.~J.}\ \bibnamefont
  {Karrila}},\ }\href@noop {} {\emph {\bibinfo {title} {Microhydrodynamics}}}\
  (\bibinfo  {publisher} {Butterworth-Heinemann New York:},\ \bibinfo {year}
  {1991})\BibitemShut {NoStop}%
\bibitem [{\citenamefont {Landau}\ and\ \citenamefont
  {Lifshitz}(1986{\natexlab{c}})}]{landau1986elasticity}%
  \BibitemOpen
  \bibfield  {author} {\bibinfo {author} {\bibfnamefont {L.}~\bibnamefont
  {Landau}}\ and\ \bibinfo {author} {\bibfnamefont {E.}~\bibnamefont
  {Lifshitz}},\ }\href@noop {} {\emph {\bibinfo {title} {Theory of
  Elasticity}}},\ Course of Theoretical Physics\ (\bibinfo  {publisher}
  {Butterworth-Heinemann},\ \bibinfo {year} {1986})\BibitemShut {NoStop}%
\bibitem [{\citenamefont {{Oswald}}(2009)}]{2009rheo.book.....O}%
  \BibitemOpen
  \bibfield  {author} {\bibinfo {author} {\bibfnamefont {P.}~\bibnamefont
  {{Oswald}}},\ }\href@noop {} {\emph {\bibinfo {title} {Rheophysics}}}\
  (\bibinfo  {publisher} {Cambridge University Press, Cambridge, UK},\ \bibinfo
  {year} {2009})\BibitemShut {NoStop}%
\bibitem [{\citenamefont {{Keramidas Charidakos}}\ \emph
  {et~al.}(2014)\citenamefont {{Keramidas Charidakos}}, \citenamefont
  {{Lingam}}, \citenamefont {{Morrison}}, \citenamefont {{White}},\ and\
  \citenamefont {{Wurm}}}]{2014arXiv1407.3884K}%
  \BibitemOpen
  \bibfield  {author} {\bibinfo {author} {\bibfnamefont {I.}~\bibnamefont
  {{Keramidas Charidakos}}}, \bibinfo {author} {\bibfnamefont {M.}~\bibnamefont
  {{Lingam}}}, \bibinfo {author} {\bibfnamefont {P.~J.}\ \bibnamefont
  {{Morrison}}}, \bibinfo {author} {\bibfnamefont {R.~L.}\ \bibnamefont
  {{White}}}, \ and\ \bibinfo {author} {\bibfnamefont {A.}~\bibnamefont
  {{Wurm}}},\ }\href@noop {} {\bibfield  {journal} {\bibinfo  {journal} {ArXiv
  e-prints}\ } (\bibinfo {year} {2014})},\ \Eprint
  {http://arxiv.org/abs/1407.3884} {arXiv:1407.3884 [physics.plasm-ph]}
  \BibitemShut {NoStop}%
\bibitem [{\citenamefont {{Tsang}}\ \emph {et~al.}()\citenamefont {{Tsang}},
  \citenamefont {{Galley}}, \citenamefont {{Stein}},\ and\ \citenamefont
  {{Gourgouliatos}}}]{MHDfuture}%
  \BibitemOpen
  \bibfield  {author} {\bibinfo {author} {\bibfnamefont {D.}~\bibnamefont
  {{Tsang}}}, \bibinfo {author} {\bibfnamefont {C.~R.}\ \bibnamefont
  {{Galley}}}, \bibinfo {author} {\bibfnamefont {L.~C.}\ \bibnamefont
  {{Stein}}}, \ and\ \bibinfo {author} {\bibfnamefont {K.~N.}\ \bibnamefont
  {{Gourgouliatos}}},\ }\href@noop {} {}\bibinfo {note} {(in
  prepartion)}\BibitemShut {NoStop}%
\bibitem [{\citenamefont {Onsager}\ and\ \citenamefont
  {Machlup}(1953)}]{PhysRev.91.1505}%
  \BibitemOpen
  \bibfield  {author} {\bibinfo {author} {\bibfnamefont {L.}~\bibnamefont
  {Onsager}}\ and\ \bibinfo {author} {\bibfnamefont {S.}~\bibnamefont
  {Machlup}},\ }\href {\doibase 10.1103/PhysRev.91.1505} {\bibfield  {journal}
  {\bibinfo  {journal} {Phys.~Rev.}\ }\textbf {\bibinfo {volume} {91}},\
  \bibinfo {pages} {1505} (\bibinfo {year} {1953})}\BibitemShut {NoStop}%
\bibitem [{\citenamefont {Landau}\ and\ \citenamefont
  {Lifshitz}(1957)}]{LL:hydroflucs}%
  \BibitemOpen
  \bibfield  {author} {\bibinfo {author} {\bibfnamefont {L.}~\bibnamefont
  {Landau}}\ and\ \bibinfo {author} {\bibfnamefont {E.}~\bibnamefont
  {Lifshitz}},\ }\href@noop {} {\bibfield  {journal} {\bibinfo  {journal}
  {JETP}\ }\textbf {\bibinfo {volume} {32}},\ \bibinfo {pages} {618} (\bibinfo
  {year} {1957})},\ \bibinfo {note} {{[{\it Sov.~Phys.~JETP} {\bf 5}, 512
  (1957)]}}\BibitemShut {NoStop}%
\bibitem [{\citenamefont {Martin}\ \emph {et~al.}(1973)\citenamefont {Martin},
  \citenamefont {Siggia},\ and\ \citenamefont {Rose}}]{Martin:1973zz}%
  \BibitemOpen
  \bibfield  {author} {\bibinfo {author} {\bibfnamefont {P.}~\bibnamefont
  {Martin}}, \bibinfo {author} {\bibfnamefont {E.}~\bibnamefont {Siggia}}, \
  and\ \bibinfo {author} {\bibfnamefont {H.}~\bibnamefont {Rose}},\ }\href
  {\doibase 10.1103/PhysRevA.8.423} {\bibfield  {journal} {\bibinfo  {journal}
  {Phys.~Rev.}\ }\textbf {\bibinfo {volume} {A8}},\ \bibinfo {pages} {423}
  (\bibinfo {year} {1973})}\BibitemShut {NoStop}%
\bibitem [{\citenamefont {De~Dominicis}\ and\ \citenamefont
  {Peliti}(1978)}]{DeDominicis:1977fw}%
  \BibitemOpen
  \bibfield  {author} {\bibinfo {author} {\bibfnamefont {C.}~\bibnamefont
  {De~Dominicis}}\ and\ \bibinfo {author} {\bibfnamefont {L.}~\bibnamefont
  {Peliti}},\ }\href {\doibase 10.1103/PhysRevB.18.353} {\bibfield  {journal}
  {\bibinfo  {journal} {Phys.~Rev.}\ }\textbf {\bibinfo {volume} {B18}},\
  \bibinfo {pages} {353} (\bibinfo {year} {1978})}\BibitemShut {NoStop}%
\bibitem [{\citenamefont {{Eyink}}(1996)}]{Eyink:1996}%
  \BibitemOpen
  \bibfield  {author} {\bibinfo {author} {\bibfnamefont {G.~L.}\ \bibnamefont
  {{Eyink}}},\ }\href {\doibase 10.1103/PhysRevE.54.3419} {\bibfield  {journal}
  {\bibinfo  {journal} {Phys.~Rev.}\ }\textbf {\bibinfo {volume} {E54}},\
  \bibinfo {pages} {3419} (\bibinfo {year} {1996})}\BibitemShut {NoStop}%
\bibitem [{\citenamefont {{Eyink}}(1998)}]{Eyink:1998}%
  \BibitemOpen
  \bibfield  {author} {\bibinfo {author} {\bibfnamefont {G.~L.}\ \bibnamefont
  {{Eyink}}},\ }\href {\doibase 10.1143/PTPS.130.77} {\bibfield  {journal}
  {\bibinfo  {journal} {Progress of Theoretical Physics Supplement}\ }\textbf
  {\bibinfo {volume} {130}},\ \bibinfo {pages} {77} (\bibinfo {year}
  {1998})}\BibitemShut {NoStop}%
\bibitem [{\citenamefont {Kovtun}\ \emph {et~al.}(2014)\citenamefont {Kovtun},
  \citenamefont {Moore},\ and\ \citenamefont {Romatschke}}]{Kovtun:2014hpa}%
  \BibitemOpen
  \bibfield  {author} {\bibinfo {author} {\bibfnamefont {P.}~\bibnamefont
  {Kovtun}}, \bibinfo {author} {\bibfnamefont {G.~D.}\ \bibnamefont {Moore}}, \
  and\ \bibinfo {author} {\bibfnamefont {P.}~\bibnamefont {Romatschke}},\
  }\href {\doibase 10.1007/JHEP07(2014)123} {\bibfield  {journal} {\bibinfo
  {journal} {JHEP}\ }\textbf {\bibinfo {volume} {1407}},\ \bibinfo {pages}
  {123} (\bibinfo {year} {2014})},\ \Eprint {http://arxiv.org/abs/1405.3967}
  {arXiv:1405.3967 [hep-ph]} \BibitemShut {NoStop}%
\bibitem [{\citenamefont {{Israel}}\ and\ \citenamefont
  {{Stewart}}(1979)}]{1979AnPhy.118..341I}%
  \BibitemOpen
  \bibfield  {author} {\bibinfo {author} {\bibfnamefont {W.}~\bibnamefont
  {{Israel}}}\ and\ \bibinfo {author} {\bibfnamefont {J.~M.}\ \bibnamefont
  {{Stewart}}},\ }\href {\doibase 10.1016/0003-4916(79)90130-1} {\bibfield
  {journal} {\bibinfo  {journal} {Annals of Physics}\ }\textbf {\bibinfo
  {volume} {118}},\ \bibinfo {pages} {341} (\bibinfo {year}
  {1979})}\BibitemShut {NoStop}%
\bibitem [{\citenamefont {Spivak}(1970)}]{spivak1970comprehensive}%
  \BibitemOpen
  \bibfield  {author} {\bibinfo {author} {\bibfnamefont {M.}~\bibnamefont
  {Spivak}},\ }\href {http://books.google.com/books?id=HFTvAAAAMAAJ} {\emph
  {\bibinfo {title} {A comprehensive introduction to differential geometry}}},\
  \bibinfo {series} {A Comprehensive Introduction to Differential Geometry}\
  No.\ \bibinfo {number} {v. 1}\ (\bibinfo  {publisher} {Brandeis University},\
  \bibinfo {year} {1970})\BibitemShut {NoStop}%
\bibitem [{\citenamefont {Schutz}(1980)}]{schutz1980geometrical}%
  \BibitemOpen
  \bibfield  {author} {\bibinfo {author} {\bibfnamefont {B.}~\bibnamefont
  {Schutz}},\ }\href {http://books.google.com/books?id=HAPMB2e643kC} {\emph
  {\bibinfo {title} {Geometrical Methods of Mathematical Physics}}}\ (\bibinfo
  {publisher} {Cambridge University Press},\ \bibinfo {year}
  {1980})\BibitemShut {NoStop}%
\bibitem [{\citenamefont {Wald}(1984)}]{wald}%
  \BibitemOpen
  \bibfield  {author} {\bibinfo {author} {\bibfnamefont {R.}~\bibnamefont
  {Wald}},\ }\href {http://books.google.com/books?id=9S-hzg6-moYC} {\emph
  {\bibinfo {title} {General Relativity}}}\ (\bibinfo  {publisher} {University
  of Chicago Press},\ \bibinfo {year} {1984})\BibitemShut {NoStop}%
\bibitem [{\citenamefont {{Thiffeault}}(2001)}]{2001JPhA...34.5875T}%
  \BibitemOpen
  \bibfield  {author} {\bibinfo {author} {\bibfnamefont {J.-L.}\ \bibnamefont
  {{Thiffeault}}},\ }\href {\doibase 10.1088/0305-4470/34/29/309} {\bibfield
  {journal} {\bibinfo  {journal} {Journal of Physics A Mathematical General}\
  }\textbf {\bibinfo {volume} {34}},\ \bibinfo {pages} {5875} (\bibinfo {year}
  {2001})},\ \Eprint {http://arxiv.org/abs/nlin/0102038} {nlin/0102038}
  \BibitemShut {NoStop}%
\bibitem [{\citenamefont {{Oldroyd}}(1950)}]{1950RSPSA.200..523O}%
  \BibitemOpen
  \bibfield  {author} {\bibinfo {author} {\bibfnamefont {J.~G.}\ \bibnamefont
  {{Oldroyd}}},\ }\href {\doibase 10.1098/rspa.1950.0035} {\bibfield  {journal}
  {\bibinfo  {journal} {Royal Society of London Proceedings Series A}\ }\textbf
  {\bibinfo {volume} {200}},\ \bibinfo {pages} {523} (\bibinfo {year}
  {1950})}\BibitemShut {NoStop}%
\bibitem [{\citenamefont {Feynman}\ and\ \citenamefont
  {Vernon}(1963)}]{FeynmanVernon:AnnPhys24}%
  \BibitemOpen
  \bibfield  {author} {\bibinfo {author} {\bibfnamefont {R.}~\bibnamefont
  {Feynman}}\ and\ \bibinfo {author} {\bibfnamefont {F.}~\bibnamefont
  {Vernon}},\ }\href@noop {} {\bibfield  {journal} {\bibinfo  {journal}
  {Ann.~Phys.}\ }\textbf {\bibinfo {volume} {24}},\ \bibinfo {pages} {118}
  (\bibinfo {year} {1963})}\BibitemShut {NoStop}%
\bibitem [{\citenamefont {Schwinger}(1961)}]{Schwinger:JMathPhys2}%
  \BibitemOpen
  \bibfield  {author} {\bibinfo {author} {\bibfnamefont {J.}~\bibnamefont
  {Schwinger}},\ }\href@noop {} {\bibfield  {journal} {\bibinfo  {journal}
  {J.~Math.~Phys.}\ }\textbf {\bibinfo {volume} {2}},\ \bibinfo {pages} {407}
  (\bibinfo {year} {1961})}\BibitemShut {NoStop}%
\bibitem [{\citenamefont {Keldysh}(1964)}]{Keldysh:JEPT20}%
  \BibitemOpen
  \bibfield  {author} {\bibinfo {author} {\bibfnamefont {L.~V.}\ \bibnamefont
  {Keldysh}},\ }\href@noop {} {\bibfield  {journal} {\bibinfo  {journal} {Zh.
  Eksp. Teor. Fiz.}\ }\textbf {\bibinfo {volume} {47}},\ \bibinfo {pages}
  {1515} (\bibinfo {year} {1964})},\ \bibinfo {note} {[English translation,
  Sov. Phys. JEPT {\bf 20}, 1018 (1965)]}\BibitemShut {NoStop}%
\bibitem [{\citenamefont {Jordan}(1986)}]{Jordan:PRD33}%
  \BibitemOpen
  \bibfield  {author} {\bibinfo {author} {\bibfnamefont {R.~D.}\ \bibnamefont
  {Jordan}},\ }\href@noop {} {\bibfield  {journal} {\bibinfo  {journal}
  {Phys.~Rev.}\ }\textbf {\bibinfo {volume} {D33}},\ \bibinfo {pages} {444}
  (\bibinfo {year} {1986})}\BibitemShut {NoStop}%
\bibitem [{\citenamefont {Calzetta}\ and\ \citenamefont
  {Hu}(1987)}]{CalzettaHu:PRD35}%
  \BibitemOpen
  \bibfield  {author} {\bibinfo {author} {\bibfnamefont {E.}~\bibnamefont
  {Calzetta}}\ and\ \bibinfo {author} {\bibfnamefont {B.~L.}\ \bibnamefont
  {Hu}},\ }\href@noop {} {\bibfield  {journal} {\bibinfo  {journal}
  {Phys.~Rev.}\ }\textbf {\bibinfo {volume} {D35}},\ \bibinfo {pages} {495}
  (\bibinfo {year} {1987})}\BibitemShut {NoStop}%
\bibitem [{\citenamefont {Calzetta}\ \emph {et~al.}(2003)\citenamefont
  {Calzetta}, \citenamefont {Roura},\ and\ \citenamefont
  {Verdaguer}}]{CalzettaRouraVerdaguer:PhysicaA319}%
  \BibitemOpen
  \bibfield  {author} {\bibinfo {author} {\bibfnamefont {E.}~\bibnamefont
  {Calzetta}}, \bibinfo {author} {\bibfnamefont {A.}~\bibnamefont {Roura}}, \
  and\ \bibinfo {author} {\bibfnamefont {E.}~\bibnamefont {Verdaguer}},\
  }\href@noop {} {\bibfield  {journal} {\bibinfo  {journal} {Physica A
  (Amsterdam)}\ }\textbf {\bibinfo {volume} {319}},\ \bibinfo {pages} {188}
  (\bibinfo {year} {2003})}\BibitemShut {NoStop}%
\bibitem [{\citenamefont {Kuwahara}\ \emph {et~al.}(2013)\citenamefont
  {Kuwahara}, \citenamefont {Nakamura},\ and\ \citenamefont
  {Yamanaka}}]{kuwahara2013classical}%
  \BibitemOpen
  \bibfield  {author} {\bibinfo {author} {\bibfnamefont {Y.}~\bibnamefont
  {Kuwahara}}, \bibinfo {author} {\bibfnamefont {Y.}~\bibnamefont {Nakamura}},
  \ and\ \bibinfo {author} {\bibfnamefont {Y.}~\bibnamefont {Yamanaka}},\
  }\href@noop {} {\bibfield  {journal} {\bibinfo  {journal} {Physics Letters
  A}\ }\textbf {\bibinfo {volume} {377}},\ \bibinfo {pages} {3102} (\bibinfo
  {year} {2013})}\BibitemShut {NoStop}%
\bibitem [{\citenamefont {Polonyi}(2014)}]{Polonyi:2014rpa}%
  \BibitemOpen
  \bibfield  {author} {\bibinfo {author} {\bibfnamefont {J.}~\bibnamefont
  {Polonyi}},\ }\href@noop {} {\  (\bibinfo {year} {2014})},\ \Eprint
  {http://arxiv.org/abs/1407.6526} {arXiv:1407.6526 [hep-th]} \BibitemShut
  {NoStop}%
\end{thebibliography}%

\end{document}